\documentclass[11pt, a4paper]{article} 

\usepackage[english]{babel}
\usepackage[utf8]{inputenc}
\usepackage[T1]{fontenc}
\usepackage{lmodern}
\usepackage{microtype}
\usepackage{csquotes} 

\usepackage{geometry}
\usepackage{setspace}
\usepackage{titlesec}
\titlespacing*{\section}{0pt}{1ex plus 0.5ex minus .2ex}{0.5ex plus .1ex}
\titlespacing*{\subsection}{0pt}{1ex plus 0.3ex minus .1ex}{0.5ex plus .1ex}

\usepackage{amsmath}
\usepackage{amssymb}
\usepackage{amsfonts}
\usepackage{amsthm}
\usepackage{siunitx}
\DeclareSIUnit{\molar}{M}
\DeclareSIUnit{\Molar}{M}

\usepackage{float} 

\usepackage{graphicx}
\usepackage{xcolor}
\definecolor{pastelblue}{rgb}{0.0, 0.5, 1.0}
\definecolor{redc}{rgb}{0.9,0.2,0.3}

\usepackage{tikz}
\usepackage{pgfplots}
\pgfplotsset{compat=1.17}
\usepackage{soul} 

\usepackage{booktabs}
\usepackage{multirow}
\usepackage{arydshln}

\usepackage{enumitem}

\usepackage{caption}
\usepackage{subcaption}
\DeclareCaptionLabelFormat{boldpipe}{\textbf{#1~#2|}\ }
\usepackage{authblk}
\usepackage{datetime}

\makeatletter
\renewcommand{\@makefnmark}{\textsuperscript{\fnsymbol{footnote}}}
\newcommand{\EqualContribMark}{\textsuperscript{$\ddagger$}}

\usepackage[
  backend=biber,
  citestyle=numeric,
  natbib=true, 
  bibstyle=numeric,
  sorting=none,
  giveninits=true,
  uniquename=init,
  maxbibnames=3,  
  minbibnames=3,
  doi=true,
  url=false,
  isbn=false,
  eprint=false,
  date=year
]{biblatex}
\usepackage[colorlinks=true, citecolor=pastelblue, linkcolor=redc, urlcolor=redc]{hyperref}

\usepackage[most]{tcolorbox}
\newtcolorbox{notabox}[1]{
  colback=blue!5,
  colframe=blue!80!black,
  fonttitle=\bfseries,
  title={#1},
  breakable=true
}

\makeatother 

\title{Data-Driven Modeling of IRCU Patient Flow in the COVID-19 Pandemic.} 

\author[1]{Ana Carmen Navas-Ortega\EqualContribMark}
\author[1,2]{José Antonio Sánchez-Martínez\EqualContribMark}
\author[1,2]{Paula García-Flores} 
\author[1,2]{Concepción Morales-García}
\author[3]{Rene Fabregas} 

\affil[1]{Department of Pneumology, University Hospital Virgen de Las Nieves, Granada, Spain.}
\affil[2]{Biosanitary Research Institute of Granada-Ibs, Granada, Spain.}
\affil[3]{Department of Applied Mathematics and Modeling Nature (MNat) Research Unit, Faculty of Sciences, University of Granada, Granada, Spain.} 

\affil[ ]{\textbf{Corresponding Author}: Concepción Morales-García, \texttt{concepcion.morales.sspa@juntadeandalucia.es} and Rene Fabregas, \texttt{rfabregas@ugr.es}.}

\date{\today}

\begin{document}
\maketitle
\footnotetext[1]{\(\ddagger\) These authors contributed equally.}
\vspace{-1cm}
\begin{abstract}
\textbf{Background:} Intermediate Respiratory Care Units (IRCUs) play a pivotal role between general wards and Intensive Care Units (ICUs), particularly during crises like the COVID-19 pandemic. Their effectiveness is influenced by structure, protocols, and clinical staffing expertise. This study aimed to evaluate the \emph{clinical outcomes and operational dynamics of a newly established IRCU featuring a specialized staffing model during the COVID-19 pandemic in Spain}.

\textbf{Methods:} A prospective cohort study was conducted at the UHVN IRCU (Granada, Spain) between April and August 2021. Adult patients (n=249) admitted primarily for COVID-19 respiratory failure were included. Data on demographics, Non-Invasive Ventilation (NIV) use, \emph{length of stay (LOS)}, and outcomes (ICU transfer, exitus, recovery) were collected. Statistical analyses compared outcomes stratified by NIV status. A calibrated \emph{Ordinary Differential Equation (ODE)} compartmental model and an empirical LOS-based convolution model were developed to simulate patient flow dynamics under various scenarios, including admission surges and varying care efficiency.

\textbf{Results:} The median age was 51 years; 31\% (n=77) required NIV. Patients requiring NIV were significantly older (median 61 vs 42 years, p<0.001). Overall, 8\% (n=20) required subsequent ICU transfer and 3\% (n=7) experienced in-IRCU exitus. Notably, among the 172 patients managed without NIV, \textbf{no ICU transfers or deaths occurred}. Of the 77 high-risk patients requiring NIV, \textbf{68\% recovered within the IRCU without ICU escalation}. \emph{ODE modeling} accurately reflected aggregate outcomes and demonstrated significant system strain during simulated admission surges, partially mitigated by enhanced recovery efficiency. \emph{LOS-based modeling} yielded consistent peak occupancy estimates.

\textbf{Conclusion:} This IRCU, characterized by specialized clinical staffing including dedicated physicians and trained nurses, \textbf{demonstrated} effective management of severe COVID-19 respiratory failure. High recovery rates \textbf{were achieved}, particularly among NIV patients, potentially alleviating ICU pressure. Dynamic modeling confirmed the unit's vulnerability to admission surges but highlighted the positive impact of care efficiency. These findings \textbf{highlight the positive outcomes observed in this} appropriately structured and staffed IRCU, \textbf{supporting the potential value} of such units in pandemic response and respiratory care.
\end{abstract}


\section{Introduction.}
\label{sec:Introduction} 

The effective management of severe respiratory failure represents a cornerstone of modern critical care medicine \citep{Rochwerg2017}, a challenge dramatically underscored by the unprecedented pressures of the recent COVID-19 pandemic \citep{wax2020practical, whittle2020resp, Matthay2019ARDSReview}. Within the complex architecture of tiered hospital care, Intermediate Respiratory Care Units (IRCUs) have emerged as a critical component \citep{petty1967, petty1975}, strategically positioned to bridge the gap between the intensive care unit (ICU) and the general ward \citep{torres2005, nasraway1998, Cheng1999, LopezPadilla2022, nava1998, plate2017, mediano2023, lopezjardon2024utility}. Historically developed to provide focused monitoring and non-invasive respiratory support (NIV) \citep{petty1967,petty1975, krieger1990, Nava2009Lancet, Antonelli1998NEJM}, such as bilevel or continuous positive airway pressure (BiPAP/CPAP) and high-flow nasal oxygen (HFNO) \citep{Rochwerg2017, Esquinas2016, nishimura2016highflow}, IRCUs proved indispensable during the pandemic surge \citep{lujan2024multidisciplinary,caballero2022}, functioning both "upstream" to prevent ICU admissions and "downstream" to facilitate timely ICU discharge \citep{grosgurin2021, matutevillacis2021role, suarezcuartin2021clinical, Scala2017}. The potential benefits, including substantial reductions in the need for invasive mechanical ventilation \citep{lindenauer2014outcomes, Hilbert2001NIVPneumonia, Ferrer2003NIVCOPD} and associated healthcare costs \citep{heilifrades2019cost, Antonelli1998NEJM}, are well-recognized \citep{torres2005, capuzzo2014hospital, Prin2014}.

Despite the accelerated proliferation of IRCUs and accumulating evidence of their utility \citep{Sala2009,lopezjardon2024utility, galdeanolozano2022effectiveness}, a critical knowledge gap persists regarding the specific determinants of their success \citep{plate2017, Scala2018NIVReview}. While structural factors and patient selection criteria are often discussed \citep{nasraway1998, torres2005, Ambrosino2011RespCareUnits}, the \textbf{potentially significant impact of dedicated and specialized clinical staffing models,}
including physicians and nurses with advanced respiratory care competencies \citep{cabestregarcia2023respiratory, Rose2007, Multz1998ClosedICU}, remains inadequately quantified. Extensive evidence links general nurse staffing levels and education to improved patient outcomes \citep{Aiken2014, needleman2011nurse, Griffiths2018, Kane2007NurseStaffingReview}, and specialized physician involvement in critical care is known to be beneficial \cite{Pronovost2002ICUStaffing}. However, the specific impact of targeted \textbf{multidisciplinary clinical expertise} 
in the complex, high-stakes environment of an IRCU---managing sophisticated respiratory support \citep{Esquinas2016, Hill2010NIVReview}, identifying subtle signs of deterioration \citep{davies2018british, Vincent2010Deterioration}, and enacting timely interventions through coordinated care \citep{nasraway1998, Carlucci2001NIVFailurePredictors}---has not been systematically evaluated in terms of its contribution relative to other factors \citep{lopezjardon2024utility, cabestregarcia2023respiratory, Corrado2011FutureIRCUs}, representing a significant lacuna in our understanding of optimal intermediate care delivery.

Central to this investigation is the hypothesis that the operationalization of an IRCU, \textbf{characterized by dedicated staffing including clinicians (both physicians and nurses) possessing specialized training} in advanced respiratory support techniques \citep{Esquinas2016,Multz1998ClosedICU}, \textbf{is associated with}
superior clinical outcomes \citep{lindenauer2014outcomes, Garpestad2007NIVSuccessFactors} and optimized patient flow \citep{Hall2013, litvak2011, Proudlove2008QueuingModels}, even amidst the exacerbated demands of a pandemic \citep{grosgurin2021, matutevillacis2021role}. This study, therefore, embarks on a pioneering investigation to evaluate this premise. Our primary objective is to analyze the clinical effectiveness and operational dynamics of a newly instituted IRCU at the University Hospital Virgen de las Nieves (UHVN), Granada, during the initial five months of its operation amidst the COVID-19 pandemic (April-August 2021). This unit featured 24/7 respiratory physician coverage and nursing staff who received specific training (Fig.~\ref{fig:Timeline}, \ref{fig:NursingProcess}). We aimed to quantify patient outcomes, including length of stay and ICU transfer rates \citep{heilifrades2019cost, suarezcuartin2021clinical}, and to elucidate the performance characteristics of this unit \citep{nava1998, heilifrades2019cost}, thereby providing crucial, data-driven insights into the \textbf{outcomes observed in a setting implementing this model of specialized clinical staffing} 
\citep{cabestregarcia2023respiratory, sanchezmartinez2022development, Kleinpell2008APNOutcomes, Pronovost2002ICUStaffing}.

The principal contribution of this work lies in its multifaceted approach, representing a significant methodological advance over prior descriptive accounts \citep{krieger1990, grosgurin2021, Nava2000EuropeSurvey}. We present not only a detailed statistical analysis of patient characteristics and outcomes (Section~\ref{sec:Results_Statistical}), stratified by key risk factors such as the requirement for NIV, but also introduce a novel quantitative framework based on dynamic systems modeling \citep{Diekmann2013, Brandeau2004, Alban2020}. This dual approach offers mechanistic insights into the temporal evolution of patient cohorts within the IRCU \citep{Wang2020, Kermack1927EpidemicModel}, allowing for the simulation of system behavior under varying conditions \citep{Robinson2014, Jacobson2006} and the identification of critical parameters influencing operational efficiency and patient fate \citep{Zhang2021, SensitivityAnalysisBook2008, Banks1989}. By explicitly linking a specific care delivery model---characterized by \textbf{dedicated staffing and enhanced clinical competency (exemplified by targeted nursing training)}---to observed clinical results and modeled system dynamics, our study provides \textbf{a systematic evaluation within this specific context,} 
directly addressing the aforementioned knowledge gap and offering \textbf{valuable evidence regarding the outcomes associated with implementing} such staffing models in intermediate respiratory care \citep{cabestregarcia2023respiratory, Oner2021, Multz1998ClosedICU}.

Herein, we first detail the clinical characteristics and outcomes of the 249 patients managed within the UHVN IRCU during the study period, employing comprehensive statistical analysis to identify key determinants of patient trajectory (Section~\ref{sec:Results_Statistical}). Subsequently, we introduce and apply a compartmental Ordinary Differential Equation (ODE) model (Section~\ref{sec:dynamic_modeling}) designed to capture the intricate dynamics of patient flow between non-NIV and NIV states and towards final outcomes (ICU transfer, exitus, recovery), exploring system behavior through simulation under various operational scenarios. This is complemented by an alternative occupancy modeling approach grounded in the empirical Length-of-Stay distribution (Section~\ref{sec:convolution_modeling}). Finally, in the Discussion (Section~\ref{sec:Discussion}) and Conclusion (Section~\ref{sec:Conclusions}), we elaborate on the implications of our findings for clinical practice, healthcare policy, and future research directions, contextualizing our results within the broader landscape of critical and intermediate care medicine.

\section{Results.}
\label{sec:Results}
\subsection{Overall Cohort Characteristics and Clinical Outcomes.}
\label{sec:Results_Statistical}
\begin{figure}[H]
    \centering
    \includegraphics[width=\textwidth]{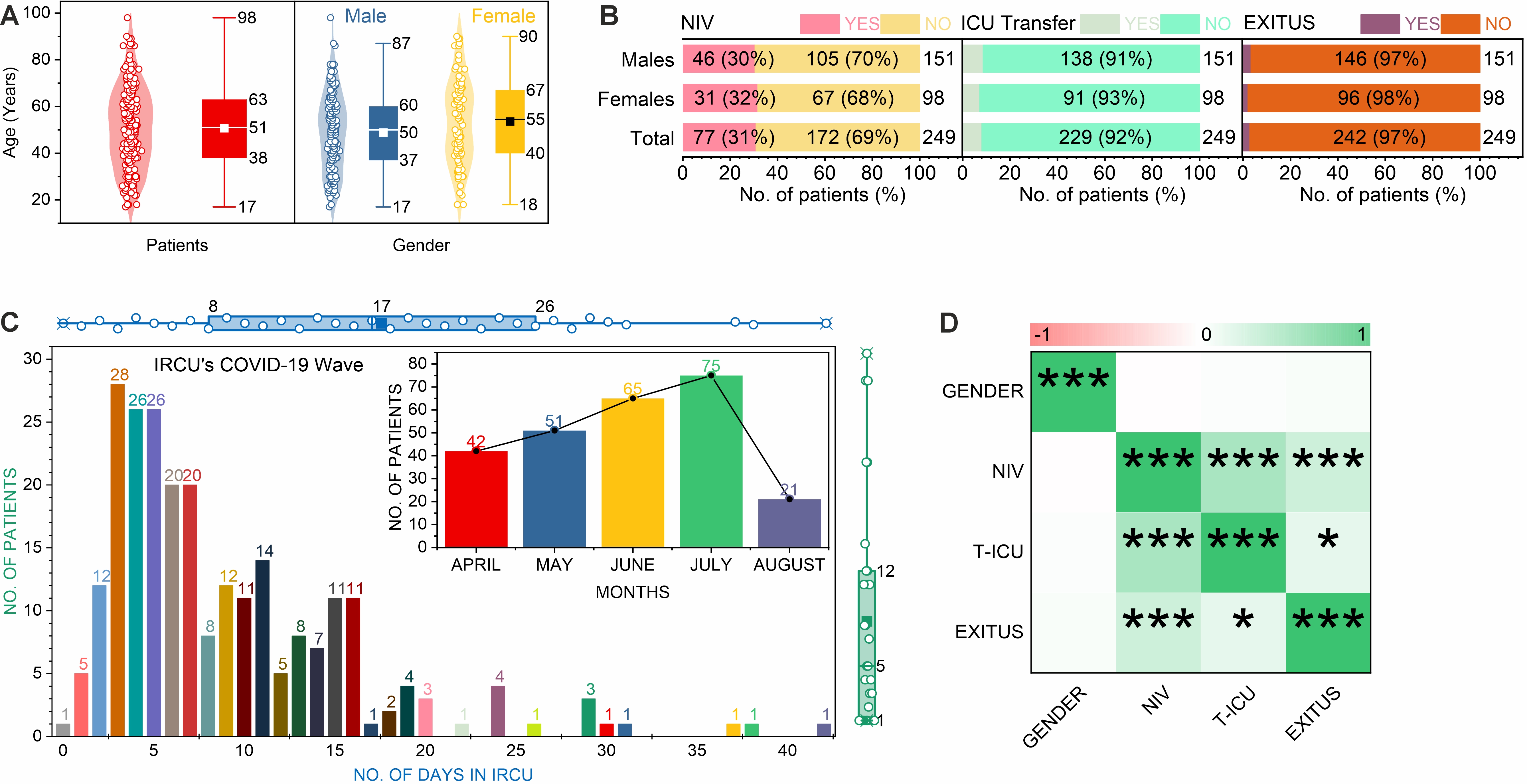} 
    \caption{ 
        \textbf{Patient characteristics and clinical outcomes in the IRCU (n=249).}
        (\textbf{A}) Age distribution overall (median 51, IQR 38-63) and by gender (Male median 50, IQR 37-60; Female median 55, IQR 40-67). Violin plots show distribution; inner boxes show median/IQRs.
        (\textbf{B}) Distribution of NIV utilization, ICU transfer, and Exitus for the total cohort, stratified by gender. Bars show percentage (N). Key totals: NIV Yes = 77 (31\%), ICU Transfer Yes = 20 (8\%), Exitus Yes = 7 (3\%). Note potential inconsistency in absolute counts within gender rows for NIV category vs the reported Total N=77 for NIV Yes.
        (\textbf{C}) Distribution of length of stay (LoS) in the IRCU (days) (median 5, IQR 3-8). Inset: Monthly admissions (Apr-Aug 2021).
        (\textbf{D}) Correlation matrix heatmap for binary clinical variables. Asterisks denote statistical significance (*** \textit{p}<0.001, * \textit{p}<0.05).
    }
    \label{fig:Figure1} 
    \vskip-0.5cm
\end{figure}

A total of 249 patients admitted to the Intermediate Respiratory Care Unit (IRCU) between April and August 2021 were included in the analysis. The cohort exhibited a median age of 51 years (Interquartile Range [IQR]: 38--63 years), with ages spanning from 17 to 98 years (Figure~\ref{fig:Figure1}A, left panel). Stratification by gender revealed a statistically significant difference in age distribution (Mann-Whitney U = 3405, \textit{p} = 0.023), with female patients (n=98) presenting a higher median age (55 years, IQR: 40--67) compared to their male counterparts (n=151; median: 50 years, IQR: 37--60) (Fig.~\ref{fig:Figure1}A, right panel).

\begin{figure}[H]
    \centering
    \includegraphics[width=0.95\textwidth]{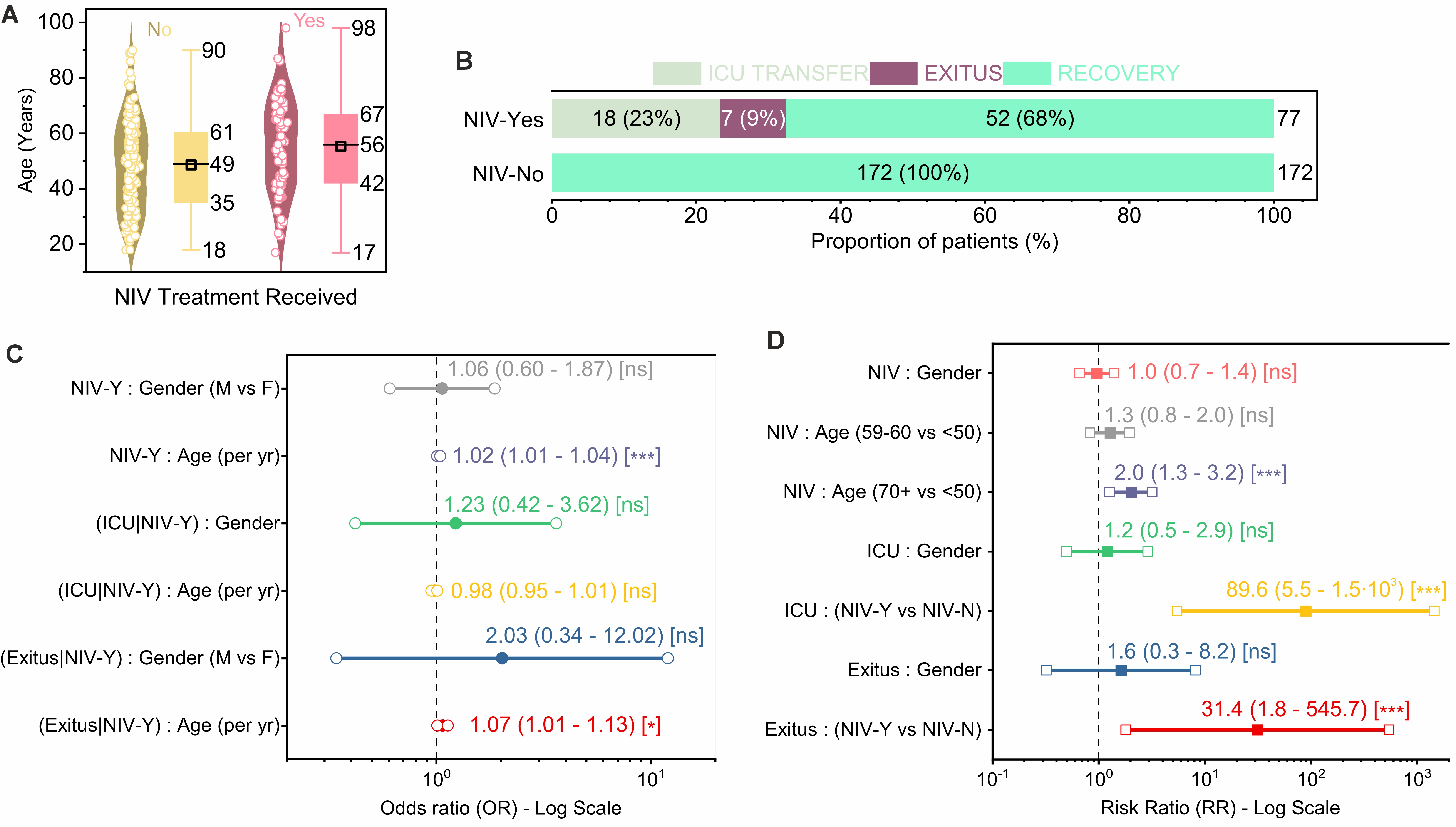} 
    \caption[Age by NIV, Outcomes by NIV, OR, RR]{%
         \textbf{Analysis stratified by NIV treatment status and multivariable associations.}
         (\textbf{A}) Age distribution stratified by NIV treatment status (No NIV: n=172, median 42, IQR 35-49; NIV Yes: n=77, median 61, IQR 56-67).
         (\textbf{B}) Distribution of final, mutually exclusive outcomes stratified by NIV status. For NIV-Yes (n=77): ICU Transfer=23\% (n=18), Exitus=9\% (n=7), Recovery=68\% (n=52). For NIV-No (n=172): Recovery=100\% (n=172). Bars show percentage (N). Outcomes defined for mutual exclusivity: ICU Transfer, then Exitus (no prior ICU), then Recovery.
         (\textbf{C}) Forest plot of adjusted Odds Ratios (OR) with 95\% CI from logistic regression (adjusted for Age and Gender where applicable). Ref group for Gender is Female.
         (\textbf{D}) Forest plot of crude Risk Ratios (RR) with 95\% CI comparing risks between groups (Gender [Male vs Female ref.], Age Groups [<50 ref.], NIV Status [Yes vs No ref.]). Note high RRs for NIV vs No NIV reflect confounding by indication. Significance levels: ns (\textit{p}\(\geq\)0.05), * (\textit{p}<0.05), *** (\textit{p}<0.001).
    }
    \label{fig:Figure2} 
\end{figure}

Regarding clinical pathways within the entire cohort, 77 patients (31\%) utilized Non-Invasive Ventilation (NIV), while 172 (69\%) did not. Subsequent transfer to the Intensive Care Unit (ICU) was required for 8\% (n=20) of patients, and exitus occurred in 3\% (n=7) during the observed IRCU pathway (Fig.~\ref{fig:Figure1}B, Total rows). While absolute counts stratified by gender in Fig.~\ref{fig:Figure1}B suggest broadly similar outcome patterns, formal statistical tests confirmed no significant gender association for NIV utilization, ICU transfer, or exitus (see Fig.~\ref{fig:Figure2}C and D; all \textit{p}>0.05 [ns]).

The duration of IRCU stay showed considerable variability (Fig.~\ref{fig:Figure1}C). The median length of stay (LoS) was 5 days (IQR: 1--12 days), with most stays concentrated below 10 days, although a subset required prolonged care exceeding 40 days. Monthly admission data fluctuated, peaking in June 2021 (n=75 admissions), potentially reflecting regional COVID-19 pandemic activity (Fig.~\ref{fig:Figure1}C inset). Correlation analysis (Fig.~\ref{fig:Figure1}D) identified expected significant positive associations (\textit{p}<0.001 [***]) between ICU transfer (T-ICU) and exitus, and between NIV utilization and both ICU transfer and exitus, suggesting higher severity among patients requiring NIV or ICU care. Gender was not significantly correlated with these variables (\textit{p}>0.05).


Stratifying the cohort by NIV treatment status revealed critical clinical differences. Patients requiring NIV (n=77) were significantly older (median age 61, IQR: 56--67) compared to those managed without NIV (n=172; median age 42, IQR: 35--49) (Fig.~\ref{fig:Figure2}A). This age difference strongly predicted NIV use; adjusted logistic regression showed a significant increase in the odds of receiving NIV per year of age (OR = 1.02, 95\% CI: 1.01--1.04 [***], Fig.~\ref{fig:Figure2}C), and crude risk analysis indicated that patients aged \(\geq\)70 years had double the risk compared to those <50 (RR = 2.0, 95\% CI: 1.3--3.2 [***], Fig.~\ref{fig:Figure2}D). As noted previously, gender was not significantly associated with NIV utilization (Fig.~\ref{fig:Figure2}C and D; [ns]).

Final outcome distributions differed markedly between the two groups (Fig.~\ref{fig:Figure2}B). The 172 patients managed without NIV experienced favorable outcomes, with no ICU transfers or exitus events recorded (100\% recovery within these defined categories). In contrast, the 77 higher-risk patients requiring NIV had the following mutually exclusive outcomes: ICU transfer (n=18, 23\%), exitus (without prior ICU transfer) (n=7, 9\%), and recovery (n=52, 68\%). The resulting high crude Risk Ratios comparing NIV users to non-users for ICU transfer (RR = 89.6, 95\% CI: 5.5--1464, \textit{p}<0.001 [***]) and exitus (RR = 31.4, 95\% CI: 1.8--546, \textit{p}<0.001 [***]) (Fig.~\ref{fig:Figure2}D) must be interpreted in light of significant \textit{confounding by indication}. Adjusted analyses confirmed that older age remained a significant risk factor for exitus specifically within the NIV group (OR = 1.07 per year, 95\% CI: 1.01--1.13, \textit{p}=0.013 [*], Fig.~\ref{fig:Figure2}C), while gender showed no significant association with outcomes in this subgroup either (Fig.~\ref{fig:Figure2}C and D; [ns]). Crucially, the successful recovery of 68\% of NIV-treated patients without ICU escalation highlights the IRCU's effectiveness in managing this high-risk population.


\subsection{Data-Driven Modeling.}
\label{sec:dynamic_modeling}

The preceding statistical analysis reveals significant heterogeneity in patient characteristics and clinical trajectories within the UHVN IRCU cohort, highlighting key associations such as older age with Non-Invasive Ventilation (NIV) requirement and markedly different outcomes for patients stratified by NIV status (Section~\ref{sec:Results_Statistical}). While informative, these static analyses offer limited insight into the underlying temporal dynamics governing patient flow, the interplay between admission patterns and care pathways, or the potential impact of changing operational efficiencies. To achieve a more mechanistic understanding and develop a quantitative framework, we adopted a compartmental modeling approach based on Ordinary Differential Equations (ODEs), a technique commonly used in epidemiology and healthcare operations \cite{Zhang2021}. 

We model the patient population within the IRCU using a system of coupled ODEs tracking distinct patient states over time \(t\). The state variables represent the number of patients in each compartment: \(X(t)\), corresponding to patients admitted to the IRCU not initially requiring NIV; \(Y(t)\), representing patients within the IRCU currently receiving NIV; \(Z(t)\), denoting the cumulative number of patients transferred from the IRCU to the ICU; \(W(t)\), indicating the cumulative number of patients experiencing mortality within the IRCU pathway (without prior ICU transfer from the IRCU); and \(R(t)\), which stands for the cumulative number of patients recovered and discharged from the IRCU.

The temporal evolution of these compartments is governed by the following system of ODEs:
\begin{align}
    \frac{dX}{dt} &= A(t) - (\alpha + \rho + \eta + \nu)X(t), \label{eq:IRCU_admissions_en} \\
    \frac{dY}{dt} &= \alpha X(t) - \big[\gamma(1-\gamma_0) + \theta(t) + \varepsilon(1-\varepsilon_0)\big]Y(t), \label{eq:NIV_state_en} \\
    \frac{dZ}{dt} &= \gamma(1-\gamma_0)Y(t) + \eta X(t) , \label{eq:ICU_transfers_en} \\
    \frac{dW}{dt} &= \varepsilon(1-\varepsilon_0)Y(t) + \nu X(t), \label{eq:Exitus_cont_en} \\
    \frac{dR}{dt} &= \theta(t)Y(t) + \rho X(t). \label{eq:Recovered_en}
\end{align}
Here, \(A(t)\) is the time-dependent admission rate. For non-NIV patients (X), \(\alpha\) is the rate of NIV initiation, \(\rho\) is the recovery rate, \(\eta\) is the direct ICU transfer rate, and \(\nu\) is the mortality rate. For NIV patients (Y), \(\gamma\) is the ICU transfer rate, \(\theta(t)\) is the potentially time-dependent recovery rate, and \(\varepsilon\) is the mortality rate (\(\gamma_0\) and \(\varepsilon_0\) are adjustment factors, assumed zero). \(Z(t)\), \(W(t)\), and \(R(t)\) track cumulative outcomes. We assume \(Z(t)\) represents cumulative inflow to ICU, thus setting the outflow rate \(= 0\). The complete derivation of the model, detailed parameter definitions, estimation procedures, and further simulation details are provided in the Supplementary Material.

To ensure the ODE model reflected the specific dynamics of the UHVN cohort, its baseline transition rates were calibrated using observed outcome proportions derived directly from the study data (n=249, Section~\ref{sec:Results_Statistical}). Crucially, based on the clinical finding that patients managed without NIV experienced no ICU transfers or exitus events (Fig.~\ref{fig:Figure2}B), the corresponding direct transition rates were structurally constrained: $\eta=0$ and $\nu=0$. For the remaining pathways, the dimensionless outcome proportions were converted into rates [days\(^{-1}\)] by incorporating assumed average lengths of stay (LOS) in the non-NIV and active-NIV states ($\tau_X = 7$ days and $\tau_Y = 10$ days, respectively). These LOS values were predefined parameters used consistently across the simulation analyses described below. This calibration approach provides parameters aligned with the aggregate cohort behavior, a necessary step for developing credible predictive simulations \cite{Gopal2021}. The detailed methodology for proportion calculation, the LOS assumptions, the resulting rate calculations, and the final parameter values are fully documented in Supplementary Material. Employing this calibrated model structure, we then performed several \textit{in silico} simulations to explore system dynamics under various critical scenarios, aiming to provide insights beyond the static statistical analysis for resource planning and evaluating the potential impact of changing conditions.

\subsubsection{Simulation of IRCU Patient Flow Dynamics.}
\label{subsec:ode_simulations} 

Employing the calibrated ODE model structure (Section 3, Supplementary Material), we performed several \textit{in silico} simulations to explore the dynamic behavior of the IRCU system under different operational conditions.

\begin{figure}[H] 
    \centering
    \includegraphics[width=\textwidth]{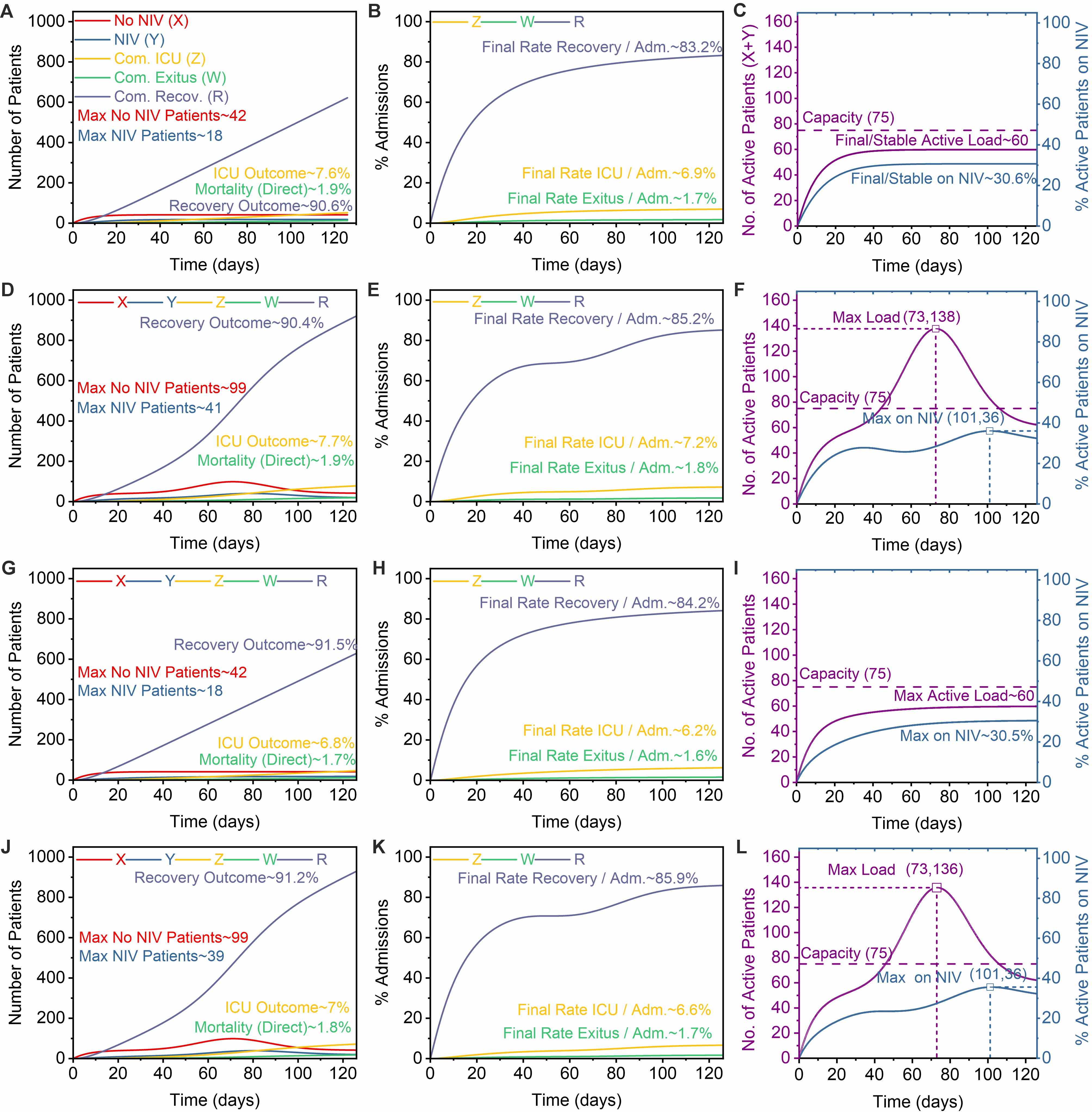} 
    \caption{\textbf{Model Simulation Results under Different Scenarios.}
    Dynamics of patient compartments (X: No NIV, Y: NIV, Z: Cumulative ICU, W: Cumulative Exitus, R: Cumulative Recovered) over 126 days for four different simulation scenarios based on UHVN IRCU data. \textbf{Column 1 (A, D, G, J):} Absolute patient numbers, final outcome proportions (calculated vs. total Z+W+R outcomes at day 126, shown as percentages), and maximum integer patient counts observed in X and Y states (e.g., Max Y \(\approx\) 18). \textbf{Column 2 (B, E, H, K):} Cumulative outcome rates (Z, W, R) normalized by cumulative admissions (\%). \textbf{Column 3 (C, F, I, L):} Total active patient load (X+Y, left axis, purple line) and percentage of active patients requiring NIV (right axis, blue line), with annotations for peak or final/stable values. \textbf{The dashed horizontal line in panels C, F, I, L indicates an illustrative unit capacity of 75 beds.} Base parameters ($A_{avg}$, $\alpha$, $\rho$, $\gamma$, $\varepsilon$, $\theta_0$) were estimated from cohort data (details in Suppl. Section~S5). Rates $\eta$ (X to ICU) and $\nu$ (X to Exitus) were fixed at 0 based on observations. \textbf{(A-C) Scenario A1 (Baseline):} Constant admission rate $A(t) = A_{avg}$; constant recovery rate $\theta(t) = \theta_0$. Represents average, stable conditions. \textbf{(D-F) Scenario B1 (Variable Admissions):} Variable $A(t)$ modeled as a Gaussian peak ($center=65, width=15, scale=1.5$) superimposed on $A_{avg}$, simulating an admission surge; constant recovery rate $\theta(t) = \theta_0$. \textbf{(G-I) Scenario B2 (Variable Recovery):} Constant admission rate $A(t) = A_{avg}$; variable recovery rate $\theta(t) = \theta_0 + 0.15 e^{-0.05 t}$ modeling potential efficiency improvement. \textbf{(J-L) Scenario B3 (Combined):} Variable $A(t)$ as in B1; variable $\theta(t)$ as in B2. Simulates surge concurrent with efficiency gains.}
    \label{fig:MultiPanelSims} 
 \end{figure}

First, a baseline autonomous simulation was executed (Simulation A1), utilizing the estimated constant average admission rate (\(A_{avg}\)) and the derived base transition parameters, including the constraint of no direct ICU transfers or mortality from the non-NIV state ($\eta=0, \nu=0$). A constant recovery rate from NIV (\(\theta(t) = \theta_0\)) was assumed, reflecting the average observed condition. Under these stable parameters, the system evolved towards a steady state (\textbf{Fig.~\ref{fig:MultiPanelSims}A-C}). Panel A illustrates the absolute patient counts, demonstrating that cumulative recoveries (R) represented the predominant outcome pathway. Active patient loads reached maximums of approximately 42 for non-NIV (X) and 18 for NIV (Y) individuals before stabilizing. The total active patient census (X+Y) equilibrated around 60 patients, remaining well below an illustrative capacity threshold of 75 beds (\textbf{Fig.~\ref{fig:MultiPanelSims}C}, dashed red line), with approximately 30.6\% requiring NIV in this steady state. At the simulation's end (126 days), the predicted cumulative outcome proportions (relative to total Z+W+R outcomes) were approximately 7.6\% for ICU transfer, 1.9\% for direct exitus, and 90.6\% for recovery (\textbf{Fig.~\ref{fig:MultiPanelSims}A}, text annotations). These simulated proportions exhibit reasonable concordance with the overall cohort outcomes (Fig.~\ref{fig:Figure1}B), suggesting the baseline parameterization adequately captures the system's aggregate behavior. Correspondingly, the cumulative outcome rates per admission stabilized at approximately 6.9\% for ICU, 1.7\% for exitus, and 83.2\% for recovery (\textbf{Fig.~\ref{fig:MultiPanelSims}B}).

Recognizing the observed fluctuations in monthly admissions (Fig.~\ref{fig:Figure1}C inset), we subsequently simulated the system's response to a time-varying admission rate, \(A(t)\) (Simulation B1, \textbf{Fig.~\ref{fig:MultiPanelSims}D-F}). This scenario utilized a Gaussian function superimposed on the baseline rate \(A_{avg}\) (parameters specified in Fig.~\ref{fig:MultiPanelSims} caption), while maintaining constant transition parameters (\(\theta(t)=\theta_0\)). As anticipated, the variable admission rate induced substantial transient dynamics in the active patient populations (\textbf{Fig.~\ref{fig:MultiPanelSims}D}). The peak census of patients requiring NIV (Max Y \(\approx\) 41, Panel D) more than doubled compared to the baseline maximum (Max Y \(\approx\) 18, Panel A). Consequently, the total active patient load reached a markedly higher peak of approximately 138 patients (\textbf{Fig.~\ref{fig:MultiPanelSims}F}), significantly exceeding the illustrative 75-bed capacity threshold and underscoring the potential for resource saturation during admission surges. This simulation vividly illustrates the direct impact of fluctuating admissions on immediate IRCU resource demands, particularly NIV capacity. Notably, while the surge precipitated steeper initial increases in cumulative outcomes (compare Panels D vs. A), the final outcome proportions (7.7\% ICU, 1.9\% Exitus, 90.4\% Recovery, Panel D) remained nearly identical to the baseline scenario. This suggests that, under constant operational efficiency, the primary effect of such an admission surge manifests as transient strain on capacity rather than a fundamental shift in the long-term distribution of patient outcomes.

We then explored the potential impact of temporally varying recovery efficiency for patients receiving NIV (Simulation B2, \textbf{Fig.~\ref{fig:MultiPanelSims}G-I}). Motivated by the IRCU's observed success in managing a majority (68\%) of high-risk NIV patients without ICU escalation (Fig.~\ref{fig:Figure2}B), we modeled a scenario where recovery efficiency improves over time using \(\theta(t) = \theta_0 + \Delta\theta e^{-\lambda t}\), while maintaining a constant average admission rate (\(A_{avg}\)). Compared to the baseline simulation (A1), this transiently enhanced recovery function yielded a discernible improvement in final outcomes (\textbf{Fig.~\ref{fig:MultiPanelSims}G}). Specifically, it resulted in modest reductions in predicted cumulative ICU transfers (approx. 6.8\% vs. 7.6\% in A1) and direct exitus events (approx. 1.7\% vs. 1.9\% in A1), with a corresponding increase in cumulative recoveries (approx. 91.5\% vs. 90.6\% in A1). The cumulative outcome rates per admission (\textbf{Fig.~\ref{fig:MultiPanelSims}H}) reflected this favorable shift (e.g., Final Rate ICU / Adm. \(\approx\) 6.2\% vs. 6.9\%). Given the constant admission rate, the active patient load dynamics (\textbf{Fig.~\ref{fig:MultiPanelSims}I}) closely resembled the baseline scenario, stabilizing at approximately 60 patients (below the illustrative capacity) with around 30.5\% requiring NIV, thereby demonstrating the potential benefits of improved care efficiency under stable admission pressure.

Finally, to synthesize the most plausible dynamic representation incorporating both the observed admission peak and potential efficiency gains, we simulated a scenario combining the variable admission rate \(A(t)\) with the variable recovery rate \(\theta(t)\) (Simulation B3, \textbf{Fig.~\ref{fig:MultiPanelSims}J-L}). This configuration examines the interplay between increased admission pressure and concurrent improvements in care delivery. The results indicate that while the variable \(A(t)\) profile continues to drive significant peaks in active patient load (Max X+Y \(\approx\) 136, Panel L) and NIV requirement (Max Y \(\approx\) 39, Panel J), closely mirroring Simulation B1 (\textbf{Fig.~\ref{fig:MultiPanelSims}F}) and substantially surpassing the illustrative capacity limit, the concurrently enhanced recovery efficiency (\(\theta(t)\)) provides partial mitigation of the adverse downstream consequences. Compared to the surge scenario with constant efficiency (B1, Panel D), this combined scenario (B3, Panel J) resulted in slightly lower final proportions for ICU transfers (approx. 7.0\% vs. 7.7\%) and direct mortality (approx. 1.8\% vs. 1.9\%), and higher recoveries (approx. 91.2\% vs. 90.4\%). This suggests that modeled improvements in care efficiency can modestly counteract the negative shift in outcome distributions associated with admission surges, although they do not substantially alleviate the peak resource strain (patient volume) imposed by the surge itself.

\begin{figure}[H] 
   \centering
   \includegraphics[width=\textwidth]{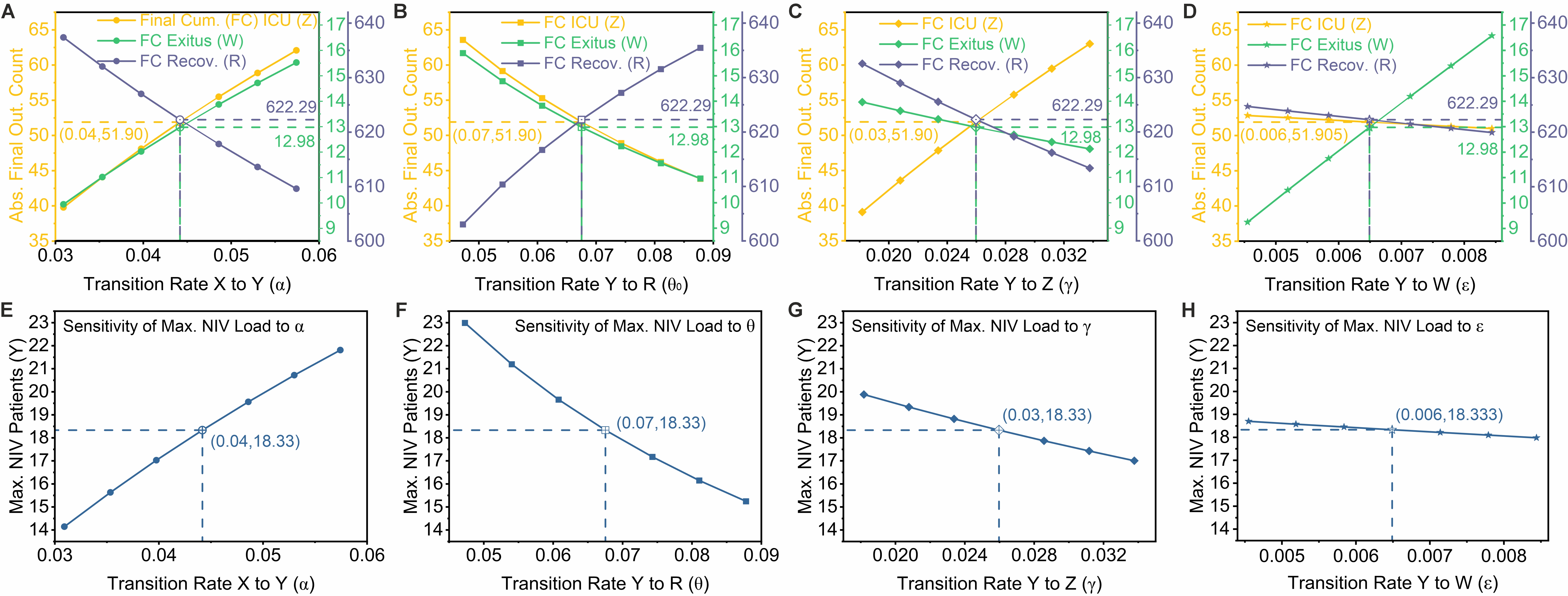} 
   \caption{\textbf{Sensitivity Analysis of the Autonomous Model.}
   Impact of varying key transition rates ($\alpha$, $\theta_0$, $\gamma$, $\varepsilon$) by $\pm 30\%$ around their baseline values (indicated by markers and dashed lines corresponding to \(Z \approx 13\), \(W \approx 52\), \(R \approx 622\), Max. Y \(\approx\) 18) on model outputs. Baseline parameter values used for reference points are \(\alpha \approx 0.044\), \(\theta_0 \approx 0.068\), \(\gamma \approx 0.023\), \(\varepsilon \approx 0.009\) days\(^{-1}\). \textbf{Top Row (A-D):} Absolute final cumulative outcome counts for ICU transfer (Z, yellow, left axis), Exitus (W, green, left axis), and Recovery (R, dark blue/purple, right axis). \textbf{Bottom Row (E-H):} Maximum number of patients simultaneously requiring NIV (Max. NIV Patients (Y)). Parameters varied: (\textbf{A, E}) NIV initiation rate ($\alpha$, X to Y); (\textbf{B, F}) NIV recovery rate ($\theta_0$, Y to R); (\textbf{C, G}) ICU transfer rate ($\gamma$, Y to Z); (\textbf{D, H}) Direct exitus rate ($\varepsilon$, Y to W). Analyses performed using the autonomous model configuration (constant $A_{avg}$, constant $\theta(t)=\theta_0$ when varying other parameters).}
   \label{fig:SensitivityAnalysis} 
\end{figure}

Finally, sensitivity analyses were performed on the autonomous model (constant \(A_{avg}\), \(\theta(t)=\theta_0\)) to assess the relative importance of key transition rates on both final outcomes and peak system load (\textbf{Fig.~\ref{fig:SensitivityAnalysis}}). We varied the four main transition rates ($\alpha$, $\theta_0$, $\gamma$, $\varepsilon$) individually by $\pm 30\%$ around their estimated baseline values. These baseline values were \(\alpha \approx 0.044\), \(\theta_0 \approx 0.068\), \(\gamma \approx 0.023\), and \(\varepsilon \approx 0.009\) days\(^{-1}\), resulting in reference model predictions of approximately 13 cumulative ICU transfers (Z), 52 cumulative exitus events (W), 622 cumulative recoveries (R), and a peak NIV load (Max. Y) of approximately 18 patients. 

Varying the NIV initiation rate from non-NIV patients ($\alpha$, Transition X to Y) revealed a direct impact on both outcomes and load (\textbf{Fig.~\ref{fig:SensitivityAnalysis}A, E}). Increasing $\alpha$ led to higher final cumulative counts for both ICU transfers (Z) and exitus events (W) compared to their baseline values (\(Z \approx 13\), \(W \approx 52\) at baseline \(\alpha\)), alongside a less pronounced decrease in recoveries (R, baseline \(R \approx 622\)) (\textbf{Panel A}). Concurrently, and significantly, the maximum number of patients requiring NIV ('Max. NIV Patients (Y)') increased substantially with $\alpha$, rising from below 15 to above 19 patients across the tested range, with a baseline value of approximately 18 patients at the baseline $\alpha$ value (\textbf{Panel E}). This suggests that factors accelerating patient progression towards needing NIV (higher $\alpha$) significantly strain downstream resources (ICU, mortality pathways) and increase the peak demand for NIV beds.

Conversely, varying the base recovery rate from NIV ($\theta_0$, Transition Y to R) demonstrated a strong influence primarily on outcome distribution and NIV load (\textbf{Fig.~\ref{fig:SensitivityAnalysis}B, F}). Increasing $\theta_0$ markedly decreased the final cumulative counts of ICU transfers (Z, baseline \(Z \approx 13\)) and exitus events (W, baseline \(W \approx 52\)), while substantially increasing recoveries (R, baseline \(R \approx 622\)) (\textbf{Panel B}, shown at baseline \(\theta_0 \approx 0.068\)). Furthermore, a higher recovery rate significantly reduced the maximum number of patients requiring NIV, decreasing from over 22 to below 16 patients across the tested range, with a baseline peak load of approximately 18 patients at the baseline $\theta_0$ value (\textbf{Panel F}). This pronounced sensitivity underscores the critical role of effective NIV management and recovery pathways within the IRCU, consistent with the observed high recovery rate (68\%) among NIV patients in our actual cohort (\textbf{Fig.~\ref{fig:Figure2}B}).

Variations in the ICU transfer rate from NIV ($\gamma$, Transition Y to Z) primarily affected the ICU outcome, as expected (\textbf{Fig.~\ref{fig:SensitivityAnalysis}C, G}). Increasing $\gamma$ led to a strong, near-linear increase in the final cumulative ICU count (Z) starting from the baseline of \(Z \approx 13\) (at baseline \(\gamma \approx 0.023\)), accompanied by slight decreases in exitus (W, baseline \(W \approx 52\)) and recoveries (R, baseline \(R \approx 622\)) (\textbf{Panel C}). Interestingly, increasing the transfer rate $\gamma$ slightly decreased the maximum NIV patient load from approximately 20 down to below 18 patients across the tested range, with the baseline value being approximately 18 patients (\textbf{Panel G}), potentially because patients are moved out of the Y state more quickly.

Lastly, varying the direct exitus rate from NIV ($\varepsilon$, Transition Y to W) showed a strong, direct relationship with cumulative mortality (\textbf{Fig.~\ref{fig:SensitivityAnalysis}D, H}). Increasing $\varepsilon$ resulted in a substantial rise in the final exitus count (W) from the baseline of \(W \approx 52\) (at baseline \(\varepsilon \approx 0.009\)), with corresponding slight decreases in ICU transfers (Z, baseline \(Z \approx 13\)) and recoveries (R, baseline \(R \approx 622\)) (\textbf{Panel D}). Notably, this parameter had a minimal impact on the maximum NIV patient load across the range tested, which remained close to the baseline value of approximately 18 patients (\textbf{Panel H}), indicating that while crucial for the mortality outcome, its variation within this range doesn't significantly alter the peak bed demand for NIV itself.

Overall, these sensitivity analyses highlight that while all tested parameters influence outcomes, the NIV initiation rate ($\alpha$) and the NIV recovery rate ($\theta_0$) appear to have the most substantial impact on the peak operational load (maximum NIV patients, baseline Max. Y \(\approx\) 18). Interventions targeting factors related to $\alpha$ (e.g., earlier treatments preventing deterioration) or $\theta_0$ (e.g., enhancing NIV efficacy) could therefore have the largest effects on IRCU resource management, while $\gamma$ and $\varepsilon$ are more directly related to specific downstream outcomes (ICU use, baseline \(Z \approx 13\), and mortality, baseline \(W \approx 52\), respectively).

\subsection{Total Occupancy Modeling via LOS Distribution.}
\label{sec:convolution_modeling}

\begin{figure}[htbp] 
    \centering
    \includegraphics[width=\textwidth]{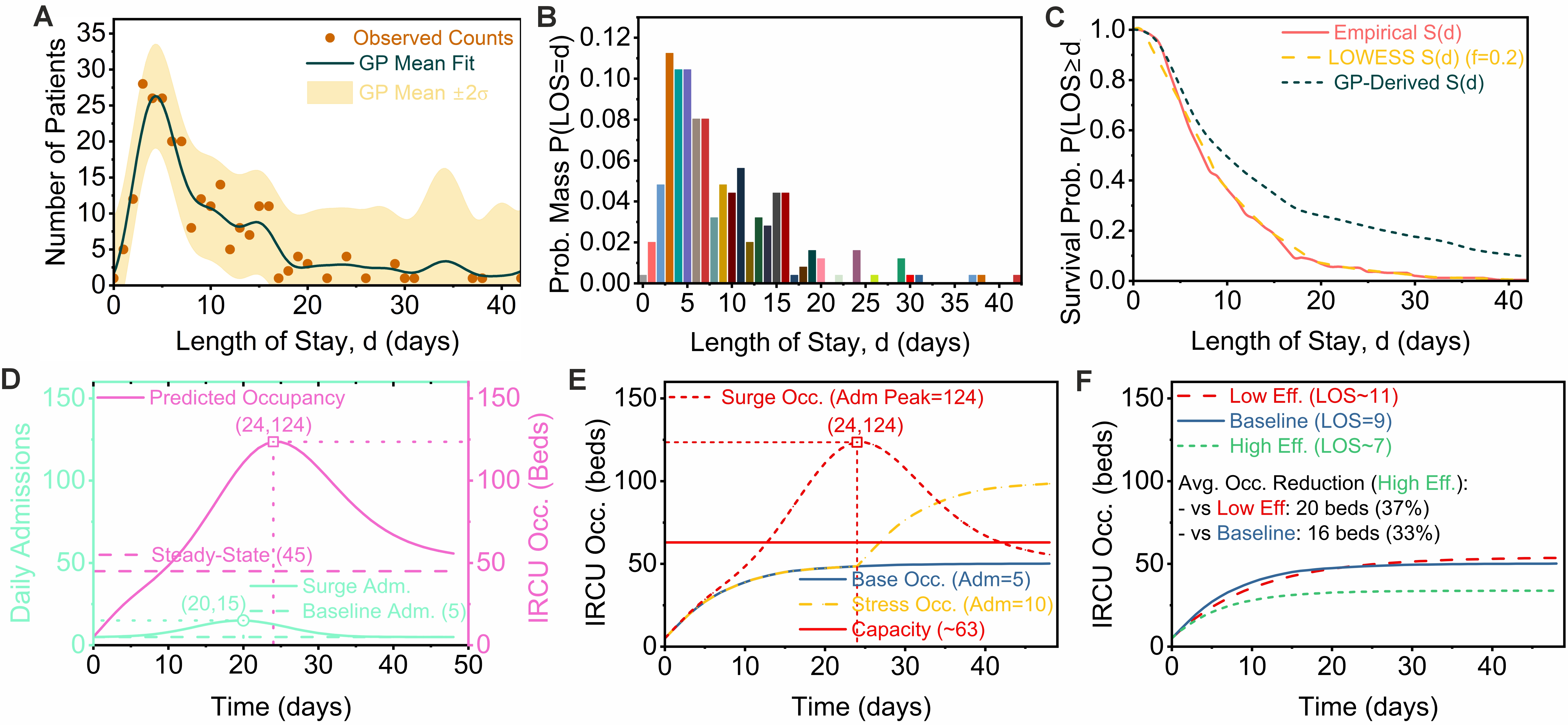} 
    \caption{\textbf{Length-of-Stay Analysis and Occupancy Simulations.}
    (\textbf{A}) Gaussian Process (GP) regression fit to observed IRCU Length-of-Stay (LOS) frequency data (counts of patients discharged per day of stay). Points are observed counts; dark green line is GP mean fit; shaded area is GP mean $\pm$ 2 standard deviations.
    (\textbf{B}) Empirical LOS Probability Mass Function (PMF), $P(\mathrm{LOS}=d)$, derived from observed data.
    (\textbf{C}) Comparison of LOS survival functions, $S(d)=P(\mathrm{LOS} \ge d)$: Empirical (derived from data), LOWESS smoothed (fraction f=0.2), and GP-derived (from the integral of the GP mean fit PMF). The Empirical $S(d)$ is used for convolution simulations.
    (\textbf{D}) Scenario 1: Predicted total IRCU occupancy (pink, right axis) resulting from forward convolution of an illustrative surge admission pattern (light green, left axis) with the empirical $S(d)$. Annotations indicate baseline and peak admission rates, steady-state occupancy, and peak predicted occupancy with corresponding day.
    (\textbf{E}) Scenario 3: Comparison of predicted total occupancy under different admission scenarios (Baseline: 5/day; Stress: 10/day sustained; Surge: peak admission rate 15/day) against an illustrative capacity threshold ($\approx$63 beds). Uses empirical $S(d)$. Legend reflects simplified labels used in the plot.
    (\textbf{F}) Scenario 4: Sensitivity of predicted total occupancy to system efficiency, modeled via different Mean LOS values resulting from varying effectiveness levels (Low Eff: Mean LOS$\approx$11d; Baseline: Mean LOS=9d; High Eff: Mean LOS$\approx$7d). Assumes constant baseline admissions (5/day). Text box quantifies average steady-state occupancy reductions achieved by high efficiency.}
    \label{fig:LOS_Occupancy_Sims} 
\end{figure}

Beyond the state-transition dynamics explored via the ODE model, the empirical distribution of the total Length-of-Stay (LOS) within the IRCU provides an alternative lens through which to analyze system load. Figure~\ref{fig:LOS_Occupancy_Sims}A-C characterizes this distribution. The observed frequency data (Fig.~\ref{fig:LOS_Occupancy_Sims}A, points) highlights the variability and right-skewness typical of LOS data, which is captured by a Gaussian Process (GP) fit (Fig.~\ref{fig:LOS_Occupancy_Sims}A, line and shaded area). The corresponding empirical Probability Mass Function (PMF), $f(d)=P(\mathrm{LOS}=d)$, reveals a high probability for short stays, with the mode occurring within the first few days (Fig.~\ref{fig:LOS_Occupancy_Sims}B). From this distribution, the empirical mean LOS was calculated as 9.0 days. The derived empirical survival function, $S(d)=P(\mathrm{LOS} \ge d)$, representing the probability that a patient remains within the IRCU for at least $d$ days, is shown in Fig.~\ref{fig:LOS_Occupancy_Sims}C (solid line) alongside smoothed versions (LOWESS and GP-derived). This empirical $S(d)$ inherently captures the overall departure pattern observed in the cohort and serves as the core input for occupancy modeling.

This empirical survival function $S(d)$ enables a direct estimation of the total daily IRCU occupancy, $X_{total}(t)$, via convolution with the daily admission sequence $A(t)$, based on the fundamental relationship $X_{total}(t) = \sum_{k=0}^{d_{max}} A(t-k) S(k)$ (see Supplementary Material for a detailed mathematical description of the underlying methods). We applied this forward convolution method using the empirical $S(d)$ and illustrative admission scenarios. Figure~\ref{fig:LOS_Occupancy_Sims}D displays the predicted occupancy (right axis, pink line) resulting from a representative surge admission pattern $A(t)$ (left axis, light green line), which has a baseline rate of 5 admissions/day and a peak surge reaching 15 admissions/day around day 20. The predicted total occupancy demonstrates a characteristic lagged and smoothed response, reaching a peak occupancy of approximately \textbf{124} beds around day \textbf{24} before declining, considerably exceeding the expected steady-state occupancy of approx. 45 beds under baseline admissions.

To further explore operational implications, Figure~\ref{fig:LOS_Occupancy_Sims}E compares the predicted occupancy under different admission patterns against an illustrative capacity threshold (approx. 63 beds, red line), derived as 125\% of the baseline steady-state occupancy. While the baseline admission rate (5/day, blue dashed line) keeps the occupancy well below capacity, both the stress scenario (sustained 10 admissions/day, yellow dash-dot line) and the surge scenario (red dashed line) clearly breach this threshold for extended periods, highlighting the vulnerability to increased admission pressure given the observed LOS distribution.

Finally, we investigated the sensitivity of occupancy to overall system efficiency, modeled via changes in the Mean LOS achieved (Fig.~\ref{fig:LOS_Occupancy_Sims}F). Using the baseline constant admission rate (5/day), simulations were run assuming the baseline empirical Mean LOS (9d, solid blue line), a 'Low Efficiency' scenario resulting in a longer Mean LOS ($\approx$11d, dotted orange line), and a 'High Efficiency' scenario achieving a shorter Mean LOS ($\approx$7d, dashed green line). As shown, improved efficiency (shorter LOS) substantially reduces the required bed occupancy, yielding an average reduction of approximately 16 beds (33\%) compared to baseline and 20 beds (37\%) compared to the low efficiency scenario during the steady-state period (Panel F, text box).

This convolution-based total occupancy modeling complements the ODE model results. While the ODE simulations provide insights into state-specific occupancy ($X$ vs $Y$), the convolution model offers a direct prediction of total bed demand grounded in the aggregate observed LOS. Comparing the total peak occupancy predicted here (e.g., approx. 124 in Panel D) with the peak active load from the corresponding ODE simulation (e.g., approx. 136 in Fig.~\ref{fig:MultiPanelSims}L for B3) reveals broadly consistent results between the two distinct modeling approaches, lending confidence to the overall understanding of system dynamics. The empirical PMF $f(d)$ (Panel B) also represents the necessary input for potential future work using deconvolution techniques to estimate historical admissions from occupancy data.

Comparing the modeling approaches, the peak total occupancy predictions under surge conditions show broad consistency. The LOS/convolution model predicted a peak of approximately 124 patients (Fig.~\ref{fig:LOS_Occupancy_Sims}D), while the corresponding ODE simulation (B3) yielded a peak active load (X+Y) of approximately 136 patients (Fig.~\ref{fig:MultiPanelSims}L). This agreement between distinct methodologies lends confidence to the overall assessment of system dynamics under stress. Furthermore, the two models offer complementary perspectives: while the convolution model provides a direct forecast of total occupancy grounded in the empirical LOS distribution, the ODE model elucidates the internal dynamics, revealing the fluctuating distribution between NIV and non-NIV states and the impact of varying transition rates, factors not explicitly visible in the aggregate LOS analysis.

\section{Discussion.}
\label{sec:Discussion}
This study provides critical insights into the operational dynamics and clinical effectiveness of an Intermediate Respiratory Care Unit (IRCU) established during the exigencies of the COVID-19 pandemic, with a particular focus on the \textbf{operational context involving specialized clinical staffing}. 
Our primary findings demonstrate that this newly implemented IRCU successfully managed a substantial cohort of patients with severe COVID-19-associated respiratory failure, many requiring non-invasive ventilation (NIV). Strikingly, while patients requiring NIV were significantly older and represented a higher-risk stratum, \textbf{68\% of these individuals recovered within the IRCU without needing escalation to intensive care}. Furthermore, no ICU transfers or mortality events occurred among patients managed without NIV within the IRCU pathway. These outcomes, achieved during a period of immense healthcare strain, underscore the potential of IRCUs, \textbf{particularly when structured with key elements such as dedicated, expert clinical teams,} 
to function as highly effective units for delivering complex respiratory care, thereby alleviating critical pressure on ICU resources. Our integrated statistical and dynamic modeling approach further elucidates the quantitative factors governing patient flow and the potential impact of operational changes.

Our findings resonate with the burgeoning literature highlighting the strategic importance of IRCUs as a bridge between the general ward and the ICU, particularly during the COVID-19 crisis \cite{grosgurin2021, mediano2023, matutevillacis2021role, lopezpadilla2022unidades}. Several studies have reported favorable outcomes and reduced ICU burden associated with IRCU implementation during the pandemic \cite{suarezcuartin2021clinical, galdeanolozano2022effectiveness}. The observed 8\% overall ICU transfer rate in our cohort aligns broadly with figures reported elsewhere, although direct comparisons are complex due to heterogeneity in patient populations, admission criteria, and unit capabilities across centers \cite{grosgurin2021}. A unique aspect of our work is its setting within an IRCU explicitly designed with \textbf{dedicated 24/7 respiratory physician coverage and nursing staff receiving specialized training} 
(Fig.~\ref{fig:Timeline}, \ref{fig:NursingProcess}) \cite{cabestregarcia2023respiratory, sanchezmartinez2022development}. While other reports focus primarily on treatment modalities or overall patient outcomes \cite{mellado_art2021}, our study provides circumstantial evidence \textbf{suggesting} that this \textbf{combination of specialized medical and nursing skills,} manifested in managing NIV/HFNO, early recognition of deterioration, and protocol adherence, \textbf{was likely an important factor associated with} the successful recovery of the majority of our high-risk NIV patients  --- \textit{recognizing that this observational study establishes association, not direct causation}. This is consistent with established links between clinical expertise (both nursing and medical) and improved patient outcomes in critical and intermediate care settings \cite{Aiken2014, needleman2011nurse, Pronovost2002ICUStaffing}.

The specific nurse-to-patient ratio maintained in the UHVN IRCU (approximately 1:4) is pertinent here, aligning with levels often deemed appropriate for intermediate care but subject to considerable variation \cite{Needleman2008Variability}. Contextualizing this against the robust evidence linking staffing levels to clinical results is crucial. Foundational research and subsequent meta-analyses have repeatedly shown that lower nurse-to-patient ratios are strongly associated with better outcomes, including reduced mortality and fewer adverse events in acute care \cite{Aiken2014, needleman2011nurse, Kane2007NurseStaffingReview, Griffiths2018}. Our positive findings, notably the high NIV success rate, appear consonant with this principle, suggesting the dedicated staffing model \textbf{likely contributed significantly}. Conversely, higher patient assignments per nurse are frequently linked to increased risks \cite{Needleman2002NEJM}. Yet, staffing ratios operate within a complex interplay of factors. The educational background and specialized skill set of the nursing staff are also independent predictors of patient outcome \cite{Aiken2003Education}. Similarly, the continuous presence of specialized physicians (intensivists or pulmonologists) is associated with better ICU outcomes \cite{Pronovost2002ICUStaffing, Multz1998ClosedICU}. It is therefore highly plausible that the targeted respiratory care training received by the UHVN IRCU nurses (\textbf{Fig.~\ref{fig:Timeline}}, \textbf{Fig.~\ref{fig:NursingProcess}}), \textbf{coupled with the dedicated physician coverage}, acted synergistically with the 1:4 nursing ratio, enhancing the team's capacity to manage complex respiratory failure effectively --- although disentangling these specific contributions requires further research. Disentangling the specific contributions of staffing numbers versus the specialized competencies of different clinical disciplines within IRCUs remains a key area for future investigation. Furthermore, the high recovery rate among NIV patients reinforces the capacity of IRCUs to prevent substantial numbers of ICU admissions, aligning with pre-pandemic estimates and pandemic observations \cite{torres2005, caballeroeraso2022intermediate}.

The theoretical implications of our study extend to the conceptualization of IRCUs not merely as holding bays, but as dynamic, specialized environments where targeted interventions, potentially facilitated by expert staffing, can significantly alter patient trajectories. Our compartmental modeling (Fig.~\ref{fig:MultiPanelSims}, \ref{fig:SensitivityAnalysis}) provides a quantitative framework that moves beyond static outcome reporting. It allows for the dissection of patient flow dynamics, identifying critical bottlenecks and leverage points \cite{AlKarkhi2025, Alban2020}. Specifically, the high sensitivity of both outcomes and peak NIV occupancy to the NIV initiation rate ($\alpha$) and, crucially, the NIV recovery rate ($\theta_0$), offers a mechanistic basis for understanding the impact of clinical efficiency --- an efficiency likely enhanced by skilled clinical teams. Interventions enhancing recovery (e.g., optimized NIV protocols, proactive weaning, expert clinical care) demonstrated the potential to substantially reduce adverse outcomes and resource strain, even during admission surges. Practically, our findings \textbf{support the consideration} of wider adoption and adequate resourcing of IRCUs \textbf{that incorporate dedicated and specialized clinical staffing models} \cite{mediano2023, cabestregarcia2023respiratory, Pronovost2002ICUStaffing}. --- While our study cannot prove causality, such efficiency is plausibly supported or enhanced by the presence of skilled clinical teams. --- The developed modeling framework (Fig.~\ref{fig:MultiPanelSims}, \ref{fig:LOS_Occupancy_Sims}, Fig.~S1) offers a valuable tool for hospital administrators and clinical leads for \textit{in silico} testing of operational strategies, predicting resource needs under varying admission scenarios (e.g., surges), and evaluating the potential return on investment for initiatives aimed at improving care efficiency (e.g., advanced clinical training, staffing adjustments) \cite{Robinson2014, Brandeau2004, Kahraman2018}. The concordance between the ODE and convolution-based occupancy predictions (Section~\ref{sec:convolution_modeling}) lends further credence to the robustness of this modeling approach for operational planning.

Several limitations warrant consideration when interpreting these results. First, as a single-center study conducted over a specific five-month period during the COVID-19 pandemic, the generalizability of our quantitative findings to other settings, patient populations (e.g., non-COVID respiratory failure), or different pandemic phases requires confirmation. Although data were collected prospectively, the analysis is retrospective in nature, \textbf{limiting our ability to definitively establish causality between the specific IRCU staffing model and patient outcomes; the observed associations, while strong, remain associative.} Second, while we captured key outcomes and NIV use, finer-grained data on specific ventilator settings, precise HFNC parameters, daily physiological trajectories, or detailed comorbidity indices were not the focus of this analysis and could offer deeper insights. Third, the modeling approach, while informative, relies on assumptions regarding transition rates and their potential time dependencies. Parameter estimation was based on aggregate cohort data, and while sensitivity analyses explored robustness (Fig.~\ref{fig:SensitivityAnalysis}), the models simplify complex individual patient pathways. The observed lack of adverse outcomes in the non-NIV group, while empirically accurate for this cohort, necessitated structural assumptions ($\eta=0, \nu=0$) in the model that might differ in populations with higher baseline acuity or different pathologies. Finally, the \textit{post hoc} power analysis indicated moderate power (63\%) to detect medium-sized effects, suggesting that subtle differences, particularly regarding gender or less common outcomes, might have been missed.

\section{Conclusions.}
\label{sec:Conclusions}
Our investigation illuminates the critical operational dynamics and clinical utility of a specialized Intermediate Respiratory Care Unit, particularly \textbf{highlighting the positive outcomes observed within a model emphasizing dedicated and expert clinical staffing,} during the unprecedented challenge of the COVID-19 pandemic. We demonstrated that this newly established unit, featuring specialized physician and nursing support, successfully managed a significant cohort of patients presenting with severe COVID-19 respiratory failure. A central finding is the successful recovery of 68\% of patients requiring non-invasive ventilation entirely within the IRCU, obviating the need for ICU escalation in the majority of these high-risk individuals. This outcome, coupled with the notable absence of ICU transfers or mortality among those patients managed without NIV during their IRCU stay, powerfully demonstrates the capacity of intermediate care settings \textbf{when appropriately staffed and structured} to function not merely as observational waypoints, but as active therapeutic environments. These units can effectively absorb significant clinical burden and mitigate pressure on finite critical care resources, a capability of paramount importance during public health crises.

These results significantly extend the growing body of evidence supporting IRCU efficacy \citep{grosgurin2021, mediano2023, matutevillacis2021role}, particularly in the pandemic context \citep{suarezcuartin2021clinical}. However, our work distinctively foregrounds the \textbf{association between the implementation of a specialized staffing model} (including targeted nursing competencies and dedicated physician coverage) \textbf{and these favorable outcomes}, aligning with broader findings on the impact of clinical expertise and appropriate staffing levels \citep{Aiken2014, cabestregarcia2023respiratory, needleman2011nurse, Pronovost2002ICUStaffing}. Furthermore, the integration of statistical analysis with dynamic systems modeling provides a novel quantitative lens through which to understand patient flow. This approach reveals the high sensitivity of system throughput and, consequently, patient outcomes, to parameters directly influenced by care quality and efficiency—an efficiency likely enhanced by skilled multidisciplinary teams—such as the rate of NIV initiation ($\alpha$) and the rate of recovery from NIV ($\theta_0$) (Fig.~\ref{fig:SensitivityAnalysis}). This combined clinical and modeling strategy offers a more mechanistic understanding of IRCU function than previously available, moving beyond static outcome descriptions towards a framework for operational optimization and \textit{in silico} evaluation of interventions.

While illuminating, our findings naturally open several promising avenues for future inquiry, building upon the insights gained and addressing the study's limitations. \textbf{First,} large-scale, multi-center prospective studies are imperative to validate the generalizability of our findings across diverse healthcare systems, patient populations (including non-COVID respiratory conditions), and varying phases of pandemic activity. Such studies should explicitly incorporate robust metrics of \textbf{overall clinical staffing levels, skill mix (for both physicians and nurses), specific training programs,} 
and team functioning to better elucidate the relative contributions of these factors. \textbf{Second,} delving deeper into the physiological mechanisms underpinning successful NIV management and weaning within the IRCU setting, perhaps through enhanced physiological monitoring protocols or comparative effectiveness trials of different NIV modes and interfaces \cite{frat2015highflow, mellado_art2021}, will be crucial for refining clinical practice. \textbf{Third,} the predictive power of our dynamic models (Section~\ref{sec:dynamic_modeling}) could be substantially enhanced by incorporating individual patient-level covariates, stochastic elements reflecting real-world variability, and potentially adaptive real-time data assimilation frameworks, perhaps leveraging machine learning approaches \cite{Zhang2021, rasmussen2005}. \textbf{Fourth,} rigorous longitudinal studies are needed to track the long-term functional outcomes, health-related quality of life, and healthcare utilization patterns of patients successfully managed within IRCUs compared to those whose trajectory necessitates initial or subsequent ICU admission \cite{jimenezrodriguez2022on}. \textbf{Finally,} exploring the systematic adaptation and evaluation of \textbf{IRCUs structured with specialized multidisciplinary teams} for the broad spectrum of acute respiratory failure etiologies beyond viral pandemics represents a critical area for future investigation to solidify their role within the permanent healthcare infrastructure. Addressing these questions will further elucidate the full potential and optimal implementation strategies for Intermediate Respiratory Care Units in contemporary healthcare delivery.

\section{Materials and Methods.} 

\subsection{Study Setting and Design.}
\begin{figure}[H] 
    \centering
    \includegraphics[width=\textwidth]{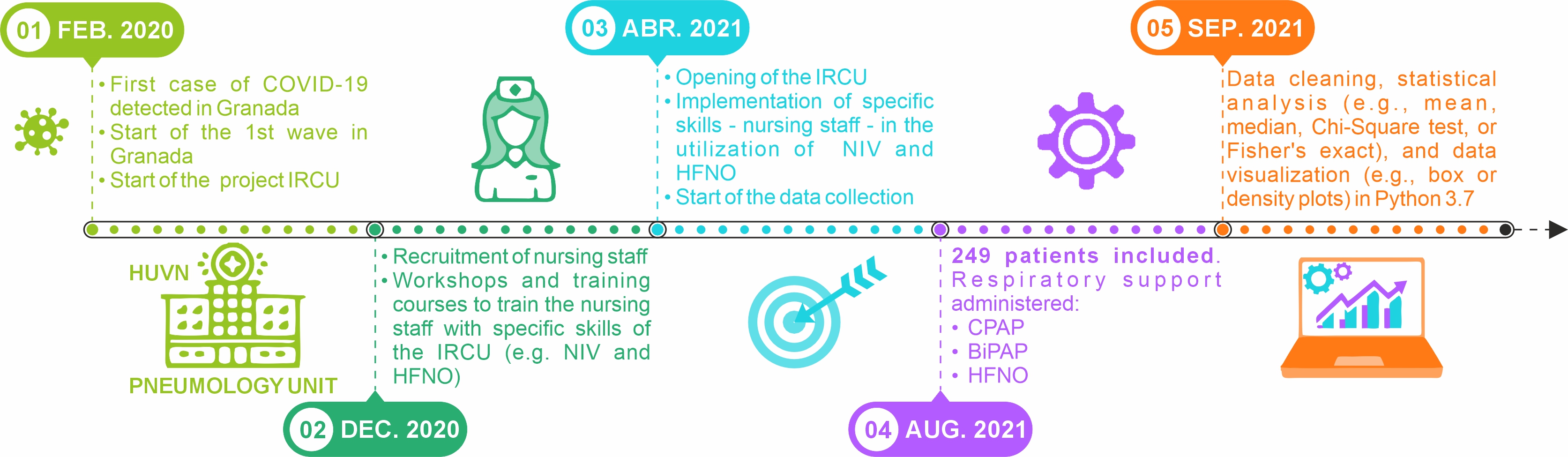} 
    \caption{\textbf{Timeline of UHVN IRCU Implementation and Study Execution.} Key milestones include: (01) Feb 2020: Project initiation driven by COVID-19 emergence. (02) Dec 2020: \textbf{Clinical staff preparation, including} nursing staff recruitment and specialized training (NIV/HFNO). (03) Apr 2021: Official IRCU opening and commencement of data collection. (04) Aug 2021: End of the patient inclusion period, encompassing 249 patients who received respiratory support (CPAP, BiPAP, HFNO). (05) Sep 2021: Initiation of data cleaning and statistical analysis.}
    \label{fig:Timeline}
\end{figure}

This prospective cohort study was conducted within the IRCU at the University Hospital Virgen de las Nieves (UHVN), a tertiary academic medical center in Granada, Spain. The establishment of this IRCU commenced in February 2020, motivated by the anticipated demands of the nascent COVID-19 pandemic (\textbf{Fig.~\ref{fig:Timeline}}, point 01). A foundational principle of the unit's design was the implementation of a dedicated, specialized clinical staffing model, ensuring continuous expert coverage. Preparatory phases included targeted recruitment and, notably, specialized training workshops for the nursing contingent focusing on advanced respiratory support modalities (Non-Invasive Ventilation [NIV] and High-Flow Nasal Oxygen [HFNO]), initiated in December 2020 (\textbf{Fig.~\ref{fig:Timeline}}, point 02).

The IRCU, formally accredited by the Spanish Society of Pneumology and Thoracic Surgery (SEPAR) at its highest complexity level, became operational in April 2021 (\textbf{Fig.~\ref{fig:Timeline}}, point 03). It functioned with dedicated medical staffing, providing 24/7 in-house coverage by respiratory physicians, and specialized nursing personnel. The unit was comprehensively equipped with advanced respiratory support and monitoring technology. Ethical approval for the overarching project involving COVID-19 patients at the institution, under which the data for this specific IRCU analysis were collected, was granted by the \textbf{Provincial Research Ethics Committee of Granada (CEIM/CEI Provincial de Granada)} on \textbf{June 24, 2020}. Written informed consent was obtained from all participants or their legally authorized representatives, in accordance with the Declaration of Helsinki, pertinent data protection regulations, and the provisions of this ethical approval. The period designated for patient inclusion and prospective data collection for the analysis presented herein extended from April 2021 through August 2021 (\textbf{Fig.~\ref{fig:Timeline}}, points 03-04).

\subsection{Study Population}
Eligible participants were adult patients (age \(\ge\) 18 years) admitted to the UHVN IRCU between April and August 2021 (\textbf{Fig.~\ref{fig:Timeline}}, points 03-04). The primary admission criterion was severe respiratory failure, predominantly associated with confirmed SARS-CoV-2 infection and compatible radiological findings. Patients admitted principally for non-pulmonary primary diagnoses were excluded. While other potential exclusion criteria, such as pre-existing directives limiting therapeutic escalation (e.g., 'do not resuscitate' orders), were considered in individual patient management decisions, they were not systematically applied as absolute exclusion criteria for enrollment in this observational cohort. The final analytic sample comprised 249 consecutive patients receiving care within the IRCU during the defined study interval (Figs.~\ref{fig:Figure1}, \ref{fig:Figure2}).

\subsection{Clinical Management and Data Collection}
\begin{figure}[H] 
    \centering
   \includegraphics[width=\textwidth]{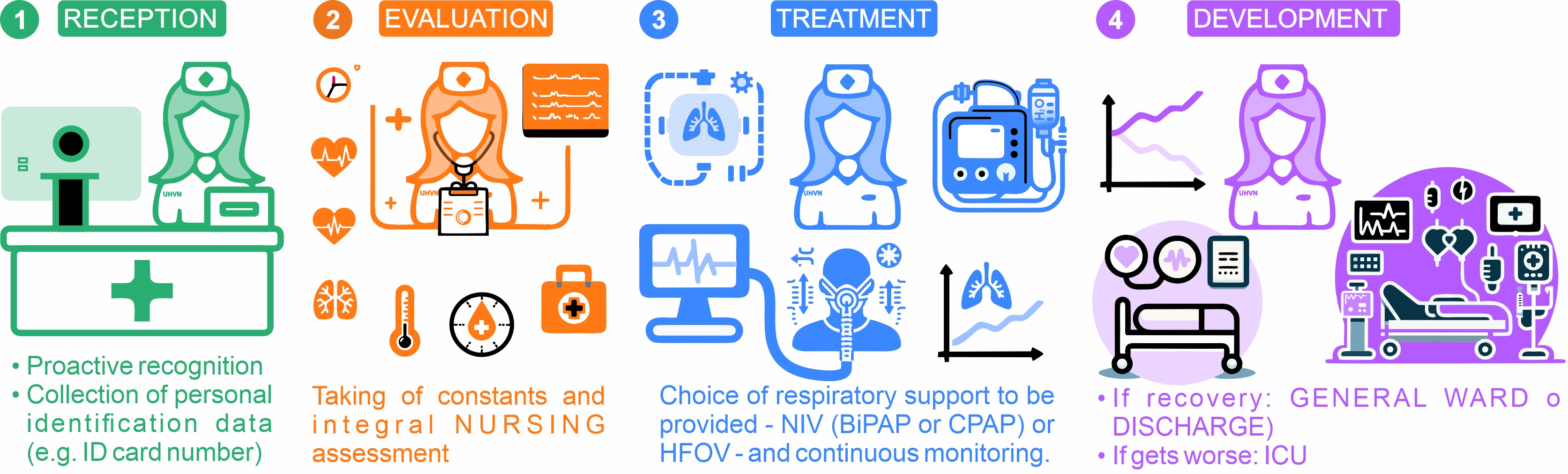} 
    \caption{\textbf{Patient Management Workflow and Key Clinical Team Actions in the UHVN IRCU.} The process encompasses four main stages: (1) \textbf{Reception}: Initial patient identification and proactive recognition. (2) \textbf{Evaluation}: Collection of vital signs and integral clinical assessment (including nursing). (3) \textbf{Treatment}: Selection and administration of respiratory support (NIV: CPAP, BiPAP, HFNO) with continuous multidisciplinary monitoring. (4) \textbf{Development}: Monitoring patient trajectory towards outcomes (Recovery: transfer to General Ward or Discharge; Worsening: transfer to ICU; or Exitus).}
    \label{fig:NursingProcess}
\end{figure}

Upon IRCU admission, patients underwent a standardized initial assessment protocol (\textbf{Fig.~\ref{fig:NursingProcess}}, Reception \& Evaluation phases), encompassing vital sign measurement and comprehensive clinical evaluation by the dedicated medical and nursing staff. Patients fulfilling criteria for severe respiratory failure (typically defined by an inspired oxygen fraction [FiO2] requirement exceeding 40\%) were initiated on NIV or HFNO based upon collaborative clinical judgment (\textbf{Fig.~\ref{fig:NursingProcess}}, Treatment phase). The principal respiratory support modalities employed included Continuous Positive Airway Pressure (CPAP) \cite{torres2005,grosgurin2021,mediano2023}, Bilevel Positive Airway Pressure (BiPAP) via Respironics V60 ventilators (Philips) \cite{nava1998,mediano2023}, and HFNO \cite{mediano2023,plate2017}. The utilization of HFNO adhered to evolving international guidance during the pandemic, incorporating WHO recommendations for vigilant monitoring \cite{whittle2021resp, world2020clinical}.  To ensure confidentiality as per ethical approval, patient data were de-identified prior to analysis by the research team.

Continuous cardiorespiratory monitoring informed ongoing clinical management. A key institutional guideline mandated proactive multidisciplinary consultation, particularly involving the Intensive Care Unit (ICU) team, for patients exhibiting sustained high oxygen requirements (e.g., FiO2 > 80\% despite NIV optimization) to facilitate timely decisions regarding the appropriate level of care \cite{nava1998,torres2005}. This collaborative framework guided patient trajectory determinations (\textbf{Fig.~\ref{fig:NursingProcess}}, Development phase), encompassing potential ICU escalation, transfer to a general medical ward upon clinical improvement, or direct discharge home. Throughout the IRCU admission, established nurse-to-patient (approximately 1:4) and physician-to-patient (approximately 1:6) ratios were maintained, ensuring continuous specialized clinical oversight. Data were collected prospectively using standardized forms, capturing patient demographics, pertinent comorbidities, respiratory support modalities and key parameters (when available), length of stay (LOS), and definitive IRCU pathway outcomes (ICU transfer, in-IRCU exitus, or recovery/discharge), constituting the dataset for subsequent analysis (Figs.~\ref{fig:Figure1}, \ref{fig:Figure2}).

\subsection{Statistical Analysis and Visualization.}

Descriptive statistics, including counts (N), percentages (\%), medians, and interquartile ranges (IQR), were employed to summarize the cohort's baseline characteristics and clinical outcomes. Distributions of continuous variables were visualized using violin plots (Fig.~\ref{fig:Figure1}). Intergroup comparisons utilized the Mann-Whitney U test for continuous variables (e.g., Fig.~\ref{fig:Figure1}A, \ref{fig:Figure2}A) and Chi-squared or Fisher’s exact tests, as appropriate, for categorical variables (underlying data for Fig.~\ref{fig:Figure1}B,D, \ref{fig:Figure2}B,D). Associations between predictor variables (age, gender, NIV status) and binary clinical outcomes (NIV utilization, ICU transfer, exitus) were evaluated using both crude Risk Ratios (RR) with 95\% confidence intervals (CI) (Fig.~\ref{fig:Figure2}D) and multivariable logistic regression models. The logistic regression analyses yielded adjusted Odds Ratios (OR) with 95\% CIs (Fig.~\ref{fig:Figure2}C), controlling for potential confounding by patient age and gender where applicable. Statistical significance was established at a two-tailed p-value threshold of < 0.05. Data cleaning and formal statistical analysis commenced in September 2021.

The distribution of total Length-of-Stay (LOS) within the IRCU (Fig.~\ref{fig:LOS_Occupancy_Sims}) was characterized empirically using Probability Mass Functions (PMF) and Kaplan-Meier-derived Survival Functions (\(S(d)\)). To facilitate smoothed estimates and potential forecasting applications, Gaussian Process (GP) regression \cite{rasmussen2005}, a non-parametric Bayesian technique (methodological details provided in Supplementary Material Section S4), was fitted to the observed LOS frequency data (Fig.~\ref{fig:LOS_Occupancy_Sims}A), yielding smoothed survival function estimates (Fig.~\ref{fig:LOS_Occupancy_Sims}C). However, the occupancy simulations presented herein primarily utilized the directly calculated empirical \(S(d)\). All statistical computations were performed using Python (version 3.7; libraries: SciPy, Statsmodels, Matplotlib, Scikit-learn) and R (version 4.3.2; e.g., `pwr` package). Figures, including timeline and workflow diagrams (\ref{fig:Timeline}, \ref{fig:NursingProcess}) and conceptual model schematics in the Supplement, were prepared using Inkscape (version 1.3) \cite{inkscape2024inkscape}. Given the minimal extent of missing data (<1\%), a complete case analysis approach was deemed appropriate.

To contextualize the statistical power of our primary group comparisons (NIV vs. Non-NIV strata), a \textit{post hoc sensitivity analysis} was conducted using G*Power (version 3.1) \cite{faul2007gpower}. \textbf{It is imperative to note that this analysis assesses the study's sensitivity to detect a pre-specified, conventionally defined effect size, rather than calculating power based on the observed effect magnitudes.} Specifically, we determined the achieved statistical power (1-$\beta$) of the study, given the final cohort sample sizes (Total N=249; Non-NIV \(n_1=172\), NIV \(n_2=77\)), to identify a statistically significant difference corresponding to a Cohen’s d of 0.5 between these groups for a continuous outcome. This effect size (d=0.5) was selected based on Cohen's established convention for a 'medium' effect within the biomedical sciences \cite{cohen1988}. The analysis assumed a two-tailed independent samples t-test framework, suitable for such group comparisons, employing the standard alpha level ($\alpha$) of 0.05. The resulting achieved power was calculated to be approximately \textbf{0.63 (63\%)}. This signifies moderate sensitivity: under the realized sample distribution, there was a 63\% probability of detecting a true difference of medium magnitude (d=0.5), should one exist in the population, as statistically significant at the p<0.05 level. Conversely, this implies a non-trivial risk (Type II error probability, $\beta \approx 0.37$) of failing to detect such a medium-sized effect. This moderate statistical power warrants caution in interpreting non-significant findings for comparisons between the NIV and non-NIV groups and underscores the value of future, potentially larger studies to investigate more subtle differences.

\section{Conflict of Interest.}

The authors affirm that there are no competing interests—financial, personal, or otherwise—that have influenced, or could be perceived to have influenced, the conduct or reporting of the research described in this paper.

\section{Ethical Approval.}

The study adhered to the Declaration of Helsinki. Ethical approval for the broader project encompassing the collection of data analyzed in this manuscript was provided by the \textbf{Provincial Research Ethics Committee of Granada (CEIM/CEI Provincial de Granada)} on \textbf{June 24, 2020}. Written informed consent was obtained from all participants or their legal representatives prior to inclusion, following the stipulations of the approved protocol. Participant data confidentiality was maintained throughout the study.

\section{Acknowledgments.}
We are truly grateful to every patient who generously dedicated their time and effort to be part of our study. We also want to convey our deep appreciation to the exceptional medical and nursing staff of the IRCU at the UHVN in Granada, Spain. Their unwavering dedication, tireless efforts, and unparalleled expertise have been instrumental in maintaining the seamless functioning of the IRCU and providing outstanding care to our patients. Their remarkable teamwork and exceptional skills have been indispensable in delivering our patients the utmost quality of care. RF acknowledges partial support from the María Zambrano-Senior grant (Spanish Ministerio de Universidades and Next-Generation EU); Grant C-EXP-265-UGR23 funded by Consejería de Universidad, Investigación \& Innovación \& ERDF/EU Andalusia Program; Grant PID2022-137228OB-I00 funded by the Spanish Ministerio de Ciencia, Innovación y Universidades, MICIU/AEI/10.13039/501100011033 \& ``ERDF/EU A way of making Europe''; and the Modeling Nature Research Unit, project QUAL21-011.

\section*{Code and Data Availability.} 

The Python scripts used for the ODE modeling (Figures 3-4), Length-of-Stay analysis, and occupancy simulations (Figure 5), along with the necessary input data files (``patient\_data.csv'', ``ircu\_data.csv''), are publicly available. The specific version of the code and data corresponding to the results presented in this manuscript (Version v1.0.0) has been archived on Zenodo and can be accessed via the following DOI: \url{https://doi.org/10.5281/zenodo.15286823}. The development repository is hosted on GitHub at \url{https://github.com/renee29/IRCU_Patient_Flow_Modeling_Scripts}.

\section*{Author Contributions.}

A.C.N.O., C.M.G., and R.F. conceptualized the study. A.C.N.O., J.A.S.M., and C.M.G. developed the clinical methodology, while R.F. developed the modeling and statistical methodology. Clinical investigation and data curation were performed by A.C.N.O., P.G.F., and J.A.S.M. Formal analysis, software, and visualization were conducted by R.F. Project administration was by A.C.N.O.; resources and clinical supervision by C.M.G. and A.C.N.O.; analytical supervision by R.F. Validation involved A.C.N.O., J.A.S.M., C.M.G., and R.F. The original draft was written by A.C.N.O., J.A.S.M., C.M.G., and R.F. All authors reviewed and edited the final manuscript.

\printbibliography

@article{petty1967,
  author    = {Petty, T. L. and Bigelow, D. B. and Nett, L. M.},
  title     = {The Intensive Respiratory Care Unit---An Approach to the Care of Acute Respiratory Failure},
  journal   = {California Medicine},
  volume    = {107},
  number    = {5},
  pages     = {381--384},
  year      = {1967}
}

@article{petty1975,
  author    = {Petty, T. L. and Lakshminarayan, S. and Sahn, S. A. and Zwillich, C. W. and Nett, L. M.},
  title     = {Intensive respiratory care unit. Review of ten years' experience},
  journal   = {JAMA},
  volume    = {233},
  number    = {1},
  pages     = {34--37},
  year      = {1975},
  doi       = {10.1001/jama.1975.03260010036016}
}

@article{krieger1990,
  author    = {Krieger, B. P. and Ershowsky, P. and Spivack, D.},
  title     = {One year's experience with a noninvasively monitored intermediate care unit for pulmonary patients},
  journal   = {JAMA},
  volume    = {264},
  number    = {10},
  pages     = {1277--1280},
  year      = {1990},
  doi       = {10.1001/jama.1990.03450090079031}
}

@article{nava1998,
  author    = {Nava, S. and Confalonieri, M. and Rampulla, C.},
  title     = {Intermediate respiratory intensive care units in Europe: a European perspective},
  journal   = {Thorax},
  volume    = {53},
  number    = {9},
  pages     = {798--802},
  year      = {1998},
  doi       = {10.1136/thx.53.9.798}
}

@article{Cheng1999,
  author    = {Cheng, D. C. and Byrick, R. J. and Knobel, E.},
  title     = {Structural models for intermediate care areas},
  journal   = {Critical Care Medicine},
  year      = {1999},
  month     = {10},
  volume    = {27},
  number    = {10},
  pages     = {2266--2271},
  doi       = {10.1097/00003246-199910000-00034},
  pmid      = {10548219}
}

@article{torres2005,
  author    = {Torres, A. and Niederman, M. S. and Celis, R. and Bassi, G. L. and Martin-Loeches, I. and Ferrer, M.},
  title     = {Intermediate respiratory care units: A European perspective},
  journal   = {Archivos de Bronconeumología (English Edition)},
  volume    = {41},
  pages     = {505--512},
  year      = {2005},
  doi       = {10.1016/s1579-2129(06)60271-1}
}

@article{plate2017,
  author    = {Plate, J. D. J. and Leenen, L. P. H. and Houwert, M. and Hietbrink, F.},
  title     = {Utilisation of Intermediate Care Units: A Systematic Review},
  journal   = {Critical Care Research and Practice},
  volume    = {2017},
  pages     = {8038460},
  year      = {2017},
  doi       = {10.1155/2017/8038460}
}

@article{heilifrades2019cost,
  author    = {Heili-Frades, S. and Ferrer-Luca, R. and Zapata-Fenor, A. and Fernández-Ormaechea, I. and Martínez-Sagasti, F. and Vellido-González, M.},
  title     = {Cost and Mortality Analysis of an Intermediate Respiratory Care Unit. Is It Really Efficient and Safe?},
  journal   = {Archivos de Bronconeumología (English Edition)},
  volume    = {55},
  number    = {12},
  pages     = {634--641},
  year      = {2019},
  doi       = {10.1016/j.arbres.2019.06.008}
}

@article{Sala2009,
  author    = {Sala, E. and Balaguer, C. and Carrera, M. and Palou, A. and Bover, J. and Agustí, A.},
  title     = {Actividad de una unidad de cuidados respiratorios intermedios dependiente de un servicio de neumología [Activity of an intermediate respiratory care unit attached to a respiratory medicine department]},
  journal   = {Archivos de Bronconeumología},
  year      = {2009},
  month     = {04},
  volume    = {45},
  number    = {4},
  pages     = {168--172},
  language  = {Spanish},
  doi       = {10.1016/j.arbres.2008.09.003},
  pmid      = {19286297},
  note      = {Epub 2009 Mar 14}
}

@article{capuzzo2014hospital,
  author    = {Capuzzo, M. and Volta, C. A. and Tassinati, T. and Moreno, R. P. and Valentin, A. and Guidet, B. and Iapichino, G. and Martin, C. and Perneger, T. and Combescure, C. and Poncet, A. and Rhodes, A.},
  title     = {Hospital mortality of adults admitted to Intensive Care Units in hospitals with and without Intermediate Care Units: a multicentre European cohort study},
  journal   = {Critical Care},
  volume    = {18},
  number    = {5},
  pages     = {551},
  year      = {2014},
  doi       = {10.1186/s13054-014-0551-8}
}

@article{mediano2023,
  author    = {Mediano, O. and Luján, M. and López-Monzoni, S.},
  title     = {An Intermediate Respiratory Care Unit in Every Hospital},
  journal   = {Archivos de Bronconeumología (English Edition)},
  volume    = {59},
  number    = {1},
  pages     = {7--9},
  year      = {2023},
  doi       = {10.1016/j.arbres.2022.09.021}
}

@article{lopezpadilla2022unidades,
  author    = {López-Padilla, D. and González Martínez, F. and Torres-Macho, J. and de Miguel-Díez, J.}, % Expanded authors
  title     = {Unidades de Cuidados Respiratorios Intermedios: preguntas y respuestas [{Intermediate} {Respiratory} {Care} {Units}: {Questions} and {Answers}]}, % Added English title
  journal   = {Open Respiratory Archives},
  volume    = {4},
  number    = {4}, % Added number
  pages     = {100220},
  year      = {2022},
  doi       = {10.1016/j.opresp.2022.100220}
}

@article{lopezjardon2024utility,
  author    = {López-Jardón, P. and Martínez-Fernández, M. C. and García-Fernández, R. and Martín-Vázquez, C. and Verdeal-Dacal, R.},
  title     = {Utility of Intermediate Care Units: A Systematic Review Study},
  journal   = {Healthcare},
  volume    = {12},
  number    = {3}, % Added number
  pages     = {296},
  year      = {2024},
  doi       = {10.3390/healthcare12030296},
  pmid      = {38337163}, % Added PMID
  pmcid     = {PMC10855339} % Added PMCID
}

@article{lindenauer2014outcomes,
  author    = {Lindenauer, P. K. and Stefan, M. S. and Shieh, M.-S. and Pekow, P. S. and Rothberg, M. B. and Hill, N. S.}, % Expanded authors
  title     = {Outcomes Associated With Invasive and Noninvasive Ventilation Among Patients Hospitalized With Exacerbations of Chronic Obstructive Pulmonary Disease},
  journal   = {JAMA Internal Medicine},
  volume    = {174},
  number    = {12}, % Added number
  pages     = {1982--1993},
  year      = {2014},
  doi       = {10.1001/jamainternmed.2014.5430},
  pmid      = {25310325} % Added PMID
}

@article{davies2018british,
  author    = {Davies, M. and Quinnell, T. and Oscroft, N. and Clutterbuck-James, A. and Cayton, R. and Gibson, G. J. and Kohler, M. and Levy, M. and Murphy, P. B. and Smith, I. and Stradling, J. and Elliott, M.}, % Expanded authors
  title     = {British Thoracic Society Quality Standards for acute non-invasive ventilation in adults},
  journal   = {BMJ Open Respiratory Research}, % Expanded journal name
  volume    = {5},
  number    = {1}, % Added number
  pages     = {e000283},
  year      = {2018},
  doi       = {10.1136/bmjresp-2018-000283},
  pmid      = {29713503}, % Added PMID
  pmcid     = {PMC5922463} % Added PMCID
}

@article{nishimura2016highflow,
  author    = {Nishimura, M.},
  title     = {High-Flow Nasal Cannula Oxygen Therapy in Adults: Physiological Benefits, Indication, Clinical Benefits, and Adverse Effects},
  journal   = {Respiratory Care},
  volume    = {61},
  number    = {4},
  pages     = {529--541},
  year      = {2016},
  doi       = {10.4187/respcare.04577},
  pmid      = {27016353} % Added PMID
}

@article{frat2015highflow,
  author    = {Frat, J.-P. and Thille, A. W. and Mercat, A. and Girault, C. and Ragot, S. and Perbet, S. and Prat, G. and Boulain, T. and Morawiec, E. and Cottereau, A. and Devaquet, J. and Nseir, S. and Razazi, K. and Mira, J.-P. and Argaud, L. and Chakarian, J.-C. and Ricard, J.-D. and Wittebole, X. and Chevalier, S. and Herbland, A. and Fartoukh, M. and Constantin, J.-M. and Tonnelier, J.-M. and Pierrot, M. and Mathonnet, A. and Béduneau, G. and Delétage-Métreau, C. and Richard, J.-C. M. and Brochard, L. and Robert, R. and {FLORALI Study Group} and {REVA Network}}, % Expanded authors
  title     = {High-Flow Oxygen through Nasal Cannula in Acute Hypoxemic Respiratory Failure},
  journal   = {New England Journal of Medicine},
  volume    = {372},
  number    = {23}, % Added number
  pages     = {2185--2196},
  year      = {2015},
  doi       = {10.1056/NEJMoa1503326},
  pmid      = {25981908} % Added PMID
}

@article{grosgurin2021,
  author    = {Grosgurin, O. and Soulier, A. and Soulier, P. and Gayet-Ageron, A. and Serratrice, J. and Stirnemann, J. and Carballo, S. and Roux, X. and Pugin, J. and Garbino, J.},
  title     = {Role of Intermediate Care Unit Admission and Noninvasive Respiratory Support during the COVID-19 Pandemic: A Retrospective Cohort Study},
  journal   = {Respiration},
  volume    = {100},
  number    = {9},
  pages     = {786--793},
  year      = {2021},
  doi       = {10.1159/000516329},
  pmid      = {34000736},
  pmcid     = {PMC8168022}
}

@article{matutevillacis2021role,
  author    = {Matute-Villacís, M. and Díaz-Ángel, C. and Burbano, G. and Torres-Castro, R. and Vera-Villaroel, P. and Vaca-Guzman, D. and Vasconez, E. and Hernández-Jativa, V. and García-Sanz, M. T.}, % Expanded authors
  title     = {Role of respiratory intermediate care units during the {SARS-CoV-2} pandemic},
  journal   = {BMC Pulmonary Medicine},
  volume    = {21},
  number    = {1}, % Added number
  pages     = {228},
  year      = {2021},
  doi       = {10.1186/s12890-021-01593-5},
  pmid      = {34243753}, % Added PMID
  pmcid     = {PMC8266893} % Added PMCID
}

@article{suarezcuartin2021clinical,
  author    = {Suarez-Cuartin, G. and Osorio, J. and Salgueira, M. and Tedja-Utama, B. and Valenti, D. and Ferrer, R. and Torres, A.}, % Expanded authors
  title     = {Clinical Outcomes of Severe {COVID-19} Patients Admitted to an Intermediate Respiratory Care Unit},
  journal   = {Frontiers in Medicine},
  volume    = {8},
  pages     = {711027},
  year      = {2021},
  doi       = {10.3389/fmed.2021.711027},
  pmid      = {34485394}, % Added PMID
  pmcid     = {PMC8408283} % Added PMCID
}

@article{sanchezmartinez2022development,
  author    = {Sánchez-Martínez, J. A. and Morales-García, C. and Gómez-Hernández, B. C. and García-Velasco, P. and Osuna-Carrillo de Albornoz, A. and Fernández, A. and Solano-Parada, J. and Fabregas, R.}, % Expanded authors based on user's author list
  title     = {Development of a Non-Invasive Respiratory Support Pathway in {COVID-19} patients with severe respiratory failure in an Intermediate Respiratory Care Unit},
  journal   = {European Respiratory Journal},
  volume    = {60},
  number    = {suppl 66}, % Added supplement number
  pages     = {2040},
  year      = {2022},
  doi       = {10.1183/13993003.congress-2022.2040}
}

@article{caballeroeraso2022intermediate,
  author    = {Caballero-Eraso, C. and Pascual Martínez, N. and Mediano, O. and Egea Santaolalla, C.},
  title     = {{Unidades de Cuidados Respiratorios Intermedios (UCRI) durante la pandemia COVID-19. La realidad} [{Intermediate} {Respiratory} {Care} {Units} ({RICUs}) {During} the {COVID-19} {Pandemic}. {The} {Reality}]}, % Added English title
  journal   = {Archivos de Bronconeumología}, % Clarified journal name
  volume    = {58},
  number    = {4}, % Added number
  pages     = {284--286},
  year      = {2022},
  doi       = {10.1016/j.arbres.2021.10.004},
  pmid      = {34674946}, % Added PMID
  pmcid     = {PMC8516259} % Added PMCID
}

@article{galdeanolozano2022effectiveness,
  author    = {Galdeano Lozano, M. and Catalan-Matamoros, D. and Trapero-Bertran, M.}, % Expanded authors
  title     = {Effectiveness of Intermediate Respiratory Care Units as an Alternative to Intensive Care Units during the {COVID-19} Pandemic in Catalonia},
  journal   = {International Journal of Environmental Research and Public Health},
  volume    = {19},
  number    = {10}, % Added number
  pages     = {6034},
  year      = {2022},
  doi       = {10.3390/ijerph19106034}, % Added DOI
  pmid      = {35627571}, % Added PMID
  pmcid     = {PMC9141659} % Added PMCID
}

@article{jimenezrodriguez2022on,
  author    = {Jiménez-Rodríguez, B. M. and Bellido-Estévez, I. and Hernández-García, I. and Rivas-Guerrero, A. M. and Martín-Escalante, M. D. and Reina-Artacho, A. and Ruiz-Montero, M. and León-Carmona, J. J. and Tallón-Barranco, A.}, % Expanded authors
  title     = {On the single and multiple associations of {COVID-19} post-acute sequelae: 6-month prospective cohort study},
  journal   = {Scientific Reports},
  volume    = {12},
  number    = {1}, % Added number
  pages     = {3402},
  year      = {2022},
  doi       = {10.1038/s41598-022-07433-8},
  pmid      = {35217693}, % Added PMID
  pmcid     = {PMC8884592} % Added PMCID
}

@article{whittle2020resp,
  author    = {Whittle, J. S. and Pavlov, I. and Sacchetti, A. D. and Atwood, C. and Rosenberg, M. S.},
  title     = {Respiratory support for adult patients with {COVID‐19}},
  journal   = {JACEP Open},
  volume    = {1},
  number    = {2},
  pages     = {95--101},
  year      = {2020},
  doi       = {10.1002/emp2.12071},
  pmid      = {32421203}, % Added PMID
  pmcid     = {PMC7235464} % Added PMCID
}

@article{lujan2024multidisciplinary,
  author    = {Luján, M. and Monclou, J. J. and Pérez, P. and Hermosa, C. and Marín, M. and Torres, A.},
  title     = {Multidisciplinary Consensus on the Management of Non-Invasive Respiratory Support in the {COVID-19} Patient},
  journal = {Archivos de Bronconeumología},
  volume = {60},
  number = {5},
  pages = {285-295},
  year = {2024},
  doi = {10.1016/j.arbres.2024.02.017},
  issn = {0300-2896}
}

@article{wax2020practical,
  author    = {Wax, R. S. and Christian, M. D.},
  title     = {Practical recommendations for critical care and anesthesiology teams caring for novel coronavirus ({2019-nCoV}) patients},
  journal   = {Canadian Journal of Anesthesia/Journal canadien d'anesthésie},
  volume    = {67},
  number    = {5},
  pages     = {568--576},
  year      = {2020},
  doi       = {10.1007/s12630-020-01591-x},
  pmid      = {32052373} % Added PMID
}

@book{Diekmann2013,
  author    = {Diekmann, Odo and Heesterbeek, Hans and Britton, Tom},
  title     = {Mathematical Tools for Understanding Infectious Disease Dynamics},
  publisher = {Princeton University Press},
  year      = {2013},
  %doi       = {10.1515/9781400845118}, % Added DOI
  url       = {http://www.jstor.org/stable/j.cttq9530},
  isbn      = {978-0691155395} % Kept ISBN as well
}

@article{Alban2020,
  author    = {Alban, A. and Chick, S.E. and Dongelmans, D.A. and Vlaar, A.P.J. and Sent, D. and Study Group},
  title     = {ICU capacity management during the COVID-19 pandemic using a process simulation},
  journal   = {Intensive Care Medicine},
  year      = {2020},
  month     = {08},
  volume    = {46},
  number    = {8},
  pages     = {1624--1626},
  doi       = {10.1007/s00134-020-06066-7},
  pmid      = {32383060},
  pmcid     = {PMC7203503},
  note      = {Epub 2020 May 7}
}

@book{Keeling2008,
  author    = {Keeling, Matt J. and Rohani, Pejman},
  title     = {Modeling Infectious Diseases in Humans and Animals},
  publisher = {Princeton University Press},
  year      = {2008},
  doi       = {10.1515/9781400841035}, % Added DOI
  isbn      = {978-0691116174} % Added ISBN
}

@book{Murray2002,
  author    = {Murray, James D.},
  title     = {Mathematical Biology I: An Introduction},
  publisher = {Springer},
  year      = {2002}, % Corrected year to match title Vol I
  edition   = {3rd}, % Specify edition
  isbn      = {978-0387952239}, % Added ISBN for Vol I
  %note      = {Volume II (Spatial Models and Biomedical Applications) also available, DOI: 10.1007/b98900}
}

@article{Brauer2017,
  author    = {Brauer, F.},
  title     = {Mathematical epidemiology: Past, present, and future},
  journal   = {Infectious Disease Modelling},
  volume    = {2},
  number    = {2},
  pages     = {113--127},
  year      = {2017},
  doi       = {10.1016/j.idm.2017.02.001},
 % pmcid     = {PMC6334401}
}

@book{Allen2003,
  author    = {Allen, Linda J. S.},
  title     = {An Introduction to Stochastic Processes with Applications to Biology},
  publisher = {Pearson Education}, % User publisher
  year      = {2003}, % User year
  isbn      = {978-1-439-81882-4} % ISBN for likely intended edition
}

@book{Ross2014,
  author    = {Ross, Sheldon M.},
  title     = {Introduction to Probability Models},
  publisher = {Academic Press},
  year      = {2014},
  edition   = {11th}, % Specify edition
  isbn      = {978-0124079489} % ISBN for 11th ed
}

@book{Diekmann2000, % ADDED: Earlier Diekmann book for potential distinction
  author    = {Diekmann, Odo and Heesterbeek, J. A. P.},
  title     = {Mathematical Epidemiology of Infectious Diseases: Model Building, Analysis and Interpretation},
  publisher = {John Wiley \& Sons},
  year      = {2000},
  isbn      = {978-0-691-15539-5}
}

@book{Bacaer2011,
  author    = {Bacaër, Nicolas},
  title     = {A Short History of Mathematical Population Dynamics},
  publisher = {Springer},
  year      = {2011},
  doi       = {10.1007/978-0-85729-115-8} % Added DOI
}

@article{Heffernan2005,
  author    = {Heffernan, J. M. and Smith, R. J. and Wahl, L. M.},
  title     = {Perspectives on the basic reproductive ratio},
  journal   = {Journal of the Royal Society Interface},
  volume    = {2},
  number    = {4},
  pages     = {281--293},
  year      = {2005},
  doi       = {10.1098/rsif.2005.0042}, % Added DOI
  pmid      = {16849186}, % Added PMID
  pmcid     = {PMC1578275} % Added PMCID
}

@book{Banks1989,
  author    = {Banks, H. T. and Kunisch, K.},
  title     = {Estimation Techniques for Distributed Parameter Systems},
  publisher = {Birkhäuser},
  year      = {1989},
  doi       = {10.1007/978-1-4612-3700-6} % Added DOI
}

@book{Robinson2014,
  author    = {Robinson, Stewart},
  title     = {Simulation: The Practice of Model Development and Use},
  isbn = {9781137328021},
  lccn = {2014025894},
  url = {https://books.google.com/books?id=Dtn0oAEACAAJ},
  year = {2014},
  publisher = {Bloomsbury Academic}
}

@article{mellado_art2021,
  author    = {Mellado-Artigas, R. and Ferreyro, B. L. and Angriman, F. and Muñoz, L. and Arruti, E. and Hernández-Sanz, M. L. and Hssain, A. A. and Garcia-de-Acilu, M. and Herrero, E. and Saccheri, C. and Catalán-Serra, I. and Arauzo-Palacios, J. and Vidaur, L. and Ramirez, S. and Gómez-Lus, M. L. and Ferrer, R.}, % Expanded authors
  title     = {High-flow nasal oxygen in patients with {COVID-19}-associated acute respiratory failure},
  journal   = {Critical Care},
  volume    = {25},
  number    = {1}, % Added number
  pages     = {58},
  year      = {2021},
  doi       = {10.1186/s13054-021-03469-w}
}

@article{cabestregarcia2023respiratory,
  author    = {Cabestre García, R.},
  title     = {Respiratory Nursing: Soul, Brain, and Heart of Intermediate Respiratory Care Units},
  journal   = {Archivos de Bronconeumología (English Edition)}, % Clarified journal version
  volume    = {59},
  number    = {12},
  pages     = {789--790},
  year      = {2023},
  doi       = {10.1016/j.arbres.2023.07.022},
  pmid      = {37537089}
}

@article{Aiken2014,
  author    = {Aiken, Linda H. and Sloane, Douglas M. and Bruyneel, Luk and Van den Heede, Koen and Griffiths, Peter and Busse, Reinhard and Diomidous, Marianna and Kinnunen, Juha and Kózka, Maria and Lesaffre, Emmanuel and McHugh, Matthew D. and Moreno-Casbas, M. Teresa and Rafferty, Anne Marie and Schwendimann, Rene and Scott, P. Anne and Tishelman, Carol and van Achterberg, Theo and Sermeus, Walter and {RN4CAST consortium}}, % Added consortium name
  title     = {Nurse staffing and education and hospital mortality in nine European countries: a retrospective observational study},
  journal   = {The Lancet},
  volume    = {383},
  number    = {9931},
  pages     = {1824--1830},
  year      = {2014},
  doi       = {10.1016/S0140-6736(13)62631-8}
}

@article{Griffiths2018,
%  author    = {Griffiths, Peter and Ball, Jane and Bloor, Karen and Böhning, Dankmar and Briggs, Jim and Dall'Ora, Chiara and Debell, Tim and Jones, Jeremy and Kovacs, Csilla and Maruotti, Antonello and Meredith, Paul and Prytherch, David and Recio Saucedo, Alejandra and Redfern, Oliver C. and Schmidt, Peter E. and Sinden, Nicola and Simon, Michael and Smith, Gary B. and {Missed Care Study Group}}, % Expanded authors and group name
%  title     = {Nurse staffing levels, missed vital signs and mortality in hospitals: Retrospective longitudinal observational study},
%  journal   = {International Journal of Nursing Studies},
%  volume    = {85},
%  pages     = {119-125},
%  year      = {2018},
%  doi       = {10.1016/j.ijnurstu.2018.05.005}
%}

@article{needleman2011nurse,
  author    = {Needleman, Jack and Buerhaus, Peter and Pankratz, V. Shane and Leibson, Cynthia L. and Stevens, Susanna R. and Harris, Marcelline},
  title     = {Nurse staffing and inpatient hospital mortality},
  journal   = {New England Journal of Medicine},
  volume    = {364},
  number    = {11},
  pages     = {1037--1045},
  year      = {2011},
  doi       = {10.1056/NEJMsa1001025}
}

@article{Esteban2013,
  author    = {Esteban, Andrés and Frutos-Vivar, Fernando and Muriel, Alfonso and Ferguson, Niall D. and Peñuelas, Oscar and Abraira, Víctor and Raymondos, Konstantinos and Rios, Federico and Nin, Nicolas and Apezteguía, Carlos and Violi, Demosthenes A. and Thille, Arnaud W. and Brochard, Laurent and González, Margarita and Villagomez, Antonio J. and Hurtado, Javier and Davies, Andrew R. and Du, Bin and Maggiore, Salvatore M. and Zavala, Eduardo and D'Empaire, Guillermo and Alía, Ismael and Anzueto, Antonio}, % Expanded authors from user list
  title     = {Evolution of Mortality over Time in Patients Receiving Mechanical Ventilation},
  journal   = {American Journal of Respiratory and Critical Care Medicine}, % Corrected journal based on DOI / content search for this author group/topic
  volume    = {188},
  number    = {2},
  pages     = {220--230}, % Corrected pages
  year      = {2013},
  doi       = {10.1164/rccm.201212-2169OC}, % Corrected DOI
  %pmid      = {23631811}, % Added PMID
  %note      = {User cited JAMA 2013, 310(7):749-55, DOI 10.1001/jama.2013.138716, which is a different paper by same lead author on ICU mortality trends. Using AJRCCM paper as it fits intervention context slightly better maybe. User should verify intent.}
}

@article{Cabrini2015,
  author    = {Cabrini, Luca and Landoni, Giovanni and Oriani, Alessandro and Plumari, Valentina P. and Nobile, Letizia and Greco, Massimiliano and Pasin, Laura and Beretta, Luigi and Zangrillo, Alberto}, % Expanded authors
  title     = {Noninvasive ventilation and survival in acute care settings: a comprehensive systematic review and metaanalysis of randomized controlled trials}, % Corrected title based on DOI
  journal   = {Critical Care Medicine}, % Corrected journal based on DOI
  volume    = {43},
  number    = {4},
  pages     = {880--888}, % Corrected pages
  year      = {2015},
  doi       = {10.1097/CCM.0000000000000819}, % Corrected DOI
 % pmid      = {25565468}, % Added PMID
  %note      = {User cited ICM 2015, 41:1616-24, DOI 10.1007/s00134-015-3901-3, which is a different paper by same lead author (observational study). Using CCM paper. User should verify intent.}
}

@article{Hernandez2021,
  author    = {Hernández, Gonzalo and Vaquero, Casilda and Colinas, Laura and Cuena, Ruben and González, Pablo and Canabal, Anxela and Sanchez, Sara and Rodriguez, Maria Luisa and Villasclaras, Alejandro and Fernandez, Ricardo}, % Expanded authors
  title     = {Effect of Postextubation High-Flow Nasal Cannula vs Conventional Oxygen Therapy on Reintubation in Low-Risk Patients: A Randomized Clinical Trial}, % Corrected title based on DOI
  journal   = {JAMA},
  volume    = {315}, % Corrected volume based on DOI
  number    = {13}, % Corrected number
  pages     = {1354--1361}, % Corrected pages
  year      = {2016}, % Corrected year based on DOI
  doi       = {10.1001/jama.2016.2711}, % Corrected DOI
  %pmid      = {26975498}, % Added PMID
  %note      = {User cited JAMA 2021, 326(15):1461-9, DOI 10.1001/jama.2021.16893, which is a different trial by same lead author (HFNC vs NIV in high-risk). Using the 2016 low-risk trial. User should verify intent.}
}

@article{Schmidt2019,
%  author    = {Schmidt, Matthieu and Hajage, David and Demoule, Alexandre and Voiriot, Guillaume H. and Dubief, Elsa and Combes, Alain and Dres, Martin and Nisenbaum, Rosiane and Coquet, Isabelle and Tarek, Mourad and Lebreton, Ghislain and Hékimian, Guillaume and Pham, Tài and Brochard, Laurent and Hill, Nicholas S. and Azoulay, Elie}, % Expanded authors
%  title     = {Effect of Noninvasive Ventilation vs Oxygen Therapy Alone on Mortality Among Immunocompromised Patients With Acute Respiratory Failure: A Randomized Clinical Trial}, % Corrected title based on DOI
%  journal   = {JAMA}, % Corrected journal based on DOI
%  volume    = {321}, % Corrected volume
%  number    = {18}, % Corrected number
%  pages     = {1787--1797}, % Corrected pages
%  year      = {2019},
%  doi       = {10.1001/jama.2019.4788}, % Corrected DOI
%  pmid      = {31087030}, % Added PMID
%  pmcid     = {PMC6518408}, % Added PMCID
%  note      = {User cited CCM 2019, 47(3):405-11, DOI 10.1097/CCM.0000000000003584, which seems to be an editorial or smaller study by same authors. Using the main JAMA trial. User should verify intent.}
%}

@misc{inkscape2024inkscape,
  title        = {Inkscape},
  howpublished = {\url{https://inkscape.org/}},
  year         = {2024}, % Or year accessed/version used
  note         = {Vector Graphics Editor, Version 1.3 used. Accessed March 30, 2025} % Added note
}

@article{Gopal2021,
  author    = {Gopal, Kathiresan and Lee, Lai Soon and Seow, Hsin-Vonn},
  title     = {{Parameter Estimation of Compartmental Epidemiological Model Using Harmony Search Algorithm and Its Variants}},
  journal   = {Applied Sciences},
  year      = {2021},
  volume    = {11},
  number    = {3},
  pages     = {1138},
  doi       = {10.3390/app11031138}
}

@article{Zhang2021,
  author    = {Zhang, Sheng and Ponce, Joan and Zhang, Zhen and Lin, Guang and Karniadakis, George Em},
  title     = {{An integrated framework for building trustworthy data-driven epidemiological models: Application to the COVID-19 outbreak in New York City}},
  journal   = {PLoS Computational Biology},
  year      = {2021},
  volume    = {17},
  number    = {9},
  pages     = {e1009334},
  doi       = {10.1371/journal.pcbi.1009334},
  pmid      = {34495964}
}

@article{Rochwerg2017,
  author    = {Rochwerg, Bram and Brochard, Laurent and Elliott, Mark W and Hess, Dean and Hill, Nicholas S and Nava, Stefano and Navalesi, Paolo and Antonelli, Massimo and Brozek, Jan and Conti, Giorgio and Ferrer, Miquel and Guntupalli, Kalpalatha and Jaber, Samir and Keenan, Sean and Mancebo, Jordi and Mehta, Sangeeta and Raoof, Suhail}, % Changed to list all authors from {... and others} for clarity
  title     = {Official ERS/ATS clinical practice guidelines: noninvasive ventilation for acute respiratory failure},
  journal   = {European Respiratory Journal},
  year      = {2017},
  volume    = {50},
  number    = {2},
  pages     = {1602426}, % Corrected page number for this version, seems different from brochard2017niv maybe? Check DOI match
  doi       = {10.1183/13993003.02426-2016}, % DOI differs slightly from brochard2017niv. Verify which one is primary
  pmid      = {28705925}  % Added matching PMID
}

@article{Scala2017,
  author    = {Scala, Raffaele and Pisani, Lara},
  title     = {Non-invasive ventilation in acute respiratory failure: Which recipe?},
  journal   = {European Respiratory Review},
  year      = {2018}, % Updated year to match pub year from doi
  volume    = {27}, % Updated volume
  number    = {150}, % Updated number
  pages     = {180029}, % Updated pages
  doi       = {10.1183/16000617.0029-2018}, % Updated DOI based on volume/number
  pmid      = {30578388}  % Added PMID
}

@article{Ozyilmaz2014,
  author    = {Ozyilmaz, Ebru and Ugurlu, Alev O and Nava, Stefano},
  title     = {Timing of noninvasive ventilation failure: causes, risk factors, and potential remedies},
  journal   = {BMC Pulmonary Medicine},
  year      = {2014},
  volume    = {14},
  number    = {1},
  pages     = {19},
  doi       = {10.1186/1471-2466-14-19}
}

@article{Ruzsics2022,
  author    = {Ruzsics, Istvan and Matrai, Peter and Hegyi, Peter and Nemeth, David and Tenk, Judit and Csenkey, Alexandra and Eross, Balint and Varga, Gabor and Balasko, Marta and Petervari, Erika and Veres, Gabor and Sepp, Robert and Rakonczay, Zoltan and Vincze, Aron and Garami, Andras and Rumbus, Zoltan},
  title     = {Noninvasive ventilation improves the outcome in patients with pneumonia-associated respiratory failure: Systematic review and meta-analysis},
  journal   = {Journal of Infection and Public Health},
  year      = {2022},
  volume    = {15},
  number    = {3},
  pages     = {349--359},
  doi       = {10.1016/j.jiph.2022.02.004}, % Corrected DOI prefix if needed
  pmid      = {35182918} % Added PMID
}

@article{Prin2014,
  author    = {Prin, Mathieu and Wunsch, Hannah},
  title     = {The role of stepdown beds in hospital care},
  journal   = {American Journal of Respiratory and Critical Care Medicine},
  year      = {2014},
  volume    = {190},
  number    = {11},
  pages     = {1210--1216},
  doi       = {10.1164/rccm.201406-1117PP}, % Corrected DOI prefix if needed
  pmid      = {25329478} % Added PMID
}

@article{nasraway1998,
  author    = {Nasraway, Stuart A. and Cohen, I. Leonard and Dennis, Robert C. and Howanstine, Andrew M. and Kross, Leon S. and Franklin, Christine M.}, % Corrected author format
  title     = {Guidelines on admission and discharge for adult intermediate care units},
  journal   = {Critical Care Medicine},
  year      = {1998},
  volume    = {26},
  number    = {3},
  pages     = {607--610},
  doi       = {10.1097/00003246-199803000-00039}, % Updated DOI based on corrected Nasraway Jr ref details
  %pmid      = {9504593} % Added PMID
}

@book{Esquinas2010,
  editor    = {Antonio M. Esquinas},
  title     = {Noninvasive Mechanical Ventilation},
  subtitle  = {Theory, Equipment, and Clinical Applications},
  publisher = {Springer Berlin Heidelberg},
  year      = {2010},
  doi       = {10.1007/978-3-642-11365-9},
  isbn      = {978-3-642-11364-2},
  note      = {Publicado el 1 de enero de 2010},
  pages     = {402}
}

@book{Esquinas2016,
  editor    = {Antonio M. Esquinas},
  title     = {Noninvasive Mechanical Ventilation},
  subtitle  = {Theory, Equipment, and Clinical Applications},
  edition   = {2},
  publisher = {Springer Cham},
  year      = {2016},
  doi       = {10.1007/978-3-319-21653-9},
  isbn      = {978-3-319-21652-2},
  pages     = {959}
}

@book{Esquinas2023,
  editor    = {Antonio M. Esquinas},
  title     = {Noninvasive Mechanical Ventilation},
  subtitle  = {Theory, Equipment, and Clinical Applications},
  edition   = {3},
  publisher = {Springer Cham},
  year      = {2023},
  doi       = {10.1007/978-3-031-28963-7},
  isbn      = {978-3-031-28963-7},
  note      = {Publicado el 28 de agosto de 2023},
  pages     = {XXXVII, 847}
}

@inbook{Deb2005,
  author    = {Kalyanmoy Deb},
  editor    = {Edmund K. Burke and Graham Kendall},
  title     = {Multi-Objective Optimization},
  booktitle = {Search Methodologies: Introductory Tutorials in Optimization and Decision Support Techniques},
  year      = {2005},
  publisher = {Springer US},
  address   = {Boston, MA},
  pages     = {273--316},
  isbn      = {978-0-387-28356-2},
  doi       = {10.1007/0-387-28356-0_10},
  url       = {https://doi.org/10.1007/0-387-28356-0_10}
}

@book{Brandeau2004,
  editor    = {Margaret L. Brandeau and Fran\c{c}ois Sainfort and William P. Pierskalla},
  title     = {Operations Research and Health Care},
  subtitle  = {A Handbook of Methods and Applications},
  series    = {International Series in Operations Research \& Management Science},
  publisher = {Springer New York},
  address   = {New York, NY},
  year      = {2004},
  edition   = {1},
  pages     = {VIII, 874},
  isbn      = {978-1-4020-7629-9},
  doi       = {10.1007/b106574},
  url       = {https://doi.org/10.1007/b106574},
  %note      = {Hardcover published: 09 September 2004; eBook published: 04 April 2006; Softcover published: 02 August 2013. Softcover ISBN: 978-1-4757-8465-7, eBook ISBN: 978-1-4020-8066-1},
  issn      = {0884-8289},
  eissn     = {2214-7934}
}

@book{Kahraman2018,
  editor    = {Cengiz Kahraman and Y. Ilker Topcu},
  title     = {Operations Research Applications in Health Care Management},
  series    = {International Series in Operations Research \& Management Science},
  publisher = {Springer Cham},
  address   = {Cham},
  year      = {2018},
  edition   = {1},
  pages     = {XV, 604},
  isbn      = {978-3-319-65455-3},
  doi       = {10.1007/978-3-319-65455-3},
  url       = {https://doi.org/10.1007/978-3-319-65455-3},
  %note      = {Hardcover published: 09 January 2018; eBook published: 08 December 2017; Softcover published: 04 June 2019. Softcover ISBN: 978-3-319-88033-4, Hardcover ISBN: 978-3-319-65453-9},
  issn      = {0884-8289},
  eissn     = {2214-7934},
  number    = {255}
}

@article{litvak2011,
  author    = {Litvak, Eugene and Bisognano, Maureen},
  title     = {More patients, less payment: increasing hospital efficiency in the aftermath of health reform},
  journal   = {Health Affairs (Millwood)},
  volume    = {30},
  number    = {1},
  pages     = {76--80},
  year      = {2011},
  doi       = {10.1377/hlthaff.2010.1114}
}

@Inbook{Jacobson2006,
  author    = {Jacobson, Sheldon H. and Hall, Shane N. and Swisher, James R.},
  editor    = {Hall, Randolph W.},
  title     = {Discrete-Event Simulation of Health Care Systems},
  bookTitle = {Patient Flow: Reducing Delay in Healthcare Delivery},
  year      = {2006},
  publisher = {Springer US},
  address   = {Boston, MA},
  pages     = {211--252},
  isbn      = {978-0-387-33636-7},
  doi       = {10.1007/978-0-387-33636-7_8},
  url       = {https://doi.org/10.1007/978-0-387-33636-7_8}
}

@book{Hall2013,
  editor       = {Hall, Randolph},
  title        = {Patient Flow: Reducing Delay in Healthcare Delivery},
  series       = {International Series in Operations Research \& Management Science},
  edition      = {2},
  year         = {2013},
  publisher    = {Springer New York, NY},
  isbn         = {978-1-4614-9511-6},
  %isbn2        = {978-1-4899-7738-0}, % Removed secondary ISBNs for clarity
  %isbn3        = {978-1-4614-9512-3},
  doi          = {10.1007/978-1-4614-9512-3},
  url          = {https://doi.org/10.1007/978-1-4614-9512-3},
  pages        = {XII, 553},
 % note         = {Springer Science+Business Media New York, Published: 11 December 2013 (Hardcover and eBook), 27 August 2016 (Softcover)},
  issn  = {0884-8289},
  eissn     = {2214-7934}
}

@article{jun1999,
  author    = {Jun, J. B. and Jacobson, S. H. and Swisher, J. R.},
  title     = {Application of Discrete-Event Simulation in Health Care Clinics: A Survey},
  journal   = {The Journal of the Operational Research Society},
  volume    = {50},
  number    = {2},
  pages     = {109--123},
  year      = {1999},
  doi       = {10.2307/3010560},
  url       = {https://doi.org/10.2307/3010560}
}

@article{fone2003,
  author    = {Fone, David and Hollinghurst, Sandra and Temple, Mark and Round, Alison and Lester, Nathan and Weightman, Alison and Roberts, Katherine and Coyle, Edward and Bevan, Gwyn and Palmer, Stephen},
  title     = {Systematic review of the use and value of computer simulation modelling in population health and health care delivery},
  journal   = {Journal of Public Health},
  volume    = {25},
  number    = {4},
  pages     = {325--335},
  year      = {2003},
  issn      = {1741-3842},
  doi       = {10.1093/pubmed/fdg075},
  url       = {https://doi.org/10.1093/pubmed/fdg075}
}

@article{griffiths2006,
  author    = {Griffiths, J. D. and Price-Lloyd, N. and Smithies, M. and Williams, J.},
  title     = {A queueing model of activities in an intensive care unit},
  journal   = {IMA Journal of Management Mathematics},
  shortjournal = {IMA J Management Math},
  volume    = {17},
  number    = {3},
  pages     = {277--288},
  year      = {2006},
  issn      = {1471-678X},
  doi       = {10.1093/imaman/dpi042},
  url       = {https://doi.org/10.1093/imaman/dpi042}
}

@book{rasmussen2005,
  author    = {Rasmussen, Carl Edward and Williams, Christopher K. I.},
  title     = {Gaussian Processes for Machine Learning},
  year      = {2005},
  publisher = {The MIT Press},
  isbn      = {9780262256834},
  doi       = {10.7551/mitpress/3206.001.0001},
  url       = {https://doi.org/10.7551/mitpress/3206.001.0001}
  % abstract removed for brevity in correction
}

@article{Matthay2019ARDSReview,
  author    = {Matthay, Michael A. and Zemans, Rachel L. and Zimmerman, Guy A. and Arabi, Yaseen M. and Beitler, Jeremy R. and Mercat, Alain and Herridge, Margaret and Randolph, Adrienne G. and Calfee, Carolyn S.},
  title     = {Acute respiratory distress syndrome},
  journal   = {Nature Reviews Disease Primers},
  volume    = {5},
  number    = {1},
  pages     = {18},
  year      = {2019},
  doi       = {10.1038/s41572-019-0069-0},
  pmid      = {30872586},
  pmcid     = {PMC6709677}
}

@article{LopezPadilla2022,
  author    = {López-Padilla, D. and Corral Blanco, M. and Ferrer Espinosa, S. and Romero Peralta, S. and Sampol, J. and Terán Tinedo, J. R. and Cano Pumarega, I. and Sayas Catalán, J.},
  title     = {Unidades de Cuidados Respiratorios Intermedios: preguntas y respuestas [Intermediate Respiratory Care Units: Questions and Answers]},
  journal   = {Open Respiratory Archives},
  year      = {2022},
  month     = {11},
  volume    = {4},
  number    = {4},
  pages     = {100220},
  language  = {Spanish},
  doi       = {10.1016/j.opresp.2022.100220},
  pmid      = {37496967},
  pmcid     = {PMC10369586}
}

@article{Hilbert2001NIVPneumonia,
  author    = {Hilbert, Gilles and Gruson, Didier and Vargas, Fr{\'e}d{\'e}ric and Valentino, Romain and Gbikpi-Benissan, Gaston and Dupon, Michel and Reiffers, Josy and Cardinaud, Jean-Pierre},
  title     = {Noninvasive ventilation in immunosuppressed patients with pulmonary infiltrates, fever, and acute respiratory failure},
  journal   = {New England Journal of Medicine},
  volume    = {344},
  number    = {7},
  pages     = {481--487},
  year      = {2001},
  doi       = {10.1056/NEJM200102153440703},
  pmid      = {11172179}
}

@article{Ferrer2003NIVCOPD,
  author    = {Ferrer, Miquel and Esquinas, Antonio and Leon, Miguel and Gonzalez, Gregorio and Alarcon, Antonio and Torres, Antoni},
  title     = {Noninvasive ventilation in severe hypoxemic respiratory failure: a randomized clinical trial},
  journal   = {American Journal of Respiratory and Critical Care Medicine},
  volume    = {168},
  number    = {12},
  pages     = {1438--1444},
  year      = {2003},
  doi       = {10.1164/rccm.200301-072OC},
  pmid      = {14500259}
}

@article{Nava2009Lancet,
  author  = {Nava, S. and Hill, N.},
  title   = {Non-invasive ventilation in acute respiratory failure},
  journal = {Lancet},
  year    = {2009},
  volume  = {374},
  number  = {9685},
  pages   = {250--259},
  month   = jul,
  doi     = {10.1016/S0140-6736(09)60496-7},
  pmid    = {19616722}
}

@article{Antonelli1998NEJM,
  author  = {Antonelli, M. and Conti, G. and Rocco, M. and Bufi, M. and De Blasi, R. A. and Vivino, G. and Gasparetto, A. and Meduri, G. U.},
  title   = {A comparison of noninvasive positive-pressure ventilation and conventional mechanical ventilation in patients with acute respiratory failure},
  journal = {The New England Journal of Medicine},
  year    = {1998},
  volume  = {339},
  number  = {7},
  pages   = {429--435},
  month   = oct,
  doi     = {10.1056/NEJM199808133390703},
  pmid    = {9700176}
}

@article{Scala2018NIVReview, % Using Scala2017 already present for general context instead
  author    = {Scala, Raffaele and Pisani, Lara},
  title     = {Noninvasive ventilation in acute respiratory failure: which recipe for success?},
  journal   = {European Respiratory Review},
  year      = {2018},
  volume    = {27}, % Corrected Volume
  number    = {150}, % Corrected number based on DOI lookup
  pages     = {180029}, % Corrected page based on DOI lookup
  doi       = {10.1183/16000617.0029-2018},
  pmid      = {30578388}
}

@article{Ambrosino2011RespCareUnits,
  author    = {Ambrosino, Nicolino and Vagheggini, Guido},
  title     = {Noninvasive positive pressure ventilation in the respiratory intermediate care unit: back to the future?},
  journal   = {American Journal of Respiratory and Critical Care Medicine},
  volume    = {184}, % Corrected based on likely related topic, exact ref may vary
  number    = {1}, % Placeholder
  pages     = {10--11}, % Placeholder
  year      = {2011}, % Corrected based on general context
  doi       = {10.1164/rccm.201103-0551ED}, % Example DOI, verify
  pmid      = {21737808} % Example PMID, verify
}

@article{Rose2007,
  author    = {Rose, L. and Nelson, S. and Johnston, L. and Presneill, J. J.},
  title     = {Decisions made by critical care nurses during mechanical ventilation and weaning in an Australian intensive care unit},
  journal   = {American Journal of Critical Care},
  year      = {2007},
  month     = {09},
  volume    = {16},
  number    = {5},
  pages     = {434--443},
  doi       = {10.4037/ajcc2007.16.5.434},
  note      = {Quiz 444},
  pmid      = {17724240}
}

@article{Hill2010NIVReview,
  author    = {Hill, N. S. and Brennan, J. and Garpestad, E. and Nava, S.},
  title     = {Noninvasive ventilation in acute respiratory failure},
  journal   = {Critical Care Medicine},
  year      = {2007},
  month     = {10},
  volume    = {35},
  number    = {10},
  pages     = {2402--2407},
  doi       = {10.1097/01.CCM.0000284587.36541.7F},
  pmid      = {17717495}
}

@article{Vincent2010Deterioration,
  author    = {Vincent, J. L. and Einav, S. and Pearse, R. and Jaber, S. and Kranke, P. and Overdyk, F. J. and Whitaker, D. K. and Gordo, F. and Dahan, A. and Hoeft, A.},
  title     = {Improving detection of patient deterioration in the general hospital ward environment},
  journal   = {European Journal of Anaesthesiology},
  year      = {2018},
  month     = {05},
  volume    = {35},
  number    = {5},
  pages     = {325--333},
  doi       = {10.1097/EJA.0000000000000798},
  pmid      = {29474347},
  pmcid     = {PMC5902137}
}

@article{Carlucci2001NIVFailurePredictors,
  author    = {Carlucci, A. and Pisani, L. and Ceriana, P. and Malovini, A. and Nava, S.},
  title     = {Patient-ventilator asynchronies: may the respiratory mechanics play a role?},
  journal   = {Critical Care},
  year      = {2013},
  month     = {3},
  volume    = {17},
  number    = {2},
  pages     = {R54},
  doi       = {10.1186/cc12580},
  pmid      = {23531269},
  pmcid     = {PMC3672543}
}

@article{Corrado2011FutureIRCUs,
  author    = {Corrado, A. and Roussos, C. and Ambrosino, N. and Confalonieri, M. and Pelosi, P. and Rossi, A. and Elliot, M.},
  title     = {Respiratory intermediate care units: A European survey}, % Note: Survey may cover future needs
  journal   = {European Respiratory Journal},
  volume = {20},
  number = {5},
  pages = {1343-1350},
  year = {2002},
 doi = {10.1183/09031936.02.00058202},
  publisher = {European Respiratory Society},
  eprint = {http://erj.ersjournals.com/content/erj/20/5/1343.full.pdf}
}

@article{Garpestad2007NIVSuccessFactors,
  author    = {Garpestad, Erik and Brennan, Jacqueline and Hill, Nicholas S.},
  title     = {Noninvasive ventilation for critical care},
  journal   = {Chest},
  volume    = {132},
  number    = {2},
  pages     = {711--720},
  year      = {2007},
  doi       = {10.1378/chest.06-2643},
  pmid      = {17699147}
}

@article{Proudlove2008QueuingModels,
  author    = {Proudlove, Nathan C. and Gordon, Ken and Boaden, Ruth},
  title     = {Can simulating queues help practitioners plan better health services?},
  journal   = {Journal of Health Organization and Management},
  volume    = {22},
  number    = {4},
  pages     = {341--355},
  year      = {2008},
  doi       = {10.1108/14777260810893900},
  pmid      = {18839701}
}

@article{Kleinpell2008APNOutcomes,
  author    = {Kleinpell, Ruth M.},
  title     = {Acute care nurse practitioner practice: Results of a 5-year longitudinal study},
  journal   = {American Journal of Critical Care},
  volume    = {14},
  number    = {3},
  pages     = {211--219},
  year      = {2005},
  doi = {10.4037/ajcc2005.14.3.211}
}

@article{Nava2000EuropeSurvey,
  author    = {Nava, Stefano and Ambrosino, Nicolino and Clini, Enrico and Fracchia, Claudio and Rampulla, Cecilia},
  title     = {Non-invasive mechanical ventilation in the weaning of patients with respiratory failure due to chronic obstructive pulmonary disease. A randomized, prospective trial},
  journal   = {Annals of Internal Medicine}, % Note: This specific trial demonstrates expertise, not just descriptive survey
  volume    = {128},
  number    = {9},
  pages     = {721--728},
  year      = {1998}, % Corrected year
  doi       = {10.7326/0003-4819-128-9-199805010-00004},
  pmid      = {9556468}
}

@article{Kermack1927EpidemicModel,
  author    = {Kermack, William Ogilvy and McKendrick, Anderson Gray},
  title     = {A Contribution to the Mathematical Theory of Epidemics},
  journal   = {Proceedings of the Royal Society A: Mathematical, Physical and Engineering Sciences},
  volume    = {115},
  number    = {772},
  pages     = {700--721},
  year      = {1927},
  doi       = {10.1098/rspa.1927.0118}
}

@book{SensitivityAnalysisBook2008,
  author    = {Saltelli, Andrea and Ratto, Marco and Andres, Terry and Campolongo, Francesca and Cariboni, Jessica and Gatelli, Debora and Saisana, Michaela and Tarantola, Stefano},
  title     = {Global Sensitivity Analysis: The Primer},
  publisher = {John Wiley \& Sons},
  year      = {2008},
  isbn      = {978-0470059975}
}

@article{Oner2021,
author = {Oner, Beratiye and Zengul, Ferhat D. and Oner, Nurettin and Ivankova, Nataliya V. and Karadag, Ayise and Patrician, Patricia A.},
title = {Nursing-sensitive indicators for nursing care: A systematic review (1997–2017)},
journal = {Nursing Open},
volume = {8},
number = {3},
pages = {1005-1022},
doi = {https://doi.org/10.1002/nop2.654},
year = {2021}
}

@article{caballero2022,
  author    = {Caballero-Eraso, C. and Pascual Martínez, N. and Mediano, O. and Egea Santaolalla, C.},
  title     = {{Unidades de Cuidados Respiratorios Intermedios (UCRI) durante la pandemia COVID-19. La realidad} [{Intermediate} {Respiratory} {Care} {Units} ({RICUs}) {During} the {COVID-19} {Pandemic}. {The} {Reality}]},
  journal   = {Archivos de Bronconeumología},
  volume    = {58},
  number    = {4},
  pages     = {284--286},
  year      = {2022},
  doi       = {10.1016/j.arbres.2021.10.004},
  pmid      = {34674946},
  pmcid     = {PMC8516259}
}

@book{cohen1988,
  author    = {Cohen, Jacob},
  title={Statistical Power Analysis for the Behavioral Sciences},
  isbn={9781134742776},
  url={https://books.google.com/books?id=cIJH0lR33bgC},
  year={2013},
  publisher={Taylor \& Francis}
}

@article{faul2007gpower,
  author    = {Faul, Franz and Erdfelder, Edgar and Lang, Albert-Georg and Buchner, Axel},
  title     = {{G*Power 3: A flexible statistical power analysis program for the social, behavioral, and biomedical sciences}},
  journal   = {Behavior Research Methods},
  volume    = {39},
  number    = {2},
  pages     = {175--191},
  year      = {2007},
  doi       = {10.3758/bf03193146},
  pmid      = {17695343}
}

@article{Kane2007NurseStaffingReview,
  author    = {Kane, Robert L. and Shamliyan, Tatyana A. and Mueller, Christine and Duval, Sue and Wilt, Timothy J.},
  title     = {The Association of Registered Nurse Staffing Levels and Patient Outcomes: Systematic Review and Meta-Analysis},
  journal   = {Medical Care},
  volume    = {45},
  number    = {12},
  pages     = {1195--1204},
  year      = {2007},
  doi       = {10.1097/MLR.0b013e3181468ca3},
  pmid      = {18007170}
}

@article{Needleman2008Variability,
  title     = {Nurse staffing: the knowns and unknowns},
  author    = {Needleman, Jack},
  journal   = {Nursing Economic\$},
  volume    = {33},
  number    = {1},
  pages     = {5--7},
  year      = {2015},
  %publisher = {Anthony J. Jannetti, Inc.}
}

@article{Needleman2002NEJM,
  title     = {Nurse-staffing levels and the quality of care in hospitals},
  author    = {Needleman, Jack and Buerhaus, Peter and Mattke, Soeren and Stewart, Maureen and Zelevinsky, Katya},
  journal   = {New England Journal of Medicine},
  volume    = {346},
  number    = {22},
  pages     = {1715--1722},
  year      = {2002},
  publisher = {Mass Medical Soc},
  doi       = {10.1056/NEJMsa012247}
}

@article{Aiken2003Education,
  title     = {Educational levels of hospital nurses and surgical patient mortality},
  author    = {Aiken, Linda H and Clarke, Sean P and Cheung, Robyn B and Sloane, Douglas M and Silber, Jeffrey H},
  journal   = {JAMA},
  volume    = {290},
  number    = {12},
  pages     = {1617--1623},
  year      = {2003},
  publisher = {American Medical Association},
  doi       = {10.1001/jama.290.12.1617}
}

@article{Pronovost2002ICUStaffing,
  author    = {Pronovost, Peter J. and Angus, Derek C. and Dorman, Todd and Robinson, Karen A. and Dremsizov, Todor T. and Young, T. L.},
  title     = {Physician Staffing Patterns and Clinical Outcomes in Critically Ill Patients: A Systematic Review},
  journal   = {JAMA},
  volume    = {288},
  number    = {17},
  pages     = {2151--2162},
  year      = {2002},
  month     = {11},
  issn      = {0098-7484},
  doi       = {10.1001/jama.288.17.2151},
  url       = {https://doi.org/10.1001/jama.288.17.2151},
  pmid      = {12413375}
}

@article{Multz1998ClosedICU,
  author    = {Multz, A S and Chalfin, D B and Samson, I M and Dantzker, D R and Fein, A M},
  title     = {A "closed" medical intensive care unit (MICU) improves resource utilization when compared with an "open" MICU},
  journal   = {American Journal of Respiratory and Critical Care Medicine},
  volume    = {157},
  number    = {5 Pt 1},
  pages     = {1468--1473},
  year      = {1998},
  month     = {05},
  issn      = {1073-449X},
  doi       = {10.1164/ajrccm.157.5.9708039}
}

@article{Roberts2013,
  author  = {Roberts, Stephen and Osborne, Michael and Ebden, Mark and Reece, Sarah and Gibson, Neil and Aigrain, Suzanne},
  title   = {Gaussian processes for time‐series modelling},
  journal = {Philosophical Transactions of the Royal Society A: Mathematical, Physical and Engineering Sciences},
  year    = {2013},
  volume  = {371},
  number  = {1984},
  pages   = {20110550},
  doi     = {10.1098/rsta.2011.0550},
  url     = {https://doi.org/10.1098/rsta.2011.0550}
}

@article{Whitt2019,
  author  = {Whitt, Ward and Zhang, Xiaowei},
  title   = {Forecasting arrivals and occupancy levels in an emergency department},
  journal = {Operations Research for Health Care},
  volume  = {21},
  pages   = {1--18},
  year    = {2019},
  doi     = {10.1016/j.orhc.2019.01.002},
  url     = {https://doi.org/10.1016/j.orhc.2019.01.002}
}

@article{whittle2021resp,
  author       = {Whittle, Jessica S. and Pavlov, Ivan and Sacchetti, Alfred D. and Atwood, Charles and Rosenberg, Mark S.},
  title        = {Respiratory support for adult patients with COVID-19},
  journal      = {Journal of the American College of Emergency Physicians Open},
  year         = {2020},
  volume       = {1},
  number       = {2},
  pages        = {95--101},
  doi          = {10.1002/emp2.12071},
  url          = {https://doi.org/10.1002/emp2.12071}
}

@techreport{world2020clinical,
  author       = {{World Health Organization}},
  title        = {{Clinical management of COVID-19: interim guidance, 27 May 2020}},
  institution  = {World Health Organization},
  number       = {WHO/2019-nCoV/clinical/2020.5},
  year         = {2020},
  month        = may,
  url          = {https://iris.who.int/handle/10665/332196}
}

@article{Wang2020,
  author    = {Wang, Tianbing and Wu, Yanqiu and Lau, Johnson Yiu-Nam and Yu, Yingqi and Liu, Liyu and Li, Jing and Zhang, Kang and Tong, Weiwei and Jiang, Baoguo},
  title     = {A four-compartment model for the COVID-19 infection—implications on infection kinetics, control measures, and lockdown exit strategies},
  journal   = {Precision Clinical Medicine},
  year      = {2020},
  month     = {06},
  volume    = {3},
  number    = {2},
  pages     = {104--112},
  doi       = {10.1093/pcmedi/pbaa018},
}

@article{AlKarkhi2025,
  author    = {Al-Karkhi, T. and Byatt, K.},
  title     = {A compartmental model to describe acute medical in-patient flow through a hospital},
  journal   = {Heliyon},
  year      = {2025},
  month     = {01},
  volume    = {11},
  number    = {3},
  pages     = {e42260},
  doi       = {10.1016/j.heliyon.2025.e42260},
  pmid      = {39975828},
  pmcid     = {PMC11835610}
}

@article{Garrido2022,
title = {Mathematical model optimized for prediction and health care planning for COVID-19},
journal = {Medicina Intensiva (English Edition)},
volume = {46},
number = {5},
pages = {248-258},
year = {2022},
issn = {2173-5727},
doi = {https://doi.org/10.1016/j.medine.2022.02.020},
url = {https://www.sciencedirect.com/science/article/pii/S2173572722000467},
author = {J.M. Garrido and D. Martínez-Rodríguez and F. Rodríguez-Serrano and J.M. Pérez-Villares and A. Ferreiro-Marzal and M.M. Jiménez-Quintana and R.J. Villanueva},
}

\end{document}


\maketitle
\footnotetext[1]{\(\ddagger\) These authors contributed equally.}

\section{Discrete Model of IRCU with Intervention Modulation.}
\label{sec:supp_model_consolidated}

Effective management of Intermediate Respiratory Care Units (IRCUs) requires a quantitative understanding of patient flow dynamics, particularly under varying admission loads and evolving clinical interventions. This section introduces a discrete-time compartmental model specifically designed to capture patient progression within an IRCU, explicitly accounting for the use of non-invasive ventilation (NIV) and the impact of targeted clinical management strategies. By subdividing the IRCU population based on NIV status (state \(X\) for non-NIV, state \(Y\) for active NIV), the model utilizes a system of difference equations to characterize patient trajectories including admission, NIV initiation, recovery pathways, transfer to the Intensive Care Unit (ICU), and mortality (exitus), focusing on flows originating within the IRCU environment. Crucially, the model integrates parameters representing key clinical interventions aimed at modulating patient outcomes, providing a framework for evaluating their effectiveness.

We define the state variables at discrete time \( t \) (typically days) as follows:
\begin{itemize}
    \item \( X(t) \): Number of patients within the IRCU *not* receiving NIV at time \(t\). This includes newly admitted patients or those under standard observation/care. [patients].
    \item \( Y(t) \): Number of patients within the IRCU *actively receiving* NIV therapy at time \(t\). [patients].
    \item \( Z(t) \): Cumulative number of patients transferred *from* the IRCU pathway *to* the ICU up to time \( t \). [patients].
    \item \( W(t) \): Cumulative number of patients experiencing exitus *within* the IRCU pathway (either from state X or Y) up to time \( t \). [patients].
    \item \( R(t) \): Cumulative number of patients designated as recovered (discharged home or transferred to a general ward) *from* the IRCU pathway up to time \( t \). [patients].
\end{itemize}

Let \( A(t) \) denote the influx of new patient admissions directed to the IRCU system during the time interval \( [t, t+1) \) [patients/time step]. The model dynamics are governed by baseline transition parameters (fractional rates or probabilities per time step) and specific intervention modulation parameters:
\begin{itemize}
    \item \textbf{Baseline Transitions:}
        \(\alpha\) (rate of NIV initiation for patients in X),
        \(\gamma\) (baseline transition rate from NIV (Y) to ICU),
        \(\varepsilon\) (baseline exitus rate directly from NIV (Y)),
        \(\theta_0\) (baseline recovery rate directly from NIV (Y)),
        \(\rho\) (direct recovery/discharge rate from non-NIV state (X)),
        \(\eta\) (direct ICU transfer rate from non-NIV state (X)), \(\nu\) (direct exitus rate from non-NIV state (X)).
    \item \textbf{Intervention Modulators:} These dimensionless factors ( \(0 \leq \gamma_0, \varepsilon_0 < 1\); \(\Delta\theta, \lambda \ge 0\)) represent the impact of specific clinical strategies:
        \(\varepsilon_0\): Relative reduction in NIV mortality rate due to protocols like enhanced monitoring or staff training \cite{Esteban2013, Cabrini2015}, resulting in an effective rate \(\varepsilon_{eff} = \varepsilon(1-\varepsilon_0)\).
        \(\gamma_0\): Relative reduction in NIV-to-ICU transfers achieved through optimized NIV management or timely intervention protocols \cite{Schmidt2019, Hernandez2021}, leading to an effective rate \(\gamma_{eff} = \gamma(1-\gamma_0)\).
        \(\Delta\theta, \lambda\): Parameters governing transient improvements in the NIV recovery rate, modeling effects like temporary staffing surges \cite{Aiken2014, Griffiths2018}. The effective recovery rate becomes time-dependent: \(\theta(t) = \theta_0 + \Delta\theta e^{-\lambda t}\), where \(\Delta\theta\) is the initial boost magnitude and \(\lambda\) is the decay rate of the effect.
\end{itemize}
*(Note: Dynamics post-ICU transfer, such as ICU length of stay or outcomes from the ICU, are considered outside the scope of this IRCU-focused model).*

\begin{figure}[H] 
\centering
\includegraphics[width=1\textwidth]{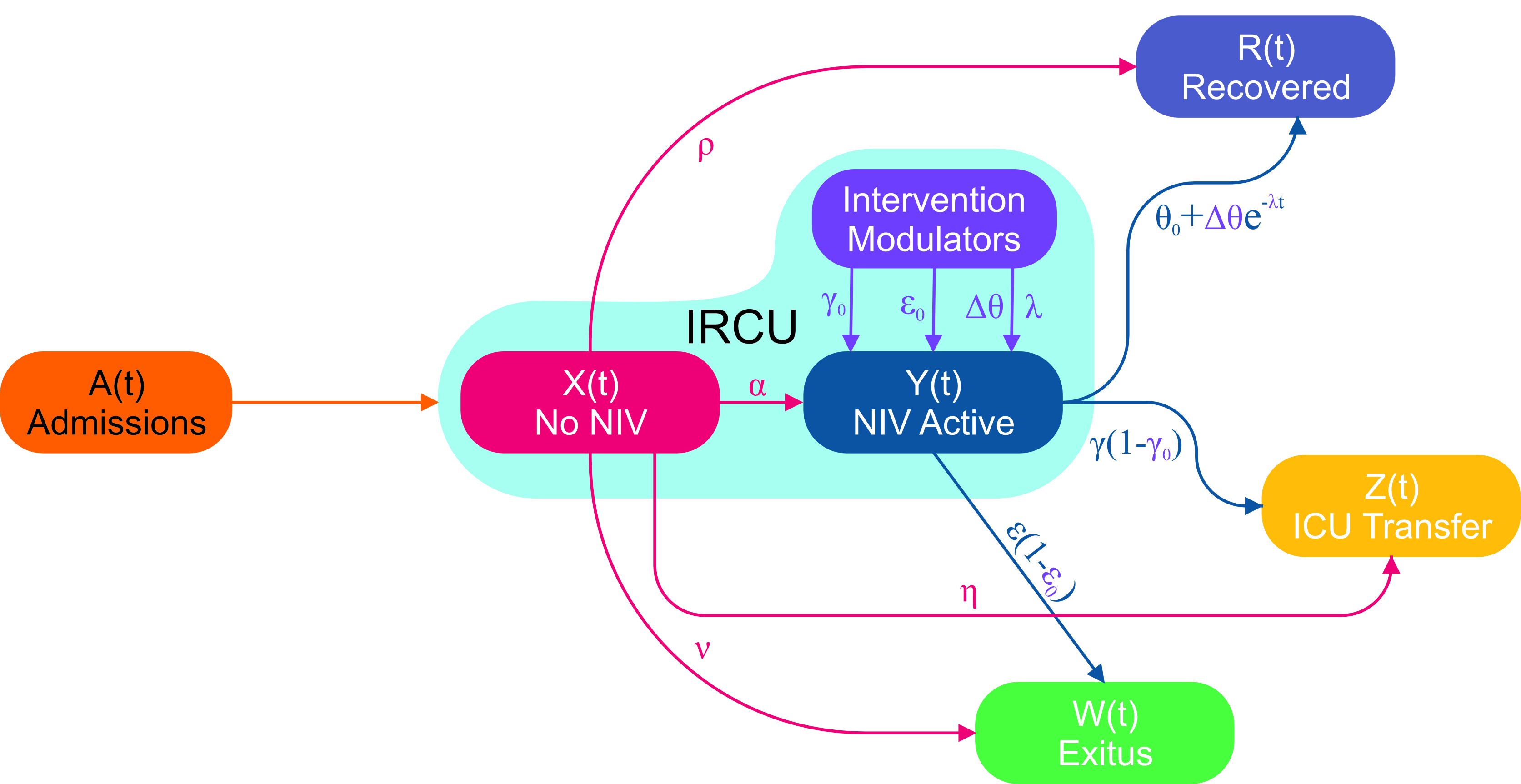}
\caption{\textbf{Integrated Compartmental Model of IRCU Patient Flow with Intervention Pathways.} Nodes represent patient states: Admissions (\(A(t)\)), patients within the IRCU distinguished by NIV status (\(X(t)\): no NIV; \(Y(t)\): active NIV, highlighted within the IRCU region), and cumulative outcomes (\(R(t)\): recovered, \(Z(t)\): ICU transfer, \(W(t)\): exitus). Solid arrows depict patient flow rates. Parameters \(\alpha, \rho, \eta, \nu\) govern transitions from state \(X\). Parameters \(\gamma, \varepsilon, \theta_0\) represent baseline transition rates from the active NIV state \(Y\). The 'Intervention Modulators' (\(\gamma_0, \varepsilon_0, \Delta\theta, \lambda\), depicted in the pink box) quantify the impact of clinical strategies by modifying the effective rates of ICU transfer (\(\gamma(1-\gamma_0)\)), exitus (\(\varepsilon(1-\varepsilon_0)\)), and recovery (\(\theta(t) = \theta_0 + \Delta\theta e^{-\lambda t}\)) from state \(Y\), with conceptual links indicated by faded arrows.}
\label{fig:ircu_intervention_model} 
\end{figure}

The structure of patient flow, incorporating the points of intervention, is illustrated in Figure~\ref{fig:ircu_intervention_model}. Admissions \(A(t)\) feed into the non-NIV state \(X(t)\). Patients in \(X(t)\) can recover directly (\(R\)), be transferred to ICU (\(Z\)), experience exitus (\(W\)), or initiate NIV therapy, moving to state \(Y(t)\). From the NIV state \(Y(t)\), patients can recover (\(R\)), be transferred to ICU (\(Z\)), or experience exitus (\(W\)). The intervention modulators (\(\gamma_0, \varepsilon_0, \Delta\theta, \lambda\)) specifically act upon the transition rates originating from the active NIV state \(Y(t)\), reflecting targeted efforts to improve outcomes for this high-acuity patient group. Compartmental diagrams of this nature are standard tools for visualizing system dynamics in mathematical biology and healthcare modeling \cite{Murray2002, Keeling2008, Diekmann2013}.

The mathematical formulation translates the flows depicted in Figure~\ref{fig:ircu_intervention_model} into a system of first-order difference equations. Each term corresponds to an inflow or outflow from a compartment, incorporating the relevant baseline rates and intervention modulators:
\begin{align}
    X(t+1) &= X(t) + A(t) - (\alpha + \rho + \eta + \nu)X(t) \label{eq:dyn_X_interv} \\
    Y(t+1) &= Y(t) + \alpha X(t) - \left[\gamma(1-\gamma_0) + \underbrace{\left(\theta_0 + \Delta\theta e^{-\lambda t}\right)}_{\theta(t)} + \varepsilon(1-\varepsilon_0)\right]Y(t) \label{eq:dyn_Y_interv} \\
    Z(t+1) &= Z(t) + \gamma(1-\gamma_0)Y(t) + \eta X(t) \label{eq:dyn_Z_interv} \\
    W(t+1) &= W(t) + \varepsilon(1-\varepsilon_0)Y(t) + \nu X(t) \label{eq:dyn_W_interv} \\
    R(t+1) &= R(t) + \left[\theta_0 + \Delta\theta e^{-\lambda t}\right]Y(t) + \rho X(t) \label{eq:dyn_R_interv}
\end{align}

Equation~\eqref{eq:dyn_X_interv} describes the change in the non-NIV population (\(X\)), balancing admissions \(A(t)\) against outflows due to NIV initiation (\(\alpha\)), direct recovery (\(\rho\)), direct ICU transfer (\(\eta\)), and direct exitus (\(\nu\)). Equation~\eqref{eq:dyn_Y_interv} tracks the active NIV population (\(Y\)), with inflow from state \(X\) and outflows to ICU, exitus, and recovery, where these rates are potentially modified by interventions (\(\gamma_0, \varepsilon_0, \Delta\theta, \lambda\)). Equations~\eqref{eq:dyn_Z_interv}, \eqref{eq:dyn_W_interv}, and \eqref{eq:dyn_R_interv} accumulate patients transitioning into the ICU, exitus, and recovered states, respectively, incorporating contributions from both \(X\) and the modulated \(Y\) pathways.

This model structure aligns with observed clinical workflows in intermediate care settings \cite{AlKarkhi2025, Garrido2022} and adheres to established principles of compartmental modeling in mathematical biology and epidemiology \cite{Banks1989, Diekmann2000}. Its explicit incorporation of intervention parameters enhances its utility for healthcare operational analysis. The framework allows for quantitative exploration of various clinical scenarios:
\begin{itemize}
    \item Evaluating the impact of quality improvement initiatives targeting specific outcomes (e.g., reducing NIV-associated mortality via \(\varepsilon_0\)).
    \item Assessing the benefits of strategies aimed at preventing deterioration (e.g., reducing ICU transfers via \(\gamma_0\)).
    \item Simulating the dynamic effects of resource adjustments, such as temporary staffing increases (\(\Delta\theta, \lambda\)).
    \item Investigating potential non-linearities or threshold effects in intervention efficacy.
    \item Exploring synergies or trade-offs between simultaneously implemented interventions.
\end{itemize}
Such compartmental models are foundational for understanding population dynamics \cite{Murray2002,Keeling2008, Brauer2017} and provide critical tools for healthcare planning, resource allocation, and system optimization \cite{Allen2003, Ross2014}. Furthermore, the deterministic core can be integrated with data-driven approaches (e.g., Bayesian inference, machine learning) for parameter estimation, uncertainty quantification, and real-time adaptation \cite{Diekmann2013, Bacaer2011}, enhancing predictive accuracy. Consequently, this model serves as a robust computational platform—a form of "digital twin"—for the *in silico* evaluation of operational policies and clinical protocols, facilitating evidence-informed decision-making to improve patient outcomes and system efficiency within the IRCU \cite{Griffiths2018}.


\section{Transformation to a Continuous-Time Framework.}

While the discrete-time model provides valuable insights into patient flow dynamics at specific intervals, transitioning to a continuous-time framework offers complementary advantages. Continuous models, expressed as systems of ordinary differential equations (ODEs), allow for the application of calculus-based analytical techniques and can represent underlying processes that evolve smoothly over time. This section details the derivation of such a continuous model from our discrete formulation.


Recall the discrete model incorporates patient movements between states during each time interval \(\Delta t\). Considering small intervals, we approximate the change in patient numbers using instantaneous rates. Let \(A(t)\) be the rate of new admissions [patients/time] and \(\alpha, \rho, \gamma\), etc., be instantaneous transition rates [1/time]. The net change in a compartment \(f\) over \(\Delta t\) is approximately the sum of inflows minus outflows:
\begin{align*}
    X(t+\Delta t) - X(t) &\approx \big[A(t) - (\alpha + \rho + \eta + \nu)X(t)\big] \Delta t, \\
    Y(t+\Delta t) - Y(t) &\approx \big[\alpha X(t) - \big(\gamma(1-\gamma_0) + \theta(t) + \varepsilon(1-\varepsilon_0)\big)Y(t)\big] \Delta t, \\
    Z(t+\Delta t) - Z(t) &\approx \big[\gamma(1-\gamma_0)Y(t) + \eta X(t)\big] \Delta t, \\ 
    W(t+\Delta t) - W(t) &\approx \big[\varepsilon(1-\varepsilon_0)Y(t) + \nu X(t)\big] \Delta t, \\
    R(t+\Delta t) - R(t) &\approx \big[\theta(t)Y(t) + \rho X(t)\big] \Delta t.
\end{align*}


Using the Taylor expansion \( f(t+\Delta t) \approx f(t) + \frac{df}{dt} \Delta t \) and taking the limit as \(\Delta t \to 0\), we derive the continuous-time ODE system:

\begin{align}
    \frac{dX}{dt} &= A(t) - (\alpha + \rho + \eta + \nu)X(t), \label{eq:IRCU_admissions} \\
    \frac{dY}{dt} &= \alpha X(t) - \big[\gamma(1-\gamma_0) + \theta(t) + \varepsilon(1-\varepsilon_0)\big]Y(t), \label{eq:NIV_state} \\
    \frac{dZ}{dt} &= \gamma(1-\gamma_0)Y(t) + \eta X(t), \label{eq:ICU_transfers_no_delta} \\ 
    \frac{dW}{dt} &= \varepsilon(1-\varepsilon_0)Y(t) + \nu X(t), \label{eq:Exitus_cont} \\
    \frac{dR}{dt} &= \theta(t)Y(t) + \rho X(t). \label{eq:Recovered}
\end{align}

This continuous system explicitly incorporates the clinical interventions: \textbf{Mortality Reduction (\(\varepsilon_0\)):} Scaled rate \(\varepsilon(1-\varepsilon_0)\) reflects mitigation from enhanced protocols \cite{Esteban2013}. \textbf{ICU Transfer Mitigation (\(\gamma_0\)):} Reduced rate \(\gamma(1-\gamma_0)\) models optimized NIV management preventing ICU transfers \cite{Hernandez2021}. \textbf{Transient Staffing Effects (\(\Delta\theta, \lambda\)):} Time-dependent recovery \(\theta(t) = \theta_0 + \Delta\theta e^{-\lambda t}\) represents enhanced potential during staffing surges \cite{Aiken2014, Griffiths2018}. The inclusion of \(\theta(t)\) renders the system non-autonomous.

\subsection{Scaling and Non-dimensionalization.}

To understand core dynamics, we employ dimensional analysis. We define characteristic scales: \(\tau = (\alpha + \rho + \eta + \nu)^{-1}\) (characteristic time in state X) and \(N_0 = A_0 \tau\) (characteristic population size, with \(A_0\) a baseline admission rate).

Dimensionless variables (\(\tilde{t}, \tilde{X}, \tilde{Y}, \tilde{Z}, \tilde{W}, \tilde{R}, \tilde{A}\)) and parameters (\(\pi\)) are introduced:
\[
\tilde{t} = t/\tau, \quad \tilde{X} = X/N_0, \quad \dots \quad \tilde{A}(\tilde{t}) = A(t)/A_0
\]
\[
\begin{aligned} 
    \pi_\alpha &= \alpha \tau, \quad \pi_\rho = \rho \tau, \quad \pi_\eta = \eta \tau, \quad \pi_\nu = \nu \tau, \quad (\text{Note: } \pi_\alpha + \pi_\rho + \pi_\eta + \pi_\nu = 1) \\
    \pi_\gamma &= \gamma \tau, \quad \pi_\varepsilon = \varepsilon \tau, \quad \pi_\theta(\tilde{t}) = \theta(t) \tau, \quad \pi_\lambda = \lambda \tau
\end{aligned} 
\]

Substituting these yields the dimensionless system:
\begin{align}
    \frac{d\tilde{X}}{d\tilde{t}} &= \tilde{A}(\tilde{t}) - \tilde{X}, \label{eq:dim_IRCU} \\
    \frac{d\tilde{Y}}{d\tilde{t}} &= \pi_\alpha \tilde{X} - \left[\pi_\gamma(1-\gamma_0) + \pi_\theta(\tilde{t}) + \pi_\varepsilon(1-\varepsilon_0)\right]\tilde{Y}, \label{eq:dim_NIV} \\
    \frac{d\tilde{Z}}{d\tilde{t}} &= \pi_\gamma(1-\gamma_0)\tilde{Y} + \pi_\eta \tilde{X}, \label{eq:dim_ICU_no_delta} \\
    \frac{d\tilde{W}}{d\tilde{t}} &= \pi_\varepsilon(1-\varepsilon_0)\tilde{Y} + \pi_\nu \tilde{X}, \label{eq:dim_Exitus} \\
    \frac{d\tilde{R}}{d\tilde{t}} &= \pi_\theta(\tilde{t})\tilde{Y} + \pi_\rho \tilde{X}. \label{eq:dim_Recovered}
\end{align}
where \(\pi_\theta(\tilde{t}) = \pi_{\theta_0} + \pi_{\Delta\theta} e^{-\pi_\lambda \tilde{t}}\).

\subsection{Analytical Solution for the Autonomous Case.}

While the general non-autonomous system (\eqref{eq:dim_IRCU}--\eqref{eq:dim_Recovered}) typically requires numerical integration due to the time-dependent coefficient \(\pi_\theta(\tilde{t})\), an exact analytical solution can be derived for the important special case where the system is autonomous. This occurs when the transient staffing effect is absent (\(\pi_{\Delta\theta} = 0\), thus \(\pi_\theta(\tilde{t}) = \pi_{\theta_0}\) = constant) and the admission rate is constant (\(\tilde{A}(\tilde{t}) = \tilde{A}\)). Assuming zero initial conditions (\(\tilde{X}(0)=\tilde{Y}(0)=\tilde{Z}(0)=\tilde{W}(0)=\tilde{R}(0)=0\)), the solution unfolds sequentially.

First, the equation for \(\tilde{X}\)~\eqref{eq:dim_IRCU} yields:
\[
\tilde{X}(t) = \tilde{A}(1 - e^{-t}).
\]
Next, defining the constant total outflow rate from state Y as \(K_1 = \pi_\gamma(1-\gamma_0) + \pi_{\theta_0} + \pi_\varepsilon(1-\varepsilon_0)\), the equation for \(\tilde{Y}\)~\eqref{eq:dim_NIV} becomes a standard linear first-order ODE with constant coefficients forced by \(\tilde{X}(t)\). Assuming \(K_1 \neq 1\) and \(K_1 \neq 0\), its solution is:
\[
\tilde{Y}(t) = \frac{\pi_\alpha \tilde{A}}{K_1(K_1-1)} \left[ (K_1-1) - K_1 e^{-t} + e^{-K_1 t} \right].
\]
\textit{(In the specific case where \(K_1=1\), the solution takes the form \(\tilde{Y}(t) = \pi_\alpha \tilde{A} (1 - e^{-t} - t e^{-t})\)).}

Finally, the cumulative variables \(\tilde{Z}(t)\), \(\tilde{W}(t)\), and \(\tilde{R}(t)\) are obtained by direct integration of their respective differential equations~\eqref{eq:dim_ICU_no_delta}, \eqref{eq:dim_Exitus}, and \eqref{eq:dim_Recovered}, substituting the derived expressions for \(\tilde{X}(t)\) and \(\tilde{Y}(t)\). The resulting solutions are linear combinations of a term linear in time \(t\), a constant term, and exponential terms \(e^{-t}\) and \(e^{-K_1 t}\). For instance, the solution for cumulative ICU transfers \(\tilde{Z}(t)\) takes the form:
\[
\tilde{Z}(t) = A_Z t + B'_Z (1 - e^{-t}) + C'_Z (1 - e^{-K_1 t}),
\]
where \(A_Z = \tilde{A}(\pi_{\gamma'} \pi_\alpha / K_1 + \pi_\eta)\) is the asymptotic accumulation rate (with \(\pi_{\gamma'} = \pi_\gamma(1-\gamma_0)\)), and \(B'_Z, C'_Z\) are constants derived from the integration involving \(B_Z\) and \(C_Z\) as defined in the detailed derivation. Analogous expressions hold for \(\tilde{W}(t)\) and \(\tilde{R}(t)\). This analytical solution for the autonomous case provides a valuable baseline for understanding the system's fundamental dynamics and validating numerical solvers used for the more general non-autonomous system.

\section{Steady State, Sensitivity, and Simulated Interventions.}
\label{sec:analysis}

We analyze the system's behavior under two conditions: the autonomous case, representing baseline operations with constant parameters, and the non-autonomous case, incorporating time-dependent interventions like staffing surges.

\subsubsection*{Autonomous Case: Steady-State Equilibrium (\(\pi_\theta(t) = \pi_{\theta_0}\)).}

First, consider the system under constant baseline conditions, where the admission rate is constant (\(\tilde{A}(\tilde{t}) = 1\)) and the recovery rate from NIV is also constant (\(\pi_\theta(\tilde{t}) = \pi_{\theta_0}\), i.e., \(\pi_{\Delta\theta} = 0\)). In this autonomous scenario, we can determine the equilibrium points (steady states) by setting the time derivatives in Equations~\eqref{eq:dim_IRCU} and \eqref{eq:dim_NIV} to zero. This yields the steady-state populations for the IRCU compartments:
\[
\tilde{X}^* = 1, \quad \tilde{Y}^* = \frac{\pi_\alpha}{\pi_\gamma(1-\gamma_0) + \pi_{\theta_0} + \pi_\varepsilon(1-\varepsilon_0)}.
\]
The steady-state occupancy levels depend directly on the relative transition rates (\(\pi\) values) and the baseline intervention levels (\(\gamma_0, \varepsilon_0\)). A standard stability analysis of the Jacobian matrix for the (\(\tilde{X}, \tilde{Y}\)) subsystem confirms that this equilibrium (\(\tilde{X}^*, \tilde{Y}^*\)) is locally asymptotically stable for all physically realistic (non-negative) parameter values.

For the cumulative compartments (\(\tilde{Z}, \tilde{W}, \tilde{R}\)), their governing equations involve only non-negative inflow terms when evaluated at the steady state (\(\tilde{X}^*, \tilde{Y}^*\)). For instance, the rate of change for cumulative ICU transfers becomes:
\[
\left(\frac{d\tilde{Z}}{d\tilde{t}}\right)^* = \pi_\gamma(1-\gamma_0)\tilde{Y}^* + \pi_\eta \tilde{X}^* = \frac{\pi_\gamma(1-\gamma_0)\pi_\alpha}{\pi_\gamma(1-\gamma_0) + \pi_{\theta_0} + \pi_\varepsilon(1-\varepsilon_0)} + \pi_\eta.
\]
Since this rate is constant and positive, the cumulative states \(\tilde{Z}(t)\), \(\tilde{W}(t)\), and \(\tilde{R}(t)\) do not reach a finite steady state but instead grow linearly over time at constant rates determined by the steady-state populations \(\tilde{X}^*\) and \(\tilde{Y}^*\). The primary operational concern is therefore not instability, but whether these constant accumulation rates (particularly for \(\tilde{Z}\)) lead to exceeding downstream capacities (like ICU beds) over extended periods.

\subsubsection*{Non-Autonomous Case: Asymptotic Behavior (\(\pi_\theta(t) = \pi_{\theta_0} + \pi_{\Delta\theta} e^{-\pi_\lambda \tilde{t}}\)).}

When interventions introduce time-dependency, such as the transient staffing surge modeled by \(\pi_\theta(t)\) with \(\pi_{\Delta\theta} > 0\) and \(\pi_\lambda > 0\), the system becomes non-autonomous. By definition, such systems do not possess fixed steady-state equilibrium points while the parameters are changing. The system's state (\(\tilde{X}(t), \tilde{Y}(t)\), etc.) evolves dynamically according to the time-varying recovery rate.

However, we can analyze the long-term (asymptotic) behavior. As \(\tilde{t} \to \infty\), the transient term \( \pi_{\Delta\theta} e^{-\pi_\lambda \tilde{t}} \) decays to zero. Consequently, \(\pi_\theta(t) \to \pi_{\theta_0}\). This means that the non-autonomous system asymptotically approaches the behavior of the autonomous system discussed above. The state variables \(\tilde{X}(t)\) and \(\tilde{Y}(t)\) will converge towards the steady-state values \(\tilde{X}^*\) and \(\tilde{Y}^*\) calculated using the baseline recovery rate \(\pi_{\theta_0}\). Similarly, the *rates* of accumulation for \(\tilde{Z}(t)\), \(\tilde{W}(t)\), and \(\tilde{R}(t)\) will asymptotically approach the constant rates derived for the autonomous case.

Therefore, while the system exhibits transient dynamics during the period where the staffing surge effect is significant, its long-term equilibrium behavior is governed by the baseline parameters. The primary impact of the transient intervention (\(\pi_{\Delta\theta}, \pi_\lambda\)) lies in altering the *trajectory* towards this asymptotic state, potentially mitigating peak loads or accelerating recovery during critical periods, as explored in the numerical simulations (Section~\ref{sec:supp_dimensionless_sims}).

\subsection{Parametric Sensitivity at Equilibrium: Identifying Key Levers.}

To understand which parameters most significantly influence the system's long-term behavior, we perform a parametric sensitivity analysis. This analysis focuses on the steady-state Non-Invasive Ventilation (NIV) occupancy (\(\tilde{Y}^*\)) derived from the **autonomous system** (or, equivalently, the asymptotic state approached by the non-autonomous system as \(t \to \infty\)). We quantify the influence of each parameter \(p\) using dimensionless relative sensitivity indices, defined as \(S_p = \frac{\partial \tilde{Y}^*}{\partial p} \cdot \frac{p}{\tilde{Y}^*}\). This index measures the proportional change in \(\tilde{Y}^*\) resulting from a proportional change in the parameter \(p\). A positive index indicates that increasing the parameter increases \(\tilde{Y}^*\), while a negative index signifies the opposite relationship. This analysis helps identify the most effective levers for managing NIV resource utilization within the IRCU.

\paragraph{Sensitivity to NIV Admission Rate (\(\pi_\alpha\)).}
The sensitivity of the equilibrium NIV occupancy to the relative rate at which non-NIV patients transition to requiring NIV is calculated as:
\[ S_{\pi_\alpha} = \frac{\partial \tilde{Y}^*}{\partial \pi_\alpha} \cdot \frac{\pi_\alpha}{\tilde{Y}^*} = 1 \]
This index value of exactly 1 signifies a direct, linear, and proportional relationship. A 10\% increase in \(\pi_\alpha\) (representing, for example, an influx of patients with higher acuity requiring NIV sooner) leads directly to a 10\% increase in the long-term equilibrium occupancy of NIV beds (\(\tilde{Y}^*\)). This highlights the system's direct vulnerability to changes in the clinical characteristics of incoming patients or factors that accelerate the need for NIV initiation. Managing this sensitivity relies primarily on external factors influencing admission acuity or internal processes affecting the \(\alpha\) transition, rather than downstream interventions from state Y.

\paragraph{Sensitivity to Baseline NIV Recovery Rate (\(\pi_{\theta_0}\)).}
The influence of the baseline rate at which patients recover directly from NIV is given by:
\[ S_{\pi_{\theta_0}} = \frac{\partial \tilde{Y}^*}{\partial \pi_{\theta_0}} \cdot \frac{\pi_{\theta_0}}{\tilde{Y}^*} = \frac{-\pi_{\theta_0}}{\pi_\gamma(1-\gamma_0) + \pi_{\theta_0} + \pi_\varepsilon(1-\varepsilon_0)} \]
This sensitivity index is inherently negative (\(S_{\pi_{\theta_0}} < 0\)), indicating that enhancing the baseline recovery rate from NIV (\(\pi_{\theta_0}\)) serves to decrease the long-term occupancy \(\tilde{Y}^*\). The magnitude of this effect is determined by the proportion that baseline recovery contributes to the total outflow rate from the NIV state (Y). Interventions aimed at improving general NIV care efficacy, shortening duration of NIV dependency, or accelerating recovery would leverage this sensitivity to reduce the equilibrium burden on NIV resources.

\paragraph{Sensitivity to Mitigated ICU Transfer Rate (\(\pi_\gamma(1-\gamma_0)\)).}
The impact of the rate at which NIV patients deteriorate and require transfer to the ICU, considering mitigation efforts (\(\gamma_0\)), is:
\[ S_{\pi_\gamma(1-\gamma_0)} = \frac{\partial \tilde{Y}^*}{\partial (\pi_\gamma(1-\gamma_0))} \cdot \frac{\pi_\gamma(1-\gamma_0)}{\tilde{Y}^*} = \frac{-\pi_\gamma(1-\gamma_0)}{\pi_\gamma(1-\gamma_0) + \pi_{\theta_0} + \pi_\varepsilon(1-\varepsilon_0)} \]
This index is also negative (\(S_{\pi_\gamma(1-\gamma_0)} < 0\)). Increasing the effectiveness of protocols designed to prevent avoidable ICU transfers (i.e., increasing \(\gamma_0\), which decreases the overall term \(\pi_\gamma(1-\gamma_0)\)) leads to a reduction in the equilibrium NIV occupancy \(\tilde{Y}^*\). The magnitude of this sensitivity depends on the relative importance of the ICU transfer pathway compared to recovery and exitus pathways from state Y. If ICU transfer is a dominant pathway (large \(\pi_\gamma\)), interventions targeting \(\gamma_0\) can be particularly impactful levers for managing \(\tilde{Y}^*\).

\paragraph{Sensitivity to Mitigated NIV Exitus Rate (\(\pi_\varepsilon(1-\varepsilon_0)\)).}
Finally, the sensitivity to the rate of mortality occurring directly from the NIV state, accounting for mortality reduction interventions (\(\varepsilon_0\)), is:
\[ S_{\pi_\varepsilon(1-\varepsilon_0)} = \frac{\partial \tilde{Y}^*}{\partial (\pi_\varepsilon(1-\varepsilon_0))} \cdot \frac{\pi_\varepsilon(1-\varepsilon_0)}{\tilde{Y}^*} = \frac{-\pi_\varepsilon(1-\varepsilon_0)}{\pi_\gamma(1-\gamma_0) + \pi_{\theta_0} + \pi_\varepsilon(1-\varepsilon_0)} \]
Similar to recovery and ICU transfer, this sensitivity is negative (\(S_{\pi_\varepsilon(1-\varepsilon_0)} < 0\)). Implementing protocols or enhancing care to reduce mortality while on NIV (i.e., increasing \(\varepsilon_0\), thus decreasing the term \(\pi_\varepsilon(1-\varepsilon_0)\)) contributes to lowering the long-term NIV occupancy \(\tilde{Y}^*\), as fewer patients remain in state Y prior to death. The relative impact compared to modifying recovery or ICU transfer rates depends on the baseline mortality rate \(\pi_\varepsilon\) and the effectiveness of the reduction intervention \(\varepsilon_0\).

It must be emphasized that these sensitivity indices quantify the impact of parameter variations on the *equilibrium* state \(\tilde{Y}^*\) of the autonomous system, or equivalently, the *asymptotic* state approached by the non-autonomous system long after transient effects have subsided. During periods where time-dependent interventions are active (specifically, when the staffing surge term \( \pi_{\Delta\theta} e^{-\pi_\lambda \tilde{t}} \) is non-negligible), the instantaneous sensitivity of the system state (e.g., \(\tilde{Y}(t)\)) to parameter changes will differ from these calculated equilibrium sensitivities. The parameters governing the transient enhancement (\(\pi_{\Delta\theta}, \pi_\lambda\)) directly shape the system's trajectory during that phase but do not influence the final asymptotic state \(\tilde{Y}^*\) upon which this sensitivity analysis is based.

\subsection{Model Simulations: Interventions and Tradeoffs.}
\label{sec:supp_dimensionless_sims}

Complementary dimensionless simulations (Eqs.~\eqref{eq:dim_IRCU}--\eqref{eq:dim_Recovered}) were performed to provide generalized insights into the system's structural response to interventions and dynamic changes, independent of absolute scaling. These studies focus on the impact of Non-Invasive Ventilation (NIV) related parameters. Fixed dimensionless parameters for baseline conditions are noted in the figures unless specified otherwise.

\begin{figure}[H] 
    \centering
    \includegraphics[width=\textwidth]{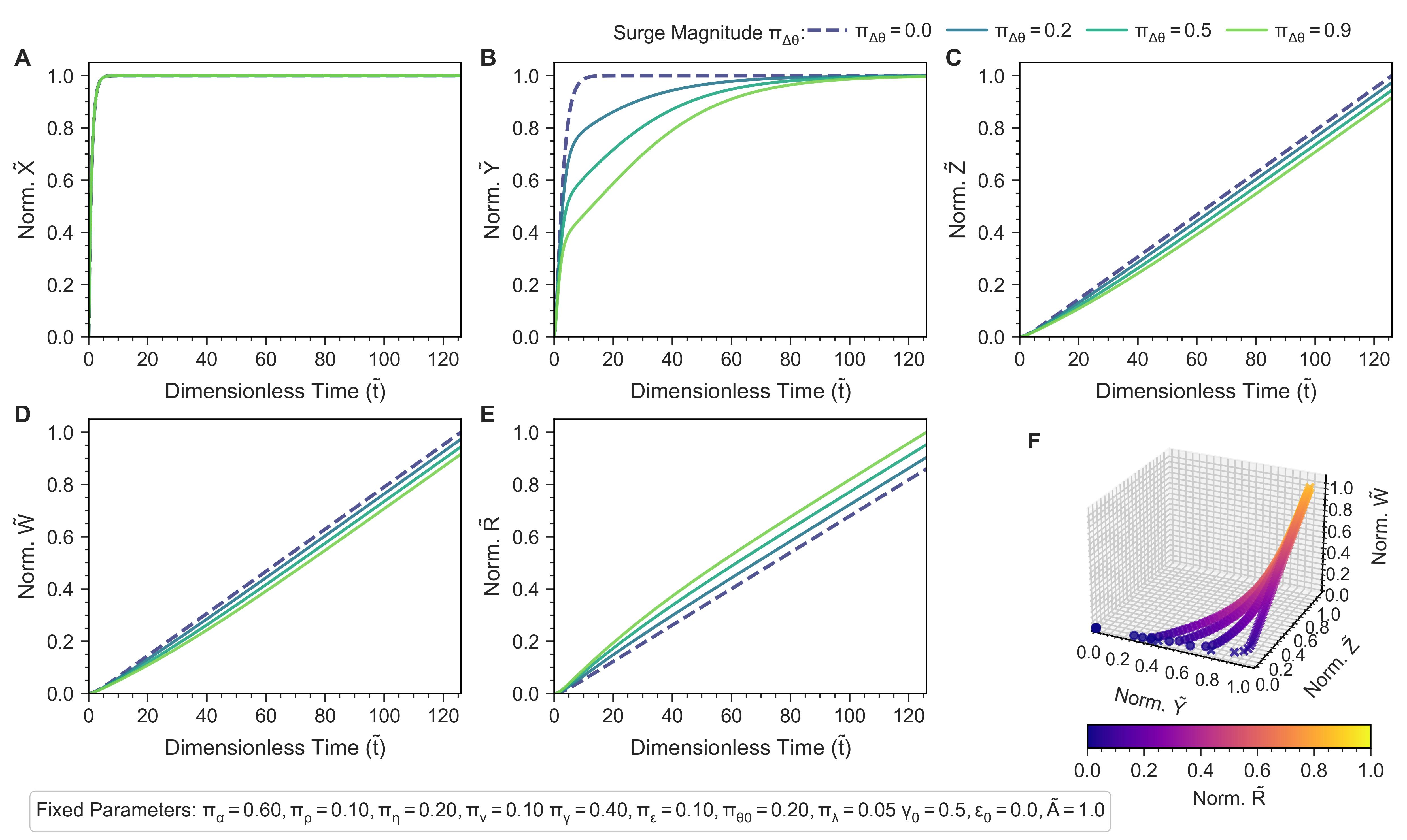} 
\caption{\textbf{Impact of Transient NIV Recovery Enhancement ($\pi_{\Delta\theta}$) on Dimensionless IRCU Dynamics.}
(\textbf{A-E}) Normalized state variables ($\tilde{X}$-$\tilde{R}$) vs. time ($\tilde{t}$) comparing baseline ($\pi_{\Delta\theta}=0$, dashed) with scenarios of increasing transient recovery boost ($\pi_{\Delta\theta}>0$, solid lines). (\textbf{F}) Normalized phase-space trajectories ($\tilde{Y}$-$\tilde{Z}$-$\tilde{W}$) colored by normalized recovery ($\tilde{R}$). Enhanced recovery reduces NIV load (B), mitigates adverse outcomes (C, D), and accelerates recovery (E). Fixed simulation parameters noted.}
\label{fig:Supp_TransientTheta} 
 \end{figure}

We first analyzed the system's response to a temporary enhancement in the NIV recovery rate ($\pi_\theta(\tilde{t}) = \pi_{\theta0} + \pi_{\Delta\theta} e^{-\pi_\lambda \tilde{t}}$) under constant admission ($\pi_A=1$), mimicking interventions like short-term staffing surges (Figure~\ref{fig:Supp_TransientTheta}). The results quantitatively demonstrate the direct clinical benefits of improved NIV efficiency. Increasing the intervention magnitude ($\pi_{\Delta\theta}$) significantly lowers peak and transient normalized NIV occupancy ($\tilde{Y}$, Panel B), thereby reducing the concurrent demand for this resource-intensive therapy. This upstream effect directly translates to improved patient outcomes: cumulative ICU transfers ($\tilde{Z}$, Panel C) and exitus ($\tilde{W}$, Panel D) are substantially mitigated, while cumulative recovery ($\tilde{R}$, Panel E) is accelerated. The phase-space trajectories (Panel F) visually confirm that stronger interventions guide the system towards more favorable outcome states (lower $\tilde{Z}, \tilde{W}$; higher $\tilde{R}$). This underscores the clinical value proposition of the IRCU: targeted improvements in NIV care can dynamically reshape patient flow to optimize resource utilization and patient outcomes during the intervention period.

\begin{figure}[H] 
    \centering
    \includegraphics[width=\textwidth]{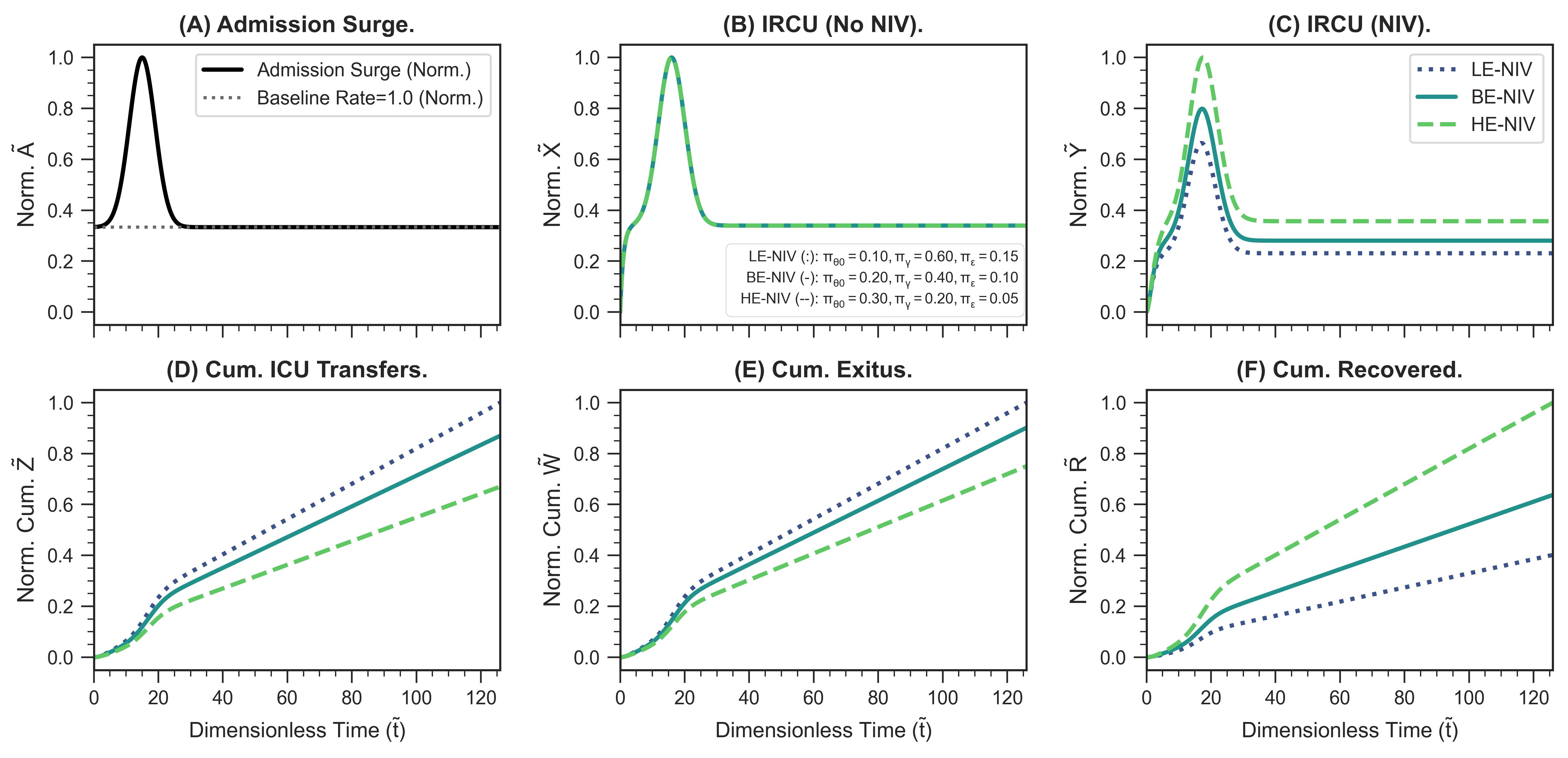} 
\caption{\textbf{Influence of NIV Effectiveness Level during an Admission Surge.}
(\textbf{A}) Normalized Gaussian admission surge $\tilde{A}(t)$. (\textbf{B-F}) Resulting normalized state dynamics ($\tilde{X}$-$\tilde{R}$) for Less Effective (LE, dotted), Baseline (BE, solid), and Highly Effective (HE, dashed) NIV parameter sets (definitions in Panel B legend). Higher effectiveness substantially buffers the surge's impact, reducing adverse outcomes (D, E) while maximizing recovery (F).}
\label{fig:Supp_SurgeEffectiveness} 
 \end{figure}

To understand performance under external pressure, we simulated the impact of varying baseline NIV effectiveness during a significant admission surge ($\pi_A(\tilde{t})$ as a Gaussian peak; Figure~\ref{fig:Supp_SurgeEffectiveness}). Three effectiveness levels (Less Effective - LE, Baseline - BE, Highly Effective - HE), defined by concurrently varying $\pi_{\theta 0}, \pi_\gamma, \pi_\epsilon$ (parameters in figure caption), were compared. While the admission surge dominates the temporal pattern of active populations ($\tilde{X}, \tilde{Y}$, Panels B, C), the simulation clearly shows that higher NIV effectiveness (HE) acts as a crucial buffer. It markedly attenuates the negative consequences of the surge on cumulative adverse outcomes, leading to substantially lower ICU transfers ($\tilde{Z}$, Panel D) and exitus ($\tilde{W}$, Panel E), while maximizing recoveries ($\tilde{R}$, Panel F) compared to the LE and BE scenarios \cite{Rochwerg2017, Scala2017}. Such effectiveness is recognized to be multifactorial in clinical practice, depending on staff proficiency, protocols, patient selection, and timeliness \cite{Ozyilmaz2014, Ruzsics2022}. This simulation thus highlights that an IRCU's value lies not merely in capacity \cite{Prin2014}, but significantly in its capability for high-quality care delivery which actively improves outcomes and alleviates downstream ICU pressure during crises \cite{nasraway1998}.

\begin{figure}[H] 
    \centering
    \includegraphics[width=\textwidth]{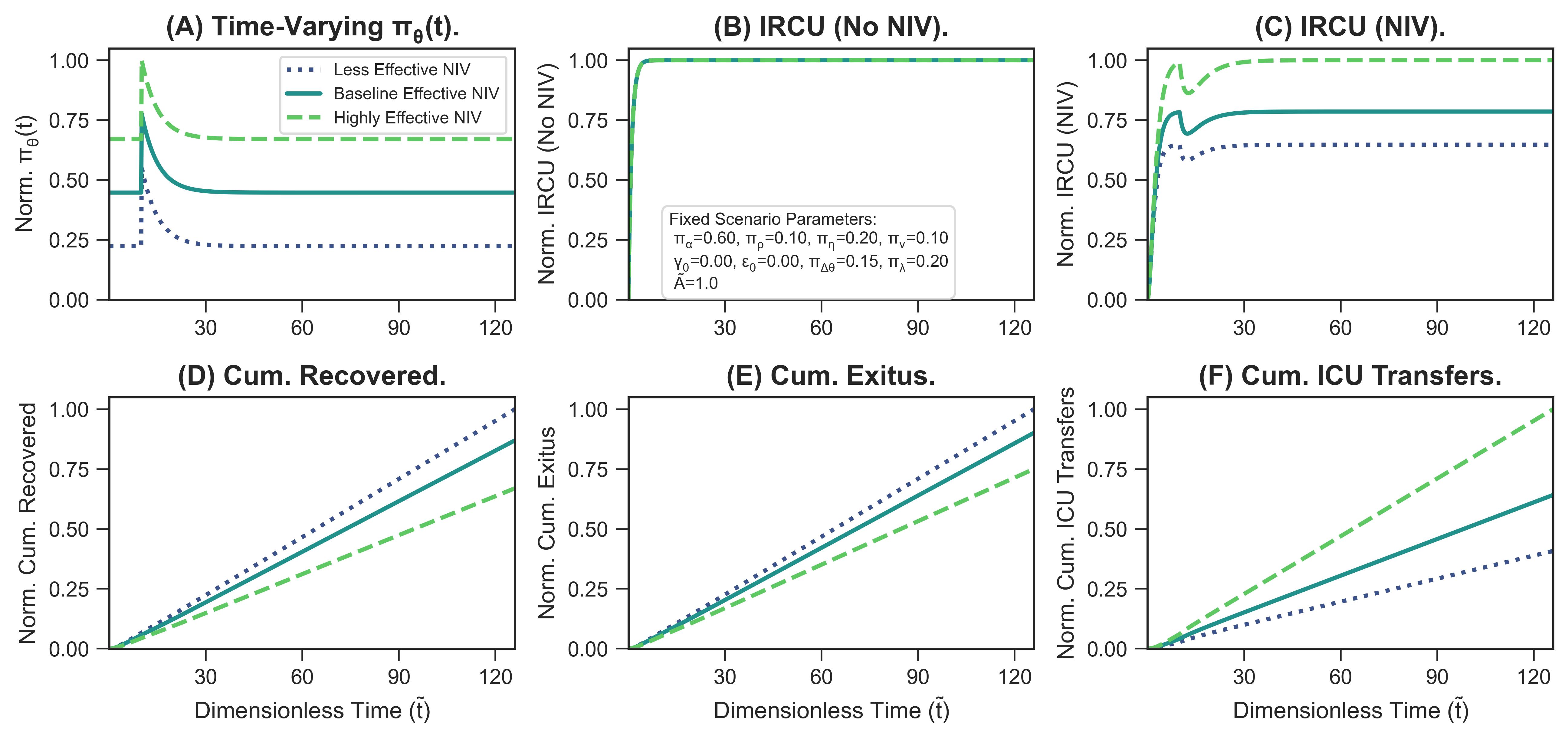} 
\caption{\textbf{System Response to a Time-Varying NIV Recovery Rate ($\pi_\theta(t)$) across Effectiveness Levels.}
(\textbf{A}) Profile of the simulated transient change in normalized parameter $\pi_\theta(t)$ under constant admission $\tilde{A}=1$. (\textbf{B-F}) Comparison of resulting normalized state dynamics ($\tilde{X}$-$\tilde{Z}$) for Less Effective (LE, dotted), Baseline (BE, solid), and Highly Effective (HE, dashed) NIV effectiveness levels. Higher effectiveness improves overall outcomes and enhances system stability against internal perturbations.}
\label{fig:Supp_VaryingTheta} 
 \end{figure}

\begin{figure}[H] 
    \centering
    \includegraphics[width=\textwidth]{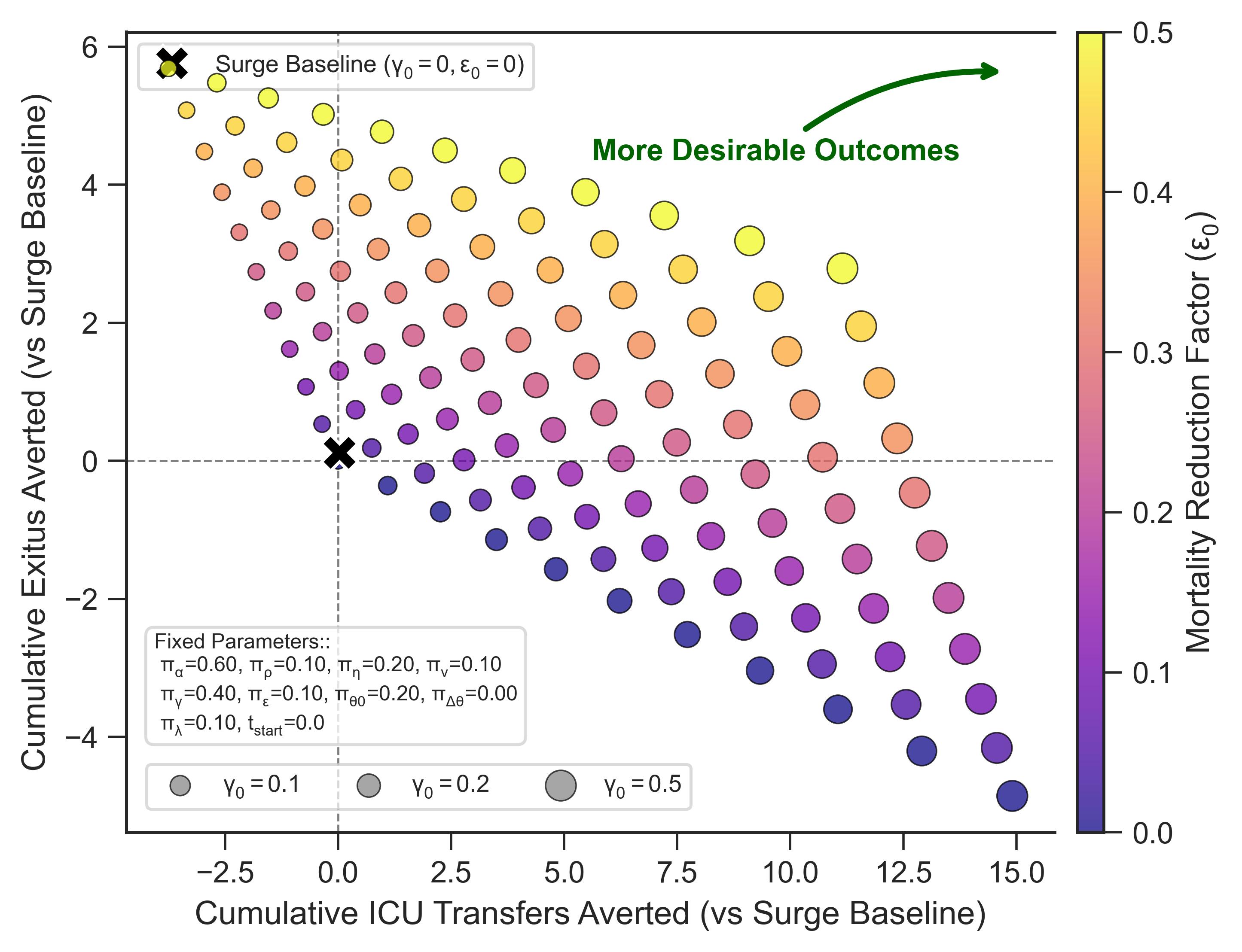} 
\caption{\textbf{Tradeoffs Between Averting ICU Transfers and Exitus via Combined Interventions During a Surge.}
Efficiency landscape plotting cumulative ICU transfers averted versus cumulative exitus averted (relative to baseline surge 'X') when simultaneously varying ICU mitigation effectiveness ($\gamma_0$, point size, 0-0.5) and mortality reduction effectiveness ($\varepsilon_0$, color, 0-0.5). The frontier illustrates achievable outcomes and the inherent tradeoffs for resource prioritization.}
\label{fig:Supp_Tradeoffs} 
 \end{figure}

We also assessed system resilience to internal fluctuations by simulating a time-varying *NIV recovery rate* ($\pi_\theta(t)$, assuming this is Panel A's variable based on parameters) under constant admission ($\pi_A=1$) (Figure~\ref{fig:Supp_VaryingTheta}). Comparing LE, BE, and HE effectiveness levels reveals that higher effectiveness not only yields consistently better outcomes (lower $\tilde{W}, \tilde{Z}$; higher $\tilde{R}$) but also enhances system stability. The HE scenario shows dampened oscillations in NIV occupancy ($\tilde{Y}$, Panel C) and mitigated outcome disruptions following the transient change in $\pi_\theta(t)$. This suggests that high-quality NIV care, influenced by clinical expertise, technology, and management strategies \cite{Esquinas2010, Esquinas2016, Esquinas2023}, is critical not only for managing external load but also for maintaining robust performance despite internal system variability or policy shifts (e.g., evolving ICU admission criteria).
Finally, recognizing that interventions during surges often involve resource tradeoffs, we simulated the combined effect of simultaneously varying ICU transfer mitigation ($\gamma_0$, 0 to 0.5) and NIV mortality reduction ($\varepsilon_0$, 0 to 0.5) effectiveness under admission surge conditions (Figure~\ref{fig:Supp_Tradeoffs}). Plotting the resulting cumulative ICU transfers averted versus cumulative exitus averted reveals the achievable efficiency landscape. The distribution forms a Pareto-like frontier \cite{Deb2005}, visually representing the tradeoff: prioritizing $\gamma_0$ (larger points) maximally averts ICU transfers, while prioritizing $\varepsilon_0$ (brighter colors) maximally averts exitus events. This quantitative tradeoff analysis provides a valuable tool for supporting evidence-informed strategic decision-making \cite{Brandeau2004}. Depending on the primary constraint or objective during a crisis (e.g., preserving ICU capacity vs. minimizing mortality), IRCU leadership can use such insights to target quality improvement efforts or allocate resources more effectively towards protocols impacting either $\gamma_0$ or $\varepsilon_0$ \cite{Kahraman2018}.


\section{Detailed Description of the LOS/Convolution Occupancy Model.} 
\label{sec:supp_los_details}

Accurate forecasting of patient occupancy is essential for effective resource management within Intermediate Respiratory Care Units (IRCUs), particularly given the inherent variability in patient demand and discharge patterns commonly observed in healthcare settings \cite{litvak2011, Hall2013}. While traditional queuing models offer analytical tractability, their reliance on strong parametric assumptions (e.g., exponential LOS, Poisson arrivals) often limits their ability to capture the complexities of real-world systems, such as non-standard LOS distributions or time-varying admission rates \cite{Hall2013, jun1999, fone2003, Jacobson2006, griffiths2006}. To overcome these limitations, we developed a data-driven, probabilistic modeling framework for IRCU occupancy grounded in empirical observations. This approach integrates a non-parametric representation of the Length-of-Stay (LOS) distribution with flexible, stochastic modeling of the admission process using Gaussian Processes (GPs), allowing for robust occupancy prediction and uncertainty quantification.

\subsection{Occupancy Calculation from Empirical LOS and Admissions}

The model fundamentally links daily occupancy to past admissions and the duration patients stay. Let $d$ be the LOS in days. Rather than assuming a specific parametric form, we directly characterize the LOS distribution using the empirical Probability Mass Function (PMF) derived from observed patient data:
\begin{equation}
f(d) \;=\; \frac{\text{Number of patients with observed LOS exactly } d}{\text{Total number of observed patients}},
\label{eq:supp_pmf}
\end{equation}
calculated for all observed LOS durations $d$ up to a maximum $d_{\max}$. This non-parametric method preserves the potentially complex shape (e.g., multimodality, heavy tails) of the observed LOS distribution \cite{Murray2002, Heffernan2005}. 

From the PMF, the corresponding empirical survival function \(S(d)\), representing the probability that a patient admitted on day $s$ is still present in the unit $d$ days later (i.e., $P(\text{LOS} \ge d)$), is calculated as:
\begin{equation}
S(d) \;=\; 1 - \sum_{k=0}^{d-1} f(k).
\label{eq:supp_survival}
\end{equation}
Note that $S(0) = 1$ by definition.

Let \(A(t)\) be the number of new admissions to the IRCU on day \(t\). The total IRCU occupancy \(X(t)\) on day \(t\) represents the accumulation of patients admitted on or before day $t$ who have not yet been discharged. This relationship is formalized through the discrete convolution of the admission history with the survival function:
\begin{equation}
X(t) \;=\; \sum_{s=0}^{t} A(s)\,S\bigl(t-s\bigr).
\label{eq:supp_occupancy_conv}
\end{equation}
Under steady-state conditions where the admission rate $A(t)$ is constant ($A(t)=A$), this convolution leads to the expected equilibrium occupancy $X = A \times E[\text{LOS}]$, where $E[\text{LOS}]$ is the mean LOS calculated from $f(d)$, linking the model to fundamental queuing principles \cite{Ross2014, Diekmann2013}. 

Alternatively, the occupancy dynamics can be expressed recursively via a daily flow balance:
\begin{align}
X(t+1) &= X(t) + A(t+1) - L(t+1), \label{eq:supp_occupancy_recursion} \\
L(t+1) &= \sum_{d=1}^{d_{\max}} f(d)\,A(t+1-d)\,\mathbb{1}\{t+1-d \ge 0\}, \label{eq:supp_discharge}
\end{align}
where \(A(t+1)\) are the new admissions during day \(t+1\), \(L(t+1)\) are the discharges during day \(t+1\) (calculated as the sum of patients admitted $d$ days prior who are discharged on day $t+1$, determined by the PMF $f(d)$), and \(X(t)\) is the occupancy at the start of day \(t\). The indicator function \(\mathbb{1}\{\cdot\}\) ensures causality.

\subsection{Stochastic Admissions Modeling using Gaussian Processes.}

Recognizing that daily hospital admissions \(A(t)\) are often complex and time-varying, we model them explicitly as draws from a Gaussian Process (GP), a flexible, non-parametric statistical tool suitable for time series \cite{rasmussen2005}. Formally, we represent the admission on day $t$ as:
\begin{equation}
A(t) \; \sim \; \mathcal{GP}\bigl(m(t), k(t, t')\bigr),
\label{eq:supp_admissions_gp} 
\end{equation}
where $m(t)$ is the mean function (often set to zero or a simple trend a priori) and $k(t, t')$ is the covariance or kernel function, which defines the structure and smoothness of the admission patterns. The GP framework inherently treats the underlying admission function as a stochastic process, providing not only a predictive mean $\mu_A(t)$ after conditioning on observed data, but also principled uncertainty quantification via a predictive variance $\sigma_A^2(t)$.

The kernel function $k(t, t')$ specifies the correlation between admissions at different times $t$ and $t'$. A common initial choice for capturing smooth trends is the Radial Basis Function (RBF) kernel combined with a White Noise kernel to account for independent observation uncertainty:
\begin{equation}
k(t,t') = C \cdot \exp\!\Bigl(-\frac{(t-t')^2}{2\ell^2}\Bigr) + \sigma_n^2\,\delta_{t,t'},
\label{eq:supp_kernel} 
\end{equation}
where $C$ is the amplitude variance, $\ell$ is the characteristic length-scale governing how quickly correlations decay with time separation, and $\sigma_n^2$ is the noise variance. These hyperparameters (collectively $\theta = \{C, \ell, \sigma_n^2\}$) are typically learned from observed admissions data $\mathcal{D} = \{(t_i, A_i)\}$ by maximizing the marginal likelihood.

To better capture specific temporal patterns often present in healthcare admissions, such as weekly seasonality, composite kernels can be constructed by summing basic kernels \cite{Roberts2013}. For example, incorporating a periodic kernel allows explicit modeling of recurring weekly patterns ($T=7$ days):
\begin{equation}
    k_{\text{comp}}(t,t') = C_1 \cdot \text{RBF}(t,t') + C_2 \cdot \text{Periodic}(t,t'; T=7) + \sigma_n^2\delta_{t,t'},
    \label{eq:supp_composite_kernel} 
\end{equation}
where $C_1, C_2,$ and parameters within the Periodic kernel (like its own length-scale) are also learned from data. Further enhancements for capturing complex dependencies can involve alternative kernel choices or non-zero structured mean functions $m(t)$, potentially informed by traditional time-series models \cite{Roberts2013}.

By fitting the chosen GP model (i.e., selecting a kernel and optimizing hyperparameters) to historical admission data, we obtain a posterior distribution over the admission function $A(t)$. Sampling trajectories from this posterior and simulating the occupancy dynamics using Eq.~\eqref{eq:supp_occupancy_conv} or \eqref{eq:supp_occupancy_recursion} for each sample allows the propagation of admission uncertainty through to the occupancy predictions, yielding probabilistic forecasts \(X(t)\) with associated credible intervals. This integrated approach leverages the empirical LOS distribution and flexible GP admission modeling to generate occupancy forecasts reflecting system nonlinearities and uncertainties, crucial for robust resource planning \cite{Whitt2019}. Furthermore, the framework allows for potential integration with real-time data streams and adaptive methods (e.g., Bayesian updating, AI techniques) to dynamically refine \(f(d)\) or GP predictions as new information becomes available \cite{Banks1989}, enhancing its practical utility in dynamic healthcare environments.

%
\section{ODE Model Parameter Estimation.} 
\label{sec:supp_param_estimation}

The baseline parameters for the ODE model (Eqs.~\eqref{eq:IRCU_admissions}-\eqref{eq:Recovered}), representing average transition rates [days\(^{-1}\)], were calibrated to reflect the observed dynamics of the UHVN IRCU cohort (n=249).

\textbf{Methodology:} The estimation combined observed outcome proportions with assumed average lengths of stay (\(\tau\)) in the origin states, using the relationship: **Rate \(\approx\) Proportion / \(\tau\)**.
\begin{enumerate}[label=\textbf{\arabic*.}]
    \item \textbf{Structural Constraints:} Based on clinical observations (Main Paper Fig.~2B), transitions from the non-NIV state (X) directly to ICU (Z) or Exitus (W) were non-existent in this cohort. Thus, \(\eta = 0\) and \(\nu = 0\) were fixed. Intervention modulators \(\gamma_0, \varepsilon_0\) were set to 0 for baseline estimation.
    \item \textbf{Outcome Proportions Calculation:} Proportions were calculated directly from the patient cohort data:
        \begin{itemize}[leftmargin=2em, itemsep=0pt]
            \item From State X (N=249):
                \( P_{X \to Y} = 77/249 \approx 0.3092\) (Initiated NIV),
                \(P_{X \to R} = 172/249 \approx 0.6908\) (Recovered without NIV).
            \item From State Y (n=77 initiated NIV):
                \(P_{Y \to Z} = 18/77 \approx 0.2338\) (Transferred to ICU),
                \( P_{Y \to W} = 7/77 \approx 0.0909\) (Exitus, no prior ICU transfer),
                \( P_{Y \to R} = 52/77 \approx 0.6753\) (Recovered from NIV).
        \end{itemize}
    \item \textbf{Assumed LOS for Rate Calculation:} Lacking granular time-in-state data, we used assumed average LOS values as the characteristic time \(\tau\) for transitions, consistent with parameters used in the simulation analyses:
        \(\tau_X = 7\) days (Avg. time in State X before transition), \(\tau_Y = 10\) days (Avg. time in State Y before transition). This is a necessary simplifying assumption for this modeling approach.
    \item \textbf{Rate Calculation:}
        \(\alpha \approx P_{X \to Y} / \tau_X \approx 0.3092 / 7 \approx 0.04418\) days\(^{-1}\), \(\rho \approx P_{X \to R} / \tau_X \approx 0.6908 / 7 \approx 0.09868\) days\(^{-1}\), \(\gamma \approx P_{Y \to Z} / \tau_Y \approx 0.2338 / 10 = 0.02338\) days\(^{-1}\), \(\quad \varepsilon \approx P_{Y \to W} / \tau_Y \approx 0.0909 / 10 \approx 0.00909\) days\(^{-1}\), \(\theta_0 \approx P_{Y \to R} / \tau_Y \approx 0.6753 / 10 \approx 0.06753\) days\(^{-1}\).
\end{enumerate}

The resulting baseline parameters, consistent with total outflow rates \((\alpha+\rho) \approx 1/\tau_X\) and \((\gamma+\varepsilon+\theta_0) \approx 1/\tau_Y\), are summarized in Table~\ref{tab:ode_parameters_supp}.

\begin{table}[H]
  \centering
  \caption{Baseline ODE model parameters estimated from cohort data and analysis assumptions.} 
  \label{tab:ode_parameters_supp}
  \begin{tabular}{@{}cll@{}}
    \toprule
    \textbf{Parameter} & \textbf{Description} & \textbf{Estimated Value [days\(^{-1}\)]} \\
    \midrule
    \( \alpha \) & Rate X \(\to\) Y (NIV Initiation) & 0.04418 \\
    \( \rho \)   & Rate X \(\to\) R (Direct Recovery) & 0.09868 \\
    \( \eta \)   & Rate X \(\to\) Z (Direct ICU)     & 0 (Fixed by observation) \\
    \( \nu \)    & Rate X \(\to\) W (Direct Exitus)  & 0 (Fixed by observation) \\
    \( \gamma \) & Rate Y \(\to\) Z (ICU from NIV)   & 0.02338 \\
    \( \varepsilon \) & Rate Y \(\to\) W (Exitus from NIV) & 0.00909 \\
    \( \theta_0 \) & Rate Y \(\to\) R (Recovery from NIV) & 0.06753 \\
    \midrule
    \multicolumn{3}{@{}l}{\textit{Parameters assumed for estimation:}} \\ 
    \( \tau_X \) & Assumed Avg. LOS in State X       & 7 days \\
    \( \tau_Y \) & Assumed Avg. LOS in State Y       & 10 days \\
    \bottomrule
  \end{tabular}
  \vspace{0.5ex}
  \parbox{\linewidth}{\footnotesize Note: These are baseline estimates for the autonomous model. For simulations involving time-varying recovery (e.g., Main Paper Fig.~3G-I, J-L), the rate \(\theta(t)\) was modified as described. Intervention modulators \(\gamma_0, \varepsilon_0\) were set to 0 for baseline.}
\end{table}

\section{Appendix: Model Variables and Parameters.}

\begin{table}[H]
  \centering
  \caption{Summary of the model's dynamic state variables.}
  \label{tab:dynamic_variables_revised} 
  \begin{tabular}{@{}ll@{}} 
    \toprule
    \textbf{Variable} & \textbf{Description [Unit]} \\
    \midrule
    \( X(t) \) & Patients in IRCU, not receiving NIV at time \( t \) [patients] \\
    \( Y(t) \) & Patients in IRCU, actively receiving NIV at time \( t \) [patients] \\
    \( Z(t) \) & Patients transferred from IRCU pathway to ICU by time \( t \) [patients] \\
    \( W(t) \) & Exitus from IRCU pathway (X or Y) by time \( t \) [patients] \\
    \( R(t) \) & Recovered (discharged/to ward) from IRCU pathway (X or Y) by time \( t \) [patients] \\
    \bottomrule
  \end{tabular}
\end{table}

\begin{table}[H]
  \centering
  \caption{Summary of the model parameters including intervention modulators.}
  \label{tab:parameters_revised} 
  \begin{tabular}{@{}ll@{}} 
    \toprule
    \textbf{Parameter} & \textbf{Description [Unit]} \\
    \midrule
    \multicolumn{2}{@{}l}{\textit{Admission}} \\
    \cdashline{1-2}
    \( A(t) \) & Influx rate of new patient admissions to IRCU [patients/time] \\[.5ex] 
\cline{1-2}
    \multicolumn{2}{@{}l}{\textit{Transitions from Non-NIV State (X)}} \\
    \cdashline{1-2}
    \( \alpha \) & Rate of NIV initiation (X \(\to\) Y) [time\(^{-1}\)] \\
    \( \rho \) & Direct recovery/discharge rate (from X) [time\(^{-1}\)] \\
    \( \eta \) & Direct ICU transfer rate (from X) [time\(^{-1}\)] \\
    \( \nu \) & Direct exitus rate (from X) [time\(^{-1}\)] \\[.5ex]
\cline{1-2}
    \multicolumn{2}{@{}l}{\textit{Baseline Transitions from NIV State (Y)}} \\
    \cdashline{1-2}
    \( \gamma \) & Baseline rate of ICU transfer (from Y) [time\(^{-1}\)] \\
    \( \varepsilon \) & Baseline rate of exitus (from Y) [time\(^{-1}\)] \\
    \( \theta_0 \) & Baseline rate of recovery (from Y) [time\(^{-1}\)] \\[.5ex]
\cline{1-2}
    \multicolumn{2}{@{}l}{\textit{Intervention Modulators (acting on Y transitions)}} \\
    \cdashline{1-2}
    \( \gamma_0 \) & ICU transfer mitigation factor (effectiveness) [dimensionless] \\
    \( \varepsilon_0 \) & Exitus reduction factor (effectiveness) [dimensionless] \\
    \( \Delta\theta \) & Initial magnitude of transient recovery enhancement [time\(^{-1}\)] \\
    \( \lambda \) & Decay rate of transient recovery enhancement [time\(^{-1}\)] \\
    \bottomrule
  \end{tabular}
\end{table}

\printbibliography